\documentclass[fleqn,10pt]{wlscirep}
\usepackage[utf8]{inputenc}
\usepackage[T1]{fontenc}
\usepackage{subcaption} %for multiple captions in single figure
\usepackage{placeins} %for \FloatBarrier
\usepackage{threeparttable} %for footnotes in tables
\usepackage{mdframed} %for making boxes
\usepackage{amsthm}

\mdfdefinestyle{normaltextstyle}{
    font=\normalfont,
    frametitlefont=\bfseries,
    nobreak=true % This prevents the box from splitting across pages
}
\theoremstyle{definition}
\newmdtheoremenv[style=normaltextstyle]{methodbox}{Box}

\title{Aging health dynamics cross a tipping point near age 75}

\author[1,2,*]{Glen Pridham}
\author[3,$\ddagger$]{Kenneth Rockwood}
\author[1,$\dagger$]{Andrew Rutenberg}

\affil[1]{{Department of Physics and Atmospheric Science}, {Dalhousie University}, {Halifax}, {B3H 4R2}, {Nova Scotia}, {Canada}}
\affil[2]{{Department of Molecular Cell Biology}, {Weizmann Institute of Science}, {Rehovot}, {Israel}} %{7610001},
\affil[3]{{Division of Geriatric Medicine}, {Dalhousie University}, {Halifax}, {B3H 2E1}, {Nova Scotia}, {Canada}}

\affil[*]{glen.pridham@dal.ca}
\affil[$\dagger$]{adr@dal.ca}


\begin{abstract}
Aging includes both continuous gradual decline, such as in physiological function, together with major deficit onset events such as morbidity, disability and ultimately death. 
These deficit events are stochastic and include non-linear feedbacks, making health trajectory forecasting challenging.
We propose a framework for modelling the gradual effects of aging together with health deficit onset events, as reflected in the frailty index (FI) -- a quantitative measure of overall age-related health.
We model damage and repair dynamics of the FI from individual health transitions within two large longitudinal studies of aging health, the Health and Retirement Study (HRS) and the English Longitudinal Study of Ageing (ELSA), which together included $N=47592$ individuals. We find that both damage resistance (robustness) and damage recovery (resilience) rates decline smoothly with both increasing age and with increasing FI, for both sexes. This leads to two distinct dynamical states: a robust and resilient young state of stable good health (low FI) and an older state that drifts towards poor health (high FI). These two health states are separated by a sharp transition near age~75. Since FI accumulation risk accelerates dramatically across this tipping point, ages 70-80 are crucial for understanding and forecasting late-life decline in health.
\end{abstract}
\begin{document}

\flushbottom
\maketitle 

%target journals:
% nature aging
%nature communications limits: 5000 words, 10 figures
%science advances limits: 15000 words, 10 figures
%npj Aging
%GeroScience

%%%%%%%%%%%%%%%%%%%%%%%%
\section*{Introduction}
%%%%%%%%%%%%%%%%%%%%%%%%%%%%%%%%% updated intro %%%%%%%%%%%%%%%%%%%
%how can aging be both?
%do they feed back?
%how do trajectories respond to gradual damage accumulation
%we have a model to understand surprising consequences of basic realities of the dynamics
Aging is a dynamical process \cite{Rattan2013-ze, Fried2021-vh, Cohen2022-gt}, characterized by both gradual decline and sudden stochastic transitions\cite{Mitnitski2007-pf}. 
Continuous decline is seen across physiology including nerve conduction speed and maximum heart rate\cite{Alon2023-lu}, apparently due to a gradual accumulation of damage from many cellular and molecular aging events \cite{Lopez-Otin2023-xw,Tarkhov2022-jd}. This is punctuated by rare major health deficit onset events: disabilities, chronic diseases, and ultimately death\cite{Fried2021-vh,Juster2010-kw}. 
These major events become much more likely with age \cite{Zenin2019-fu,Guay2014-rs}, and can feedback to increase the risk of further adverse events\cite{Mitnitski2015-ia}. Lacking is a dynamical framework for understanding organism-level health trajectories that captures these complexities of aging biology: continuous decline, non-linear feedbacks and stochastic health transitions\cite{Cohen2022-gt}.
%Without such a framework, attempts to bridge cellular and molecular drivers of aging to overall, organism-level health can be misleading, as stochasticity and feedback should be explicitly modelled to avoid errors in causal inference\cite{Cohen2022-gt}.
Such a framework for typical age-related health trajectories has the potential to enhance public health forecasting and individualized risk management by incorporating these complexities.
%by incorporating our growing understanding of the dynamical behaviour of aging\cite{Cohen2022-gt}. 
%Aging biology is known to include strong stochasticity and non-linear feedbacks\cite{Cohen2022-gt} that can create surprising dynamical behaviours not captured by standard statistical models such as regression and linear latent growth curves. 
%It would also be a first step towards connecting our growing understanding of cellular and molecular drivers of aging\cite{Lopez-Otin2023-xw} to organism-level health.
%, while offering a plausible bridge to connect health to the underlying cellular and molecular drivers of aging.
%while respecting the complex dynamics involved including strong stochasticity and feedback\cite{Cohen2022-gt}
%Unfortunately, such dynamical models of health trajectories are not well explored.
%\cite{Gladyshev2024-ed} 

%FI as summary health measure
Health can be quantified by summary ``functional'' (organism-level) measures. These are important to understand since they affect healthcare needs, drive mortality, and affect quality of life \cite{Hoogendijk2019-pk}. Here we focus on the frailty index (FI), which is defined as the average fraction of binary, age-related clinical measures that are deficit, including signs, symptoms, limitations and diagnoses  \cite{Searle2008-xi, Howlett2021-my}. The FI is an efficient representation for a battery of clinical measures \cite{Pridham2023}, is an excellent predictor of further adverse outcomes including hospitalization and death \cite{Clegg2016}, and is largely robust across studies and choice of measurement battery used \cite{Theou2023-aw}. The FI has growing clinical applications and is currently used for screening and risk stratification\cite{Kim2024-wx}, making it an important potential translational measure for public health and individualized medicine\cite{Howlett2021-my}. 
%This gives the FI strong translational potential from basic research to both public health and individualized medicine \cite{Howlett2021-my}. 
%Understanding the dynamical behaviour of the FI offers a quantitative framework for describing normative aging at the organism level, with potential applications in public health forecasting and individualized risk management.
%Deficit onset is stochastic, as is the FI  \cite{Rockwood2004, Gu2009}. 

%rephrased in medical jargon
%introduce frailty
Notably, the FI can be used to identifying frail individuals\cite{Kim2024-wx}. Frailty is a state of increased vulnerability to adverse health outcomes\cite{Fried2021-vh,Kim2024-wx} that grows exponentially in prevalence with age, exceeding 30\% around age~80.\cite{Hoogendijk2019-pk} Frail individuals can be identified by a cutoff, e.g.\ FI > 0.2, or treated as a continuous gradation using the FI directly\cite{Kim2024-wx, Howlett2021-my}.
We aim to provide a framework for accurately forecasting the FI, opening up the potential to anticipate public health needs as cohorts become increasingly frail with age. 
%Fried and colleagues have argued that both dysregulation of stress-response systems due to aging \cite{Fried2021-vh} and feedback between health deficits \cite{Fried2001-mr} provide the necessary conditions for a frail state.
%Conventional theories of frailty focused on the damage-promotes-damage paradigm\cite{Fried2001-mr,Mitnitski2015-ia}, that has subsequently expanded to consider the role of age-related changes\cite{Fried2021-vh,Kim2024-wx}.
%Lacking is a clear framework that describes when and how continuous decline from underlying aging processes create the conditions for such stochastic transitions, that is essential for predicting age-related health trajectories.
%Understanding when and how continuous decline from underlying aging processes create the conditions for such stochastic transitions is essential for predicting age-related health trajectories.
%Fried and colleagues have argued that both dysregulation of stress-response due to aging \cite{Fried2021-vh} and feedback between health deficits \cite{Fried2001-mr} provide a positive feedback that can push health into a frail state. 
%In turn, these trajectories help identify at-risk individuals (cite) and public health demands (cite).

%background on mathemathical models %gap
An important conceptual model of aging is positive feedback in the dynamics of damage accumulation \cite{Ogrodnik2018, Belikov2019}. Deficit accumulation enhances the rate of further deficit accumulation\cite{Mitnitski2015-ia, Taneja2016-bj}, i.e.\ damage-promotes-damage. The effect is generic but reflects real underlying biology, for example diabetes increases risk of macular degeneration \cite{Chen2014-xo} and poor eyesight increases risk of falls causing injury \cite{Wieczorek2024-ts}. Such positive feedback is suggested by the approximately exponential accumulation of FI with age \cite{Mitnitski2016-xc}. Indeed, dynamical models of health decline that include positive feedback but no explicit age dependence, describe the nonlinear accumulation of deficits well (see e.g. \cite{Avchaciov2022, Rutenberg2018}). Nevertheless, models of deficit accumulation that are \emph{only} age dependent can also describe frailty accumulation \cite{Farrell2022-di}. 
While these contrasting models can both describe health trajectories, they leave a gap in understanding the role of continuous decline due to age vs feedback from major stochastic health events as captured by FI. 
We address this gap by directly modelling health transition rates as functions of both age and number of accumulated deficits (FI). 

%what we did
%introduce damage and repair
%mention robustness and resilience
The FI serves dual purposes since it is both an overall measure of health and is composed of binary deficits that can be directly modelled using longitudinal data. Here we directly model stochastic transition rates. This includes damage (from healthy to deficit), repair (from deficit to healthy), and mortality. A convenient nomenclature is to describe the ability to resist damage as robustness and the ability to repair as resilience \cite{Ukraintseva2016-ca}. We model both damage/robustness and repair/resilience rates as smooth functions of age and FI (health state). Thus we simultaneously model rates of stochastic events, continuous decline with age, and health state feedback from the FI.

%specific of what we did
We analyzed large-scale human longitudinal health deficit data from the Health and Retirement study (HRS) and the English Longitudinal Study of Ageing (ELSA). We fit and select joint models of damage, repair, and mortality, and simulate them to confirm they reproduce realistic population-level behaviour. We find that individuals pass from a robust and resilient dynamical phase of good health, prior to approximately age~75, to a dynamical phase of accumulating health deficits after age~75. Thus there appears to be a tipping point in aging biology beyond which the average person can no longer maintain a stable state of health, leading to a sudden increase in the risk of accumulating health deficits.
%These distinct dynamical phases correspond to a rapid change of stability of the underlying health states, from healthy states being stable to unhealthy states being stable. 

%%%%%%%%%%%%%%%%%%%
\section*{Model}
The FI is the average number of deficit health attributes, therefore it can only increase when undamaged attributes are damaged and can only decrease when damaged attributes are repaired. This constrains the FI and its dynamical behaviour, e.g.\ restricting it to the interval [0,1]. We employ the most general model that respects these constraints to describe the mean dynamics of the FI, $f$, as a smooth function of age ($t$) and health feedback ($f$),
\begin{align}
        \frac{df}{dt} &=  (1-f)D(f,t)   -  f R(f,t), \nonumber \\
 i.e. \;\;       \text{FI velocity} &= (\text{undamaged attributes})\cdot \text{damage rate} -  (\text{damaged attributes})\cdot \text{repair rate}, \label{eq:fi}
\end{align}
where $D(f,t)$ and $R(f,t)$ are the damage and repair rates, respectively. Underlying Eq.~\ref{eq:fi} is a stochastic process of damage and repair of individual binary health attributes. Directly modelling $D$, $R$, and the survival hazard $h$ using time-to-event statistics allows us derive a log-likelihood objective function for fitting (see Methods and supplemental). A ``mean-field'' approximation of this objective function recovers Eq.~\ref{eq:fi} (see supplemental), and this becomes exact for a large number of attributes. The prefactors on the damage and repair rates, which represent the undamaged and damaged fraction of deficits, ensure that the model is not stuck at $0$ and cannot exceed $1$. This is in contrast to the classical exponential model \cite{Mitnitski2016-xc} ($df/dt=\alpha f; \alpha > 0$). The specific damage and repair functions are determined during model selection.

Prior related works using Poisson modelling suggest that transition rates should vary smoothly and approximately log-linearly with age \cite{Farrell2022-di}, and that a log-linear function of $f$ fits well \cite{Mitnitski2006-fq}. Those works neglected to analyze the dynamical behaviour of the FI as a function of both age and health, which we address here. Using a log-linear model is also convenient to ensure non-negative rates. We thus investigated rates with the general form
\begin{align}
    \ln{(\Gamma)} &= \gamma_0 + \gamma_f f + \gamma_t t + \gamma_{ft} ft + \gamma_{f^2} f^2
    \label{eq:Gamma}
\end{align}
where $\Gamma$ is a generic rate that can be used to describe each of damage, $D$, repair, $R$, or survival hazard, $h$ -- with appropriate parameters for each.  

%Observe that the model has a nullcline where the velocity $df/dt=0$ that corresponds to a balancing of damage and repair,
%\begin{align}
%        (1-f^*)D(f^*,t^*)   &= f^*R(f^*,t^*)
%\end{align}
%for some specific $f^*$ and $t^*$.

%%%%%%%%%%%%%%%%%%
\section*{Data}
We fit to longitudinal health and survival data. Health attributes are binary with 0 indicating normal and 1 indicating deficit (unhealthy). Health attributes are based on questionnaire data. Survey weights were not used. The attribute sets used are from previous publications \cite{Theou2023-aw,Rogers2017-ke} with minor modifications as described in the supplemental. For HRS, we used waves~6-15 (2002-2020) via the RAND preprocessed files \cite{hrs}. We included 34672~individuals with 189096~visits; median age: 66.8 (inter-quartile range: 58.7-76.0). Waves are measured every 2 years for both HRS and ELSA. For ELSA we used waves~1-8 (2002-2016) \cite{elsa}. We included 12920~individuals with 65261~visits; median age: 67 (inter-quartile range: 60-74). ELSA survival estimates were based on end-of-life interviews, which capture only a fraction of the deaths due to a variety of response rate and fieldwork issues \cite{NatCen_Social_Research2015-nw}. This means that we underestimate the mortality rate for ELSA because we are forced to assume that any individual without an end-of-life interview was censored instead of dying (Supplemental Section~S4 demonstrates that the model fits ELSA, with a lower baseline hazard). We excluded individuals from ELSA above age~89 since those ages are top-coded, meaning that all individuals past age~89 at baseline are labeled as age~90.
% (waves~1-5 did not XXXXX)

%%%%%%%%%%%%%%%%%%%%%%%%%%
\section*{Methods}

\begin{figure*}[!ht]
\centering
\includegraphics[width=0.85\textwidth]{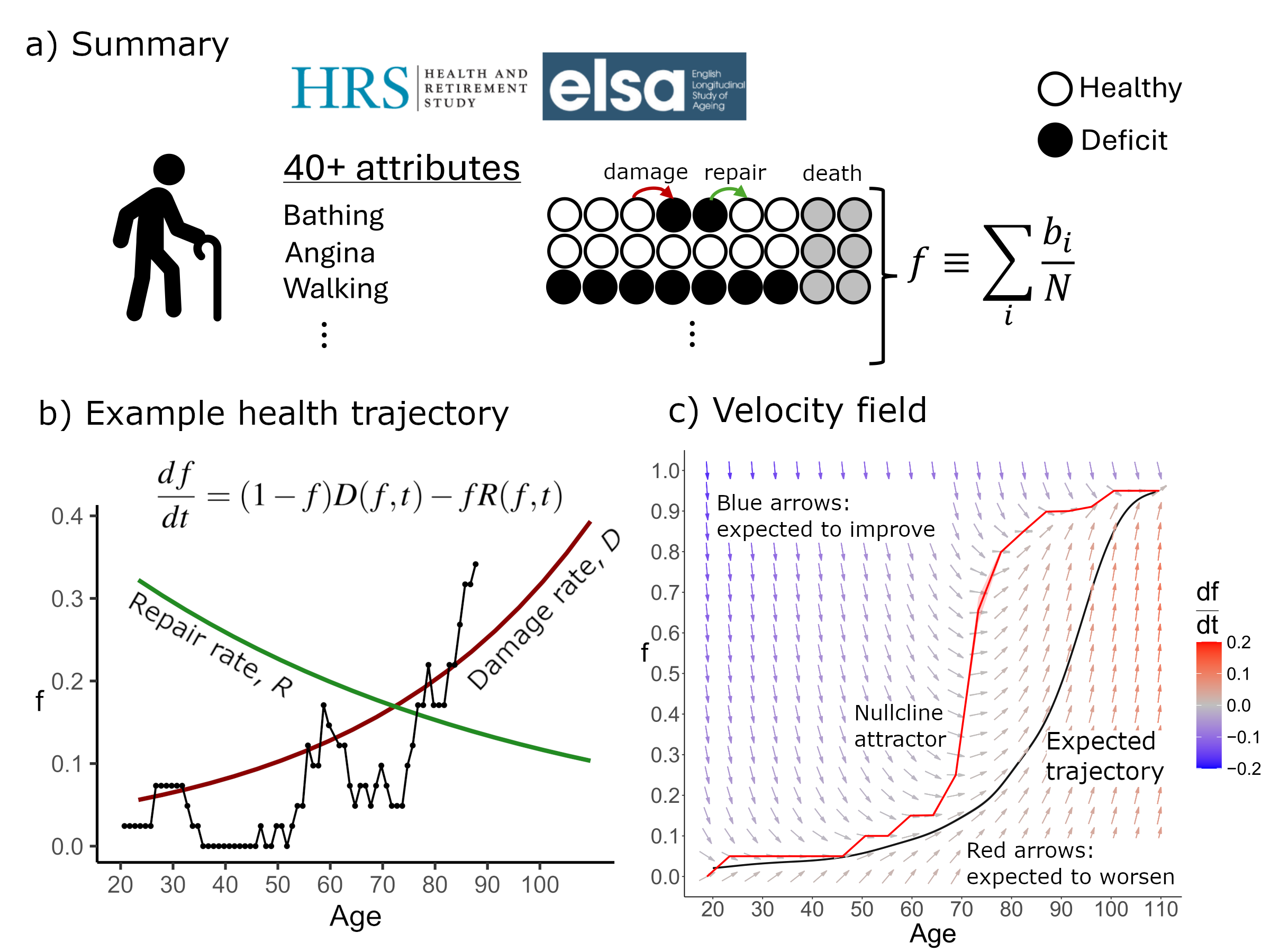} 
    \caption{\textbf{Study summary}. (a) we modelled health transition rates among a battery of health attributes: damage (healthy to deficit, $0\to1$), repair (deficit to healthy, $1\to0$), and death. We used two large scale longitudinal studies, HRS: 34672 individuals, 10 waves and ELSA: 12920 individuals, 8 waves. (b) individual health is summarized using the frailty index, $f$. Damage (red) and repair (green) rates are allowed to vary smoothly with age and $f$. Rates change continuously over time whereas individual FI change stochastically due to random events. (c) The velocity field summarizes the dynamical behaviour of our model. The velocity field describes the prevailing `wind' (slope vs time) for each individual's current pair of values (FI, age). The nullcline (red line) is an attractor that describes the stable position where the velocity is 0. The average individual trajectory (black line, without mortality) tends towards the nullcline, with a lag due to gradual deficit accumulation.} \label{fig:summary}
\end{figure*}

Figure~\ref{fig:summary} summarizes our approach. Box~\ref{box} provides key definitions.

%%%%%%%%%%%%%%%%%%%%%%%%%%%%% BOX %%%%%%%%%%%%%%%%%%%%%%%%%%%%%%%
%\begin{mdframed} 
\begin{methodbox}[Definitions]
\label{box}
%Add a short, plain-language Methods overview box:
    %what constitutes a damage/repair event; 
    %how rates depend on age and FI; 
    %what the velocity field and nullcline mean; 
    %what the tipping point means operationally
    %plus a worked example narrative.
\textbf{} 
\medskip \\
\textbf{Damage and repair events}\\
We consider clinical health attributes that are binarized as either healthy/non-deficit (0) or deficit (1), agnostic of the specifics of each attribute. A damage event is an observed health attribute that is deficit at the current measurement date after having been non-deficit at the previous measurement. A repair event is the reverse, when an observed attribute was deficit in the previous measurement date but is now non-deficit. Damage could represent a transient injury or a permanent event. Repair could represent, for example, healing or adaptation. Damage and repair could also capture fluctuations.  We use data sampled every 2~years and therefore cannot resolve changes faster than $\sim 1~\text{year}^{-1}$. \\
% We fit to event-free survival probabilities so that we don't need to know the exact event timing.  % seems better for supplemental technical 
%\medskip\\
%\textbf{We model event rates}\\
%We model event rates for damage, repair and mortality. A change in these rates translates into a change in FI velocity (slope), not FI value. This permits changes in rates to occur before an increase in FI. Rates are hypothesized to depend on robustness and resilience.
\\
\textbf{Robustness and resilience}\\
We refer to higher repair rate as higher resilience and lower damage rate as lower robustness.
%We hypothesized that changes to age-related health will robustness and resilience, rather than directly on the FI. This permits a delay between when a biological changes starts, such as a new risk factor, and when the FI increases, such as clinical emergence of a new disease.
\medskip\\
%\medskip\\
\textbf{Age and FI dependence}\\
We explored model classes where damage (robustness) and repair (resilience) change smoothly with increasing age or FI. Model selection indicated that both log-damage and log-repair are approximately linear in both age and FI.
\medskip\\
\textbf{Stochasticity}\\
We assume all (damage, repair and mortality) events occur randomly with frequencies determined by the estimated rates.
\medskip\\
\textbf{Velocity field} \\ 
The velocity field is used for visualizing how individual trajectories will evolve in the (FI,age) plane, e.g.\ Figure~\ref{fig:summary}c. The velocity field describes the slope of the FI at each point corresponding to the current value of the FI (y-axis) and the current age (x-axis). By analogy, the arrows indicate the direction of `wind' (slope vs time) pushing each individual's age and FI. 
\medskip\\
\textbf{Nullcline (i.e.\ setpoint)} \\ 
The nullcline is defined as the position where the FI velocity is exactly 0, meaning the arrows in the velocity field are pointing horizontally in the direction of increasing age (e.g., red line of Figure~\ref{fig:summary}c). The nullcline describes the FI velocity, not the FI directly. It can be used to forecast long-term behaviour since the FI will eventually move towards the nullcline, as it is an attractor where the FI (on average) will stop changing if it ever reaches that specific value in the (FI,age) plane. Points far from the nullcline typically change the fastest (nearly vertical velocity arrows). 
%Points far from the nullcline typically change the fastest (nearly vertical arrows). 
%The nullcline, delineated by the red line, is where the FI velocity is 0 -- and the arrows are pointing in the direction of increasing age.
%, meaning that the FI (on average) will stop increasing if it ever reaches that specific value in the (FI,age) plane
\medskip\\
\textbf{Tipping point}
A tipping point is a sudden change in system behaviour that occurs when a slowly varying control parameter (e.g. age) carries the system through a critical point of its underlying dynamics, and is typically associated with a fold (saddle–node)–type bifurcation \cite{strogatz2024nonlinear}. In our case, this is reflected by a sharp increase of the nullcline near age 75, where the FI velocity changes abruptly, marking a tipping point from a regime with a low-FI setpoint (stable good health) to a regime with a high-FI setpoint.

%A tipping point is a discontinuity where the nullcline suddenly increases and the system bifurcates into two qualitatively different behaviours. We observed that the nullcline increases sharply near age 75, meaning that the velocity suddenly increases at that age making it a ``tipping point'' where one regime ends (low FI nullcline) and another beings (high FI nullcline). 
%The exponential model does not have a tipping point, it has a continuously increasing slope.

%\medskip\\
%\textbf{Example narrative}  \\
%IDK
\end{methodbox}
%%%%%%%%%%%%%%%%%%%%%%%%%%%%% BOX %%%%%%%%%%%%%%%%%%%%%%%%%%%%%%%
All analyses used \texttt{R} version 4.1.1.\cite{R_Core_Team2021-uq} We fit our model using the Broyden–Fletcher–Goldfarb–Shanno (BFGS) quasi-Newton method \cite{R_Core_Team2021-uq} to numerically optimize the (survival-modified) log-likelihood. The log-likelihood, $l$, has five terms, capturing the four types of transitions in health (as indicated to the right in the following equation) together with mortality. 
\begin{align}
    l &\equiv \sum_{i=1}^N \sum_{j=1}^p\sum_{k=1}^{T-1} (1-\delta_{ik}) \bigg[  \nonumber \\
    &\phantom{+}b_{ijk}(1-b_{ijk-1})\ln{(1-S_d)} &\text{damaged ($0\to1$)}  \nonumber \\
    &+(1-b_{ijk})b_{ijk-1}\ln{(1-S_r)} &\text{repaired ($1\to0$)}\nonumber \\
    &+b_{ijk}b_{ijk-1}\ln{(S_r)} &\text{did not repair ($1\to1$)} \nonumber \\
    &+ (1-{b_{ijk}})(1-{b_{ijk-1}})\ln{(S_d))}\bigg] &\text{did not damage ($0\to0$)} \nonumber \\
    &+\sum_{i=1}^N \sum_{k=1}^{T-1} \bigg[\ln{(S(f_{ik}, f_{ik-1}, \Delta t_{ik}))}+\delta_{ik}\ln{(h(f_{ik},f_{ik-1},t_{ik}))} \bigg], &\text{survived or died at time $t_{ik}$} \label{eq:ll}
\end{align}
where $N$ is the number of individuals, $p$ is the number of variables, and $T$ is the number of time points. $b_{ijk}$ is the binary health attribute of individual $i$, attribute $j$ and time $k$. $\delta_{ik}$ is the binary mortality indicator for individual $i$ at time $k$ and is $1$ if the individual died at that time and $0$ otherwise. The $S$ are survival functions of the form, 
\begin{subequations}
\begin{align}
    S_d &\equiv \exp{\bigg( -\int_{t_{k-1}}^{t_{k}} D(f,t) dt \bigg)} %&\text{did not damage between times $t_{k-1}$ and $t_k$}
\end{align}
\begin{align}
    S_r &\equiv \exp{\bigg( -\int_{t_{k-1}}^{t_{k}} R(f,t) dt \bigg)} %&\text{did not repair between times $t_{k-1}$ and $t_k$}  
\end{align}
\begin{align}
    S &\equiv \exp{\bigg( -\int_{t_{k-1}}^{t_{k}} h(f,t) dt \bigg)}   %&\text{did not died between times $t_{k-1}$ and $t_k$}
\end{align}
\end{subequations}
where $D$, $R$ and $h$ are the damage, repair and mortality hazards, respectively. For example, $S_d$ is the probability of an attribute `surviving' damage and thus represents the (non-)transition of $0\to0$. Eq.~\ref{eq:ll} estimates the parameters of Eq.~\ref{eq:fi} in the limit of a large number of attributes. See supplemental for full details.
%for example, $S_d\equiv \exp{-\int_{t_{k-1}}^{t_k} D(f,t)dt}$ i.e.\ the probability of not damaging between times $t_{k-1}$ and $t_k$. 

When we compare models, all rates (damage, repair and survival) have the same parametric form given by Eq.~\ref{eq:Gamma} unless otherwise stated. The model likelihood must be solved numerically past linear order in $t$ (i.e. past Gompertz), hence we did not consider rates with quadratic $t^2$ dependence. We fit directly to longitudinal binary health attribute deficit data, together with survival. During model selection we determine which of the $\gamma_i$ are necessary to efficiently fit the data.
%Note that ELSA survival was based on end-of-life interviews and hence captures only a fraction of the deaths due to a variety of issues \cite{NatCen_Social_Research2015-nw}, resulting in a much lower baseline hazard. 

Missing values were uncommon. For the health attribute variables, 99.1\% (HRS) and 99.8\% (ELSA) of data were reported (available case), with 80.7\% (HRS) and 96.2\% (ELSA) of individuals having all variables measured (complete case). The least commonly measured variable that we included in the FI for HRS was difficulty climbing several flights of stairs (5.7\% missing) and for ELSA was self-reported general health (0.4\% missing). We fit to the available case data and simulated using the complete case data.

%%%%%%%%%%%%%%%%%%
\section*{Results}
\section*{Model selection leads to a (log-)linear model in both $f$ and $t$}
\begin{figure*}[!ht]
\centering
\includegraphics[width=\textwidth]{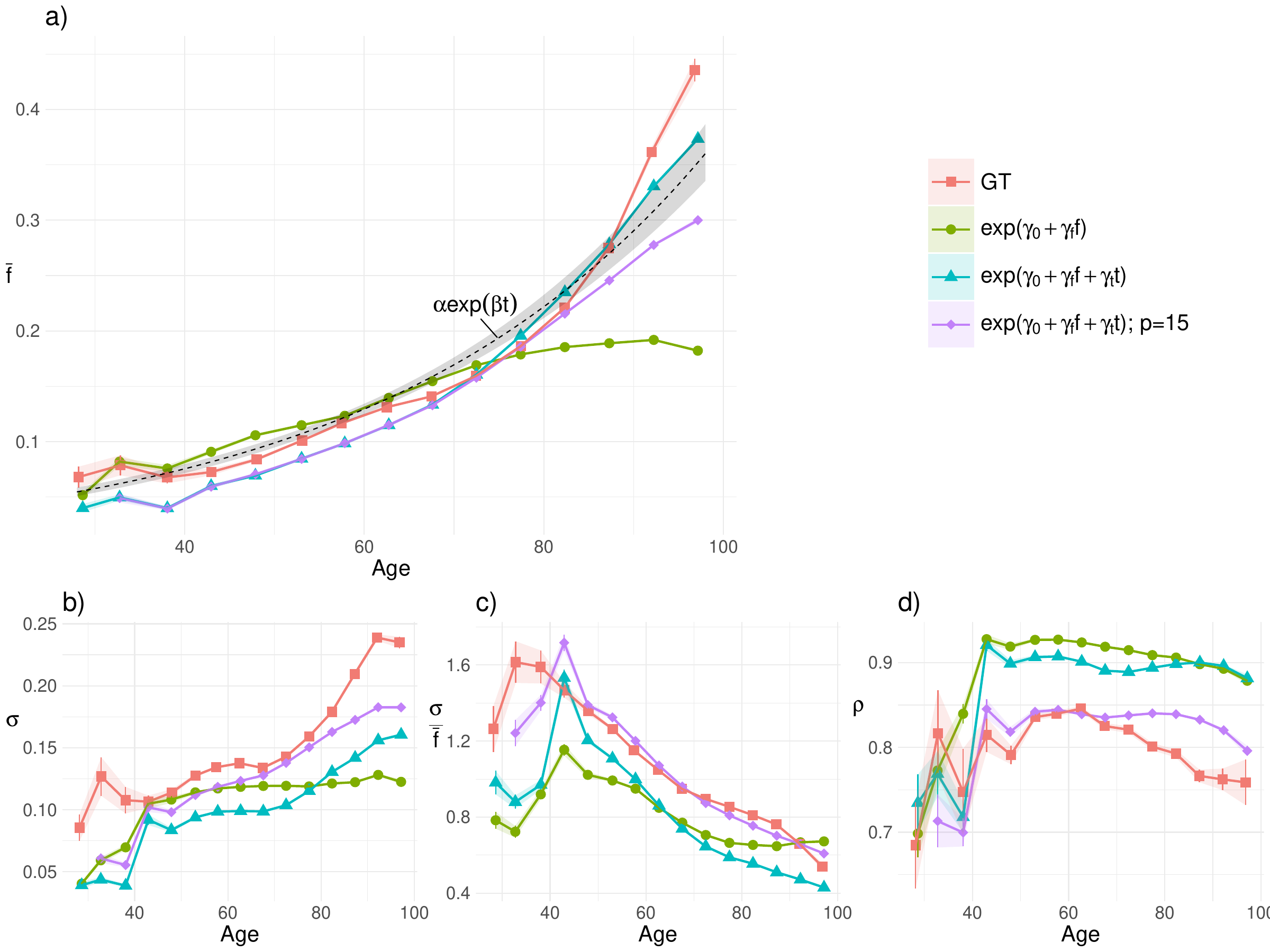} 
    \caption{\textbf{FI population-level statistics} show that a linear model including $f$ and $t$ qualitatively recapitulates the correct FI behaviour (HRS). Simulated models versus ground truth (GT). (a) mean FI, $\bar{f}$, (b) FI standard deviation, $\sigma$, (c), coefficient of variation $\sigma/\bar{f}$, and (d) FI autocorrelation, $\rho$ (lag-1). The linear model including both the FI, $f$, and age, $t$, (blue triangles) correctly captures the superlinear convexity of the mean FI (a), the complex shape of the standard deviation (b), the linearly decreasing CV past age 40 (c), and the roughly constant autocorrelation past age 40 (d). We can further improve the model fit by reducing the number of attributes from the default ($p=41$) to $p=15$ (purple diamonds), implying correlations between attributes are reducing the number of effective degrees of freedom \cite{Pridham2023}. Complete case data. Simulation was seeded with the first wave from GT. Error bars are standard errors (bootstrap, 100~repeats). Additional models in Supplemental Figure~S6. For ELSA see Supplemental Figure~S7.} \label{fig:modelphenom}
\end{figure*}

Model selection using either the Bayesian Information Criterion or the test log-likelihood showed that the optimal model complexity was linear dependence of both $f$ and $t$, see Supplemental Figure~S4. This linear model is  
\begin{subequations}
    \begin{align}
        D &= e^{d_0+d_ff+d_tt}
    \end{align}
    \begin{align}
        R &= e^{r_0+r_ff+r_tt}
    \end{align}
    \begin{align}
        h &= e^{h_0+h_ff+h_tt}
    \end{align}
\end{subequations}
for damage rate $D$, repair rate $R$, and mortality (hazard) rate $h$.
In Figure~\ref{fig:modelphenom} we show that the linear model (blue triangles) also visually fit the mean $f$ well, but other statistics only fit the qualitative trends -- as compared to the ground truth (GT, red squares). The primary culprit for this quantitative misfit appears to be our independence assumption. The $p=41$ attributes we used are strongly correlated \cite{Pridham2023} and thus the effective number of independent attributes should be smaller than 41. Reducing the number of independent attributes to $p=15$ greatly improved the visual agreement of the higher order statistics (purple diamonds, see Supplemental Figure~S9 for other $p$). Reducing $p$ did, however, depress the mean $f$ at higher ages since it increases the variance and hence the hazard at larger $f$ \cite{Pridham2024-kidney}. 
%The subsequent misfit in the mean is ostensibly due to the survival model which fits fine, but tends to be too high at young and old ages, consistent with the observed undershoots (Figure~\ref{fig:s}; higher variance increases the expected hazard,\cite{Pridham2024-kidney} thus magnifying the misfit).  
%is the best choice, and that it captures the essential behaviour of the system,
%with fit quality affected by the number of attributes.
%Note that while the age-only model, $\exp{(a+ct)}$, for the statistics reasonably well, it fit the trajectories much worse than the models that included the FI (Figure~\ref{fig:modelselection}). Including of non-linear terms showed no visual change in fit. These observations were confirmed quantitatively using both the BIC and out-of-sample log-likelihood which showed substantial improvement when $f$ and $t$ were both included but only marginal improvement for more complex models. 
%Thus we infer that an efficient model is 

%The effective number of deficits depends on the underlying structure, which prior research indicates contains nested domains \cite{Pridham2023}. This reduces the effective number of deficits compared to our model, which assumes conditional independence given $f$ and age (which should be sensitive to overall health but not domains). If we simulate fewer deficits we better agree quantitatively for the higher order statistics, Figure~\ref{fig:tuning}.

We confirmed that linear damage, repair and hazard rates are consistent with the data using a non-parametric approach in Supplemental Section S11.

%Notably, the mean fits well quantitatively whereas the higher-order statistics (b-d) tend to be the right shape but wrong scale. 
%Note that the exponential model, fit directly to the mean, does not correctly capture the curvature of the mean FI: the curvature is too large at young ages and too small at old ages (black dashed line; a). The mean FI increases slowly until around age 70-80 where it starts to rapidly increase. 

%%%%%%%%%%
\begin{figure*}[t]
\includegraphics[width=.44\textwidth]{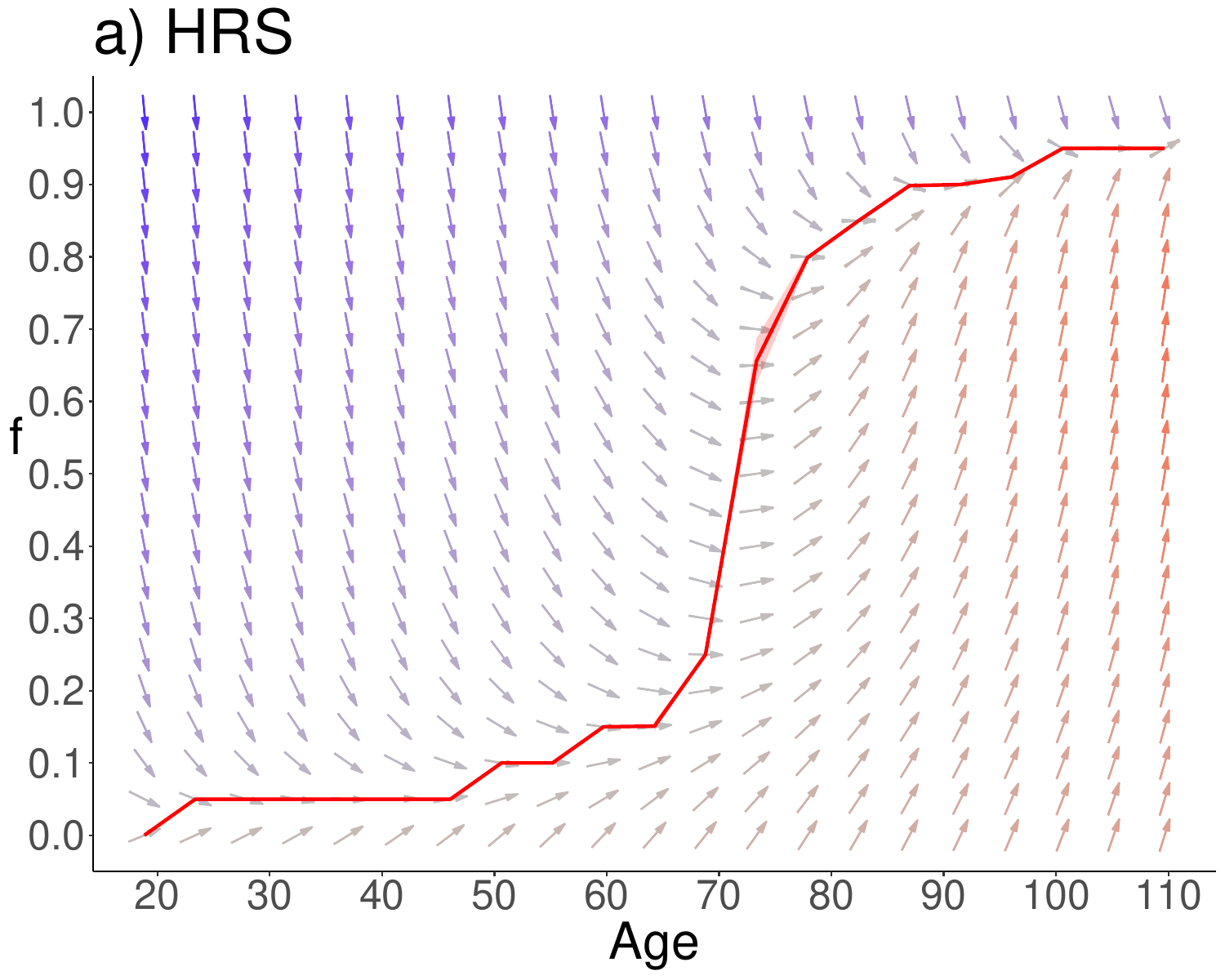} 
\includegraphics[width=.44\textwidth]{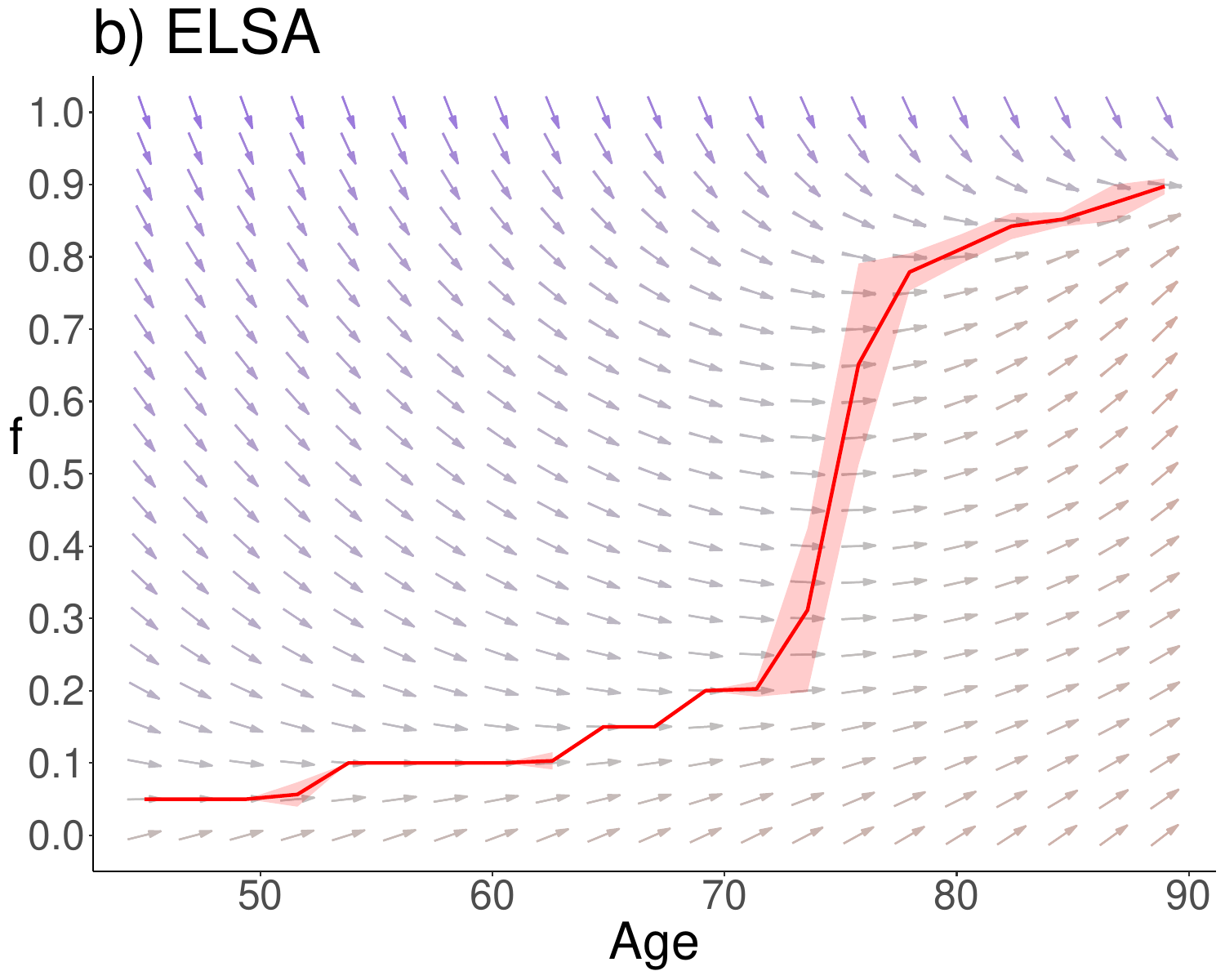} 
\includegraphics[width=.1\textwidth]{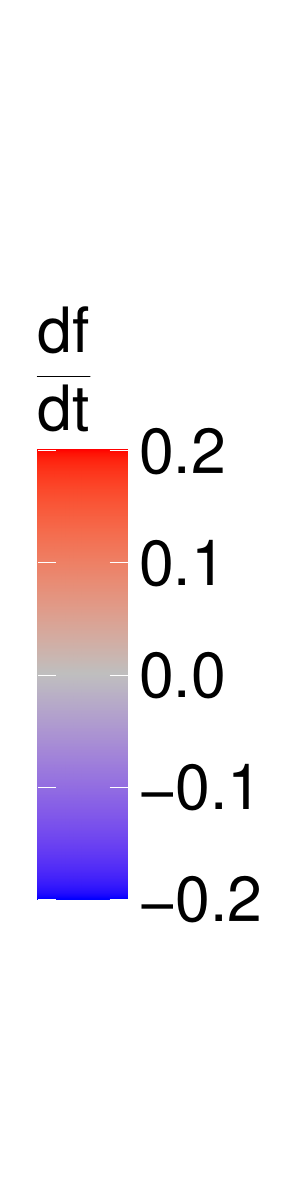} 
%    \centering 
%        \begin{subfigure}[t]{0.49\textwidth}
%        \centering
%        \includegraphics[width=\textwidth]{figures/hrs_dfdt.pdf} 
%        \caption{HRS.}
%    \end{subfigure}
%    ~
%    \begin{subfigure}[t]{0.49\textwidth}
%        \centering
%        \includegraphics[width=\textwidth]{figures/elsa_dfdt.pdf} 
%        \caption{ELSA.}
%    \end{subfigure}%
    \caption{\textbf{FI velocity field} in terms of current health and age, for both HRS (a) and ELSA (b) data. Higher FI, $f$, corresponds to worse health. Arrows represent the expected direction of individual flow  at each point. Blue arrows point down and are expected to decrease in $f$ over time, red arrows point up and are expected to increase. The nullcline (red line) is where the rate of change of $f$ is $0$ and hence the expected flow only increases age (grey arrows). Observe that between ages 65-80 the nullcline increases sharply, suggesting a tipping point. Uncertainties are included in the nullcline as bands. Uncertainties in the arrows are too small to see.} \label{fig:dfdt}
\end{figure*}
%The population occupies primarily ages 60-70 and FIs 0-0.2 and does appear to follow the field lines (Figure~\ref{fig:dfdtpop}).
%%%%%%%%%%%%%%%%%%%%%%%%%%%%%%%%%%
%\section{Robustness and resilience are lost continuously with increasing age and decreasing health}
\section*{Gradual loss of robustness and resilience with worsening health and age causes a tipping point near age 75}
We observed that damage rate increased continuously with respect to both age and FI, indicating a loss of robustness with both age and declining health (Supplemental Figure~S1). Conversely, the repair rate decreased continuously indicating a loss of resilience with both age and declining health. Death hazard rates also increased continuously with age and FI. At advanced ages and high FIs we observed hazard rates in excess of 3, indicating only a 5\% chance of surviving the next year (0.25\% chance of surviving until 2-year follow-up). Hence older individuals with high FIs will almost certainly die before followup, explaining why the FI has an empirical limit near 0.7. \cite{Mitnitski2016-xc}  

The tug-of-war between damage and repair determines the velocity field for the FI, Figure~\ref{fig:dfdt}. The velocity field is the derivative evaluated at each point in the plane, $\frac{df}{dt}(t,f)$ using Eq.~\ref{eq:fi}. An individual at any point in this plane is expected to move parallel to the local arrow. 
%By analogy, the arrows indicate the direction of `wind' (slope vs time) pushing each individual's age and FI. %moved to Box 1
%The nullcline, delineated by the red line, is where the FI velocity is 0 -- and the arrows are pointing in the direction of increasing age. %moved to Box 1
The nullcline, delineated by the red line, separates regions that are expected to improve in health (blue arrows) from regions that are expected to worsen (red arrows). Remarkably, there is a sudden increase in the nullcline near age~75 in both studies, indicating a tipping point. This means, for example, that while there may be healthy individuals over age~75, they are expected to worsen (e.g.\ Figure~\ref{fig:summary}c). In Supplemental Figure~S2 we show that the population density cloud is consistent with the velocity field.
%The population occupies primarily ages 60-70 and FIs 0-0.2 and does appear to follow the field lines (Figure~\ref{fig:dfdtpop}).
%This discontinuity represents a tipping point with respect to age.

Our model estimates yield tipping point at age $72.3\pm0.3$~years for HRS and $75.0\pm0.5$~years for ELSA (see Supplemental Section~S7 for details). In Supplemental Section~S12 we validated the tipping point using a logistic-regression changepoint model for each of the HRS deficits and observed a change in slope at age $75\pm4$ years old (median$\pm$mean absolute deviation). Combining these results using random-effects modelling we estimate the tipping point occurs at $74\pm1$ years old (95\% $71.4-76.0$ years old).\cite{metafor}

\begin{figure*}[t]
    \centering 
    %    \begin{subfigure}[t]{0.49\textwidth}
    %    \centering
    %    \includegraphics[width=\textwidth]{figures/hrs_dfdt_v2.pdf} 
    %    \caption{HRS.}
    %\end{subfigure}
    %~
    %\begin{subfigure}[t]{0.49\textwidth}
    %    \centering
    %    \includegraphics[width=\textwidth]{figures/elsa_dfdt_v2.pdf} 
    %    \caption{ELSA.}
    %\end{subfigure}
\includegraphics[width=0.42\textwidth]{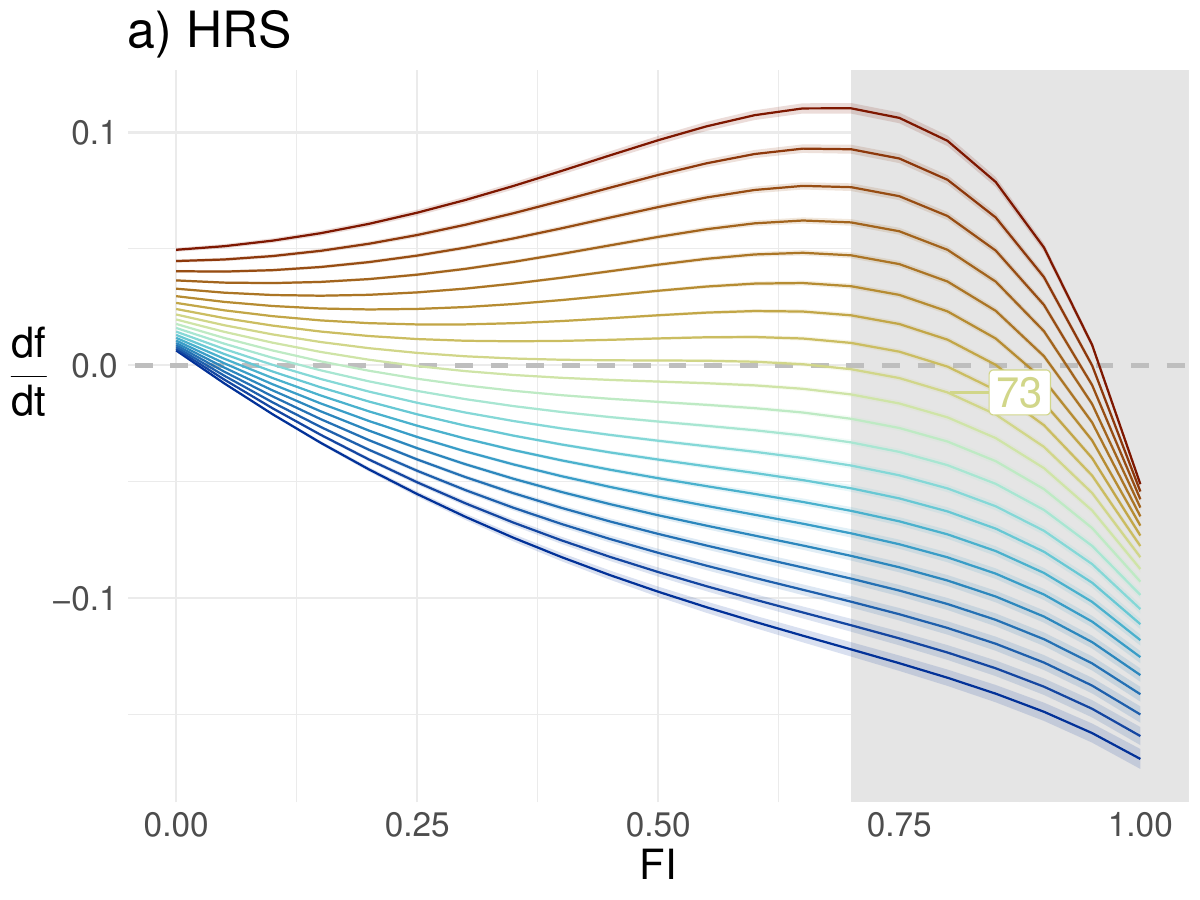} 
\includegraphics[width=0.42\textwidth]{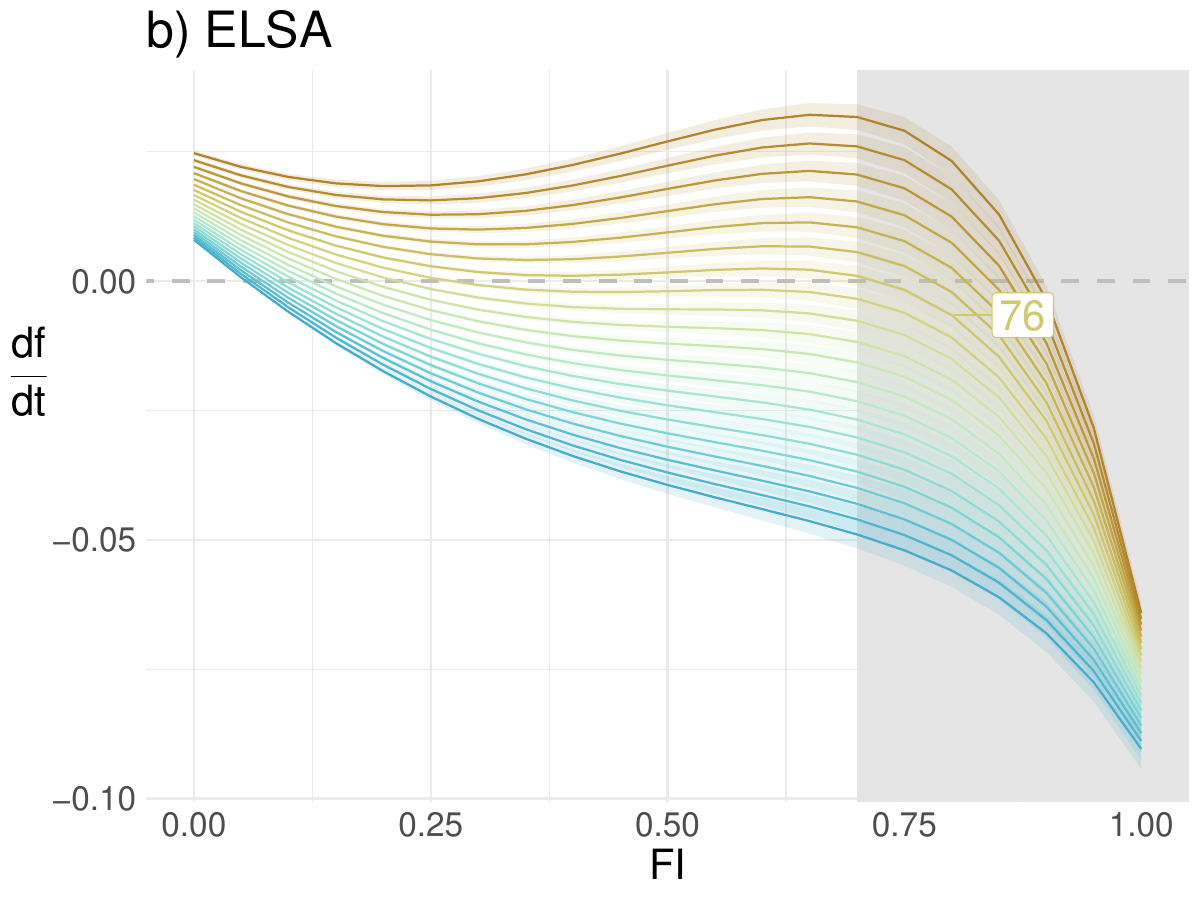} 
\includegraphics[width=0.12\textwidth]{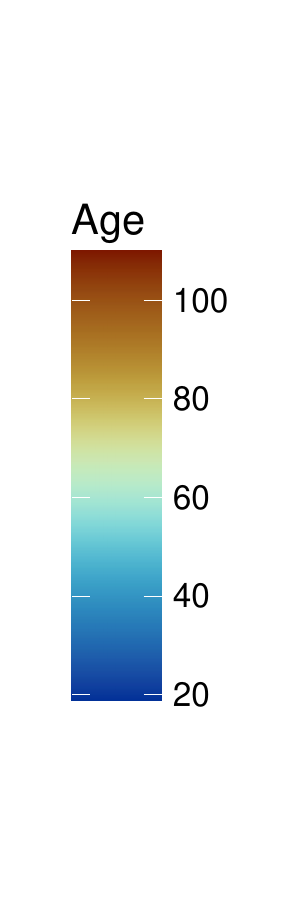} 
    \caption{\textbf{FI accumulation accelerates for older individuals (ages~$75+$) but not for younger individuals}. Individuals will tend to proceed along their age-specific line (coloured lines) towards the nullcline (grey dashed line, where the frailty velocity vanishes with $df/dt=0$). At young ages the velocity drops with increasing FI thus stabilizing at low values (blue). At older ages, 73+ for HRS and 76+ for ELSA, the velocity is constant or increasing with respect to the FI, indicating accelerating deficit accumulation (red). At very large FI, the FI saturates and $df/dt$ decreases with FI -- but by this point mortality becomes almost certain (see Supplemental Figure~S1c). FI past 0.7 was almost never observed ($f=0.7$ was the 98th percentile for HRS and 99.9th for ELSA; grey shaded). Each line is a fitted model prediction using Eq.~\ref{eq:fi}; bands are standard error from bootstrap (100 repeats).} \label{fig:dfdt2}
\end{figure*}

In Figure~\ref{fig:dfdt2} we take a closer look at the rate of change of $f$ with age ($df/dt$) as a function of FI for different ages, as indicated by the legend.  The horizontal dashed grey line is the nullcline ($df/dt=0$), which separates increasing from decreasing FI. Observe that below age 73 for HRS and age 76 for ELSA the curves are below the nullcline for all $f \geq 0.05$ and hence the velocity decreases when $f$ increases past 0.05. For these younger individuals the FI velocity slows down and reverses as the FI increases, leading these individuals to stop worsening and eventually recover (to $f=0.05$). In contrast, for older individuals, the FI velocity \textit{speeds up} as the FI increases beyond very small values, causing these individuals to get even worse until a very high FI is reached, at which point imminent mortality becomes a near certainty (indicated by the greyed background, for $f \gtrsim 0.7$). As we will show, a tipping point separates these young and old regimes and is caused by a rapid loss of robustness and resilience with increasing $f$.
%At such a high FI mortality becomes likely, making high FIs difficult to observe. 

%%%%%%%%%%%%%%%%

\begin{figure*}[t]
    \centering 
    \includegraphics[width=\textwidth]{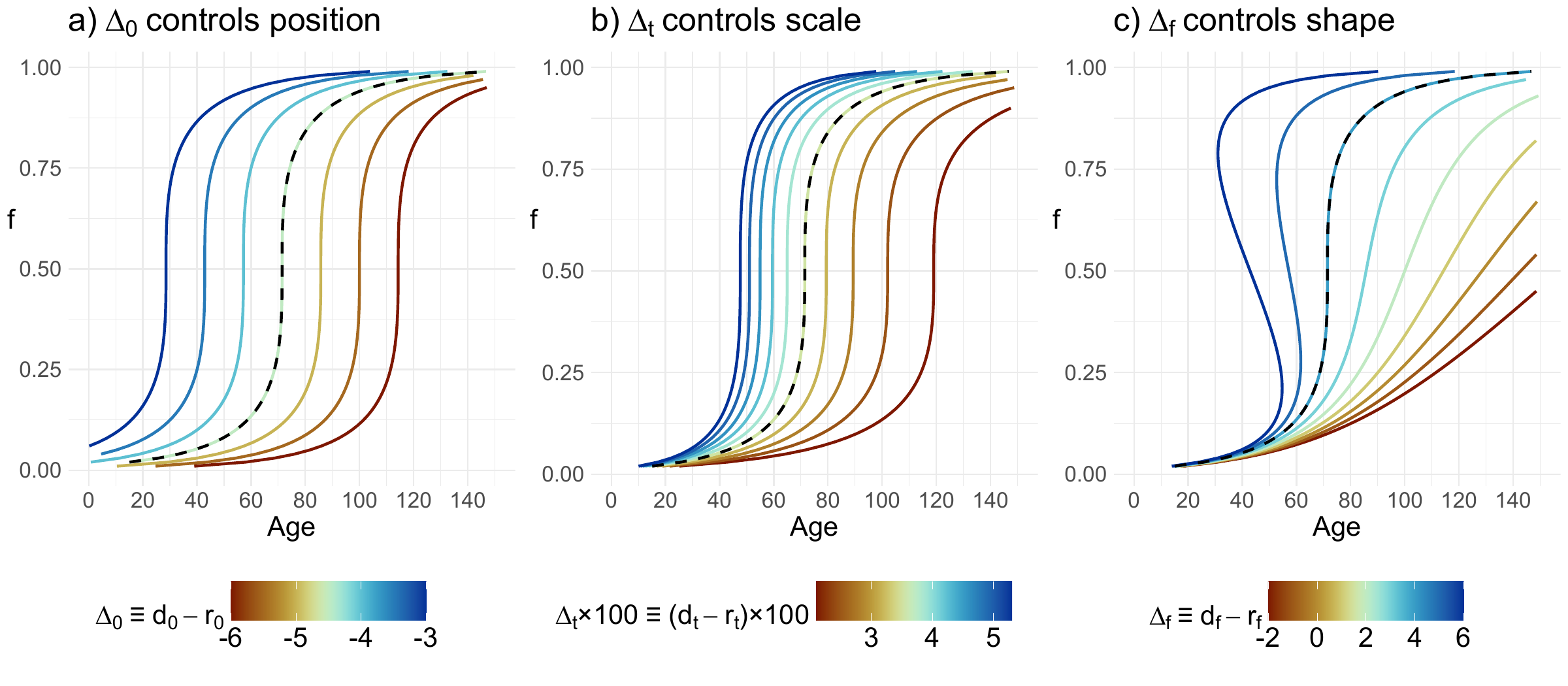} 
    \caption{\textbf{The differences between parameters control the position ($\Delta_0 = d_0-r_0$), scale ($\Delta_t=d_t-r_t)$, and shape ($\Delta_f=d_f-r_f$) of the nullcline}, Eq.~\ref{eq:tf}. Black dashed lines are the model fit estimates (average across studies). Notably, there is a discontinuity in the nullcline at $\Delta_f^\ast=4$: smaller values increase gradually and larger values have a discontinuity and bifurcation (meaning there are two stable FIs: one low and one high; c). The model fit is almost exactly $\Delta_f=4$ leading to a nearly vertical tipping point (black dashed line in c).  Individuals are expected to move towards the curve at their specific age. y-axis is the value of $f$ on the nullcline.} \label{fig:nullcline}
\end{figure*}

The rate at which robustness and resilience are lost determines how sharply the nullcline changes and thus individual prognosis. We derive and analyze the average nullcline behavior in the supplemental, and obtain the nullcline equation
\begin{align}
     t^* &= -\frac{\Delta_0}{\Delta_t}-\frac{\Delta_f}{\Delta_t}f^* + \frac{1}{\Delta_t}\ln{\bigg(\frac{f^*}{1-f^*}\bigg)}, \label{eq:tf}
\end{align}
where $t^*$ and $f^*$ are the times and FIs that lay on the nullcline. This nullcline curve is plotted in Figure~\ref{fig:nullcline}, for various values of  $\Delta_0\equiv d_0-r_0$, $\Delta_t \equiv d_t-r_t$ and $\Delta_f\equiv d_f-r_f$. Baseline robustness and resilience, $\Delta_0\equiv d_0-r_0$, controls the position of the nullcline in time. Aging rate, $\Delta_t\equiv d_t-r_t$, controls the scale of the curve, with large values compressing the curve with respect to time. And finally health sensitivity, $\Delta_f \equiv d_f-r_f$, controls the shape of the nullcline. The simple and characteristic behaviour controlled by these parameter combinations make them useful to consider for population health statistics.

%%%%%%%%%%%%%%%%%%%%%%%%

The nullcline has a discontinuity, with $df/dt \to \infty$ for
\begin{align}
    \Delta_f = \Delta_f^* &\equiv  4.
\end{align}
Below, $\Delta_f<4$, there is no discontinuity, and above, $\Delta_f>4$, the discontinuity bifurcates and gradually spreads to younger ages with increasing $\Delta_f$. For $\Delta_f<4$ there is no tipping point, but instead a smooth superlinear increase. Above, $\Delta_f>4$, in addition to the tipping point, there is a trap where young people who acquire a high FI are expected stay high without recovering. Remarkably, the model fits for both HRS and ELSA $\Delta_f \equiv d_f-r_f$ yielded estimates very close to 4 ($3.93 \pm 0.03$ for HRS and $4.15 \pm 0.06$ for ELSA), suggesting a tipping point with no bifurcation (Supplemental Figure~S3). This also confirms that the loss of robustness and resilience with increasing $f$ leads to the observed tipping point.

Our results are qualitatively identical for males and females, but the specific parameter estimates differed (Supplemental Figures~S10 and S11). In particular, across studies males had lower baseline damage $d_0$, but were more sensitive to increasing damage with worsening health $d_f$ and advanced age $d_t$, and experienced less of a drop in repair with worsening health $r_f$. In terms of $\Delta$s, in both studies males showed an earlier tipping point (smaller $\Delta_0$), a faster aging rate (larger $\Delta_t$), and faster tipping point (larger $\Delta_f$). Males also had higher mortality rates $h_0$ and higher risk of death with increasing FI, $h_f$. Overall males had better robustness and resilience at young ages but worse at older ages. This led to a lower nullcline at young ages and a higher nullcline at older ages (Supplemental Figure~S12).

%%%%%%%%%
\section*{Consistent with the model, sub-populations that are older and have higher FI tend to have higher damage and lower repair rates}\label{sec:subpop}
\begin{figure*}[!ht]
\centering
\includegraphics[width=0.99\textwidth]{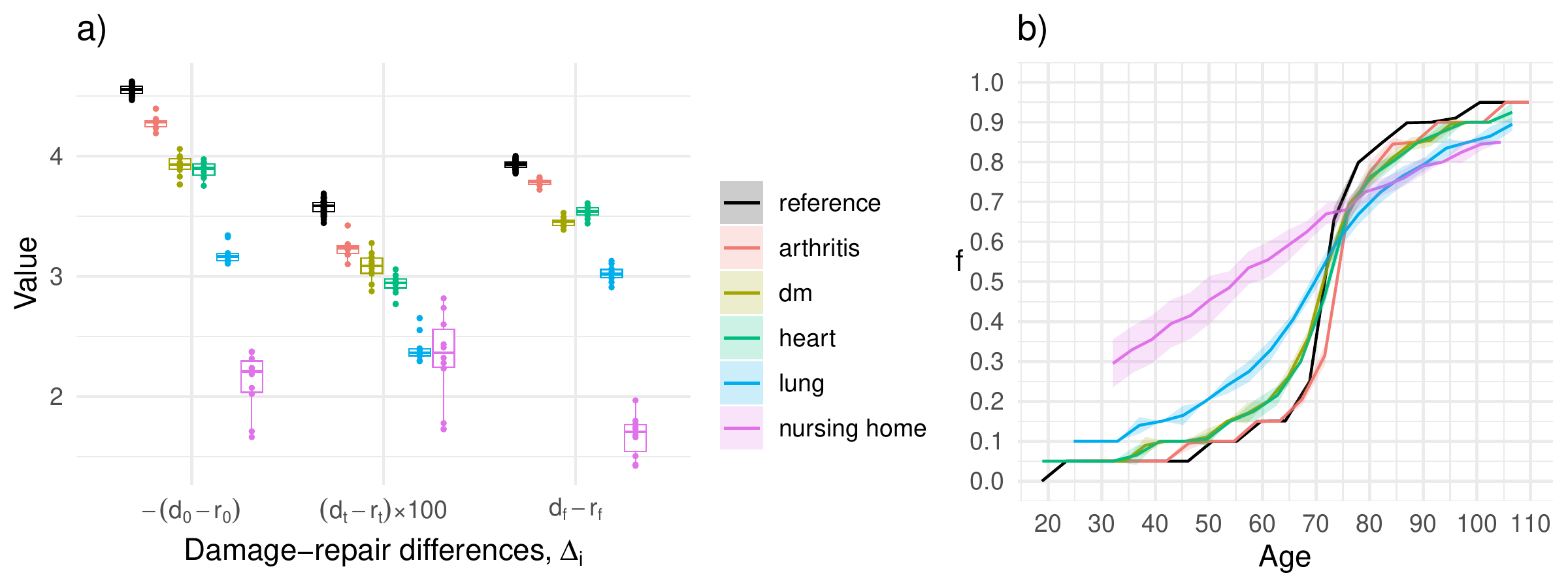} 
    \caption{\textbf{Parameter estimates for sub-populations of interest -- HRS}. a) Parameter differences show a clear trend of worsening from reference group (everyone, highest) to living in a nursing home at time of interview (lowest). This ordering reflects differences in FI and age, that make it qualitatively consistent with our model (see text). b) Nullcline estimates indicate that the nullcline remains at approximately age 75 but flattens with worsening trend. Each point is a bootstrap replicate (100 for reference population, 10 for all other populations). Line is median, box is interquartile range and whiskers are $1.5\times$IQR.} \label{fig:subpoppar}
\end{figure*}

We fit to sub-populations of interest by stratifying by a specific health variable (that was then excluded from fitting). %We stratified by: individuals with arthritis (arthre=1), diabetes (diabe=1), heart problems (hearte=1), lung disease (lunge=1), or living in a nursing home(nhmliv=1). 
We stratified by individuals currently: with arthritis, diabetes, heart problems, lung disease, or living in a nursing home.
The parameter fits show a clear trend in the damage-repair differences, Figure~\ref{fig:subpoppar}a, with an apparent worsening of the damage-repair difference in order of: reference (highest repair), arthritis, diabetes, heart problems, lung disease and then living in a nursing home (highest damage). This ordering between sub-populations is strongly correlated with differences captured by the FI (Spearman's r=-1.00, -0.94 and -0.94 for $\Delta_0$, $\Delta_t$ and $\Delta_f$, respectively, p<0.05) and moderately-strongly correlated with different ages (Spearman's r=-0.89 (p=0.03), -0.77 (p=0.10) and -0.71 (p=0.10) for $\Delta_0$, $\Delta_t$ and $\Delta_f$, respectively). The observed ordering is qualitatively consistent with our model that encodes damage-promotes-damage and age-promotes-damage.
%non-linear decrease in repair rate and increase in damage rate, with both increasing FI and age -- 

We also considered the nullcline shape in these subpopulations, Figure~\ref{fig:subpoppar}b. We found that that the tipping point does not move it remains in the age range of 70-80 years old, but the nullcline does appear to flatten from sharp to linear as the sub-population gets older and has higher median FI.  The sub-population of those living in a nursing home showed no apparent tipping point, instead only a linear increase in the nullcline. Since the nullcline is already high at earlier ages, this population could be considered to be already past any tipping point and into an end-stage of life. 
%See Supplemental Section S13 for full details.

%%%%%%%%%%%%%%%

\section*{We performed sensitivity analysis to validate our key results and assumptions}
We performed sensitivity analysis of our tipping point results using changepoint logistic regression model. We fitted attribute deficit prevalence curves and permitted a changepoint in slope. This recapitulated: (i) the existence of a tipping point at age 75, \cite{muggeo2008segmented} (ii) the consistency of age 75 across sub-populations, and (iii) the weakening of the changepoint in the nursing home sub-population (Supplemental Sections S12 and S13).

We also performed sensitivity analysis on the types of health attributes used. Some of the included attributes cannot repair by definition, the ``ever diagnosed'' variables in HRS never repaired while the chronic disease diagnoses in ELSA only repair on rare occasions (all $< 1$\%) reflecting errors in self-reporting or changes in diagnosis. In Supplemental Section S14 we restricted our analysis to only repairable attributes and observed higher repair rates but similar qualitative behaviour, including the tipping point around age 75.

%%%%%%%%%%%%%%%%%%%%%%%%%%%%%%
\FloatBarrier
\section*{Discussion}
We dynamically modelled health trajectories from transitions in health attributes during aging. We directly assessed age and health effects in both robustness and damage, and resilience and repair, for binary health deficits. We observed that both robustness and resilience decreased continuously with both increasing age and worsening health, as measured by a summary measure of aging health -- the FI ($f$). The behavior was exemplified by the behavior of the nullcline, which determines how stable values of FI change with age, i.e.\ the population-level FI setpoint. At the stable nullcline, the average FI does not change. At young ages, robustness and resilience were sufficiently high that damage was uncommon and repair was swift, causing the nullcline to rest near $f\approx0.05$, leading to an accordingly small FI. At older ages, both robustness and resilience were lower and the nullcline instead tended towards very large values, $f\approx 0.9$ (mortality truncation masks these large values in observed data.) The separation between young and old was well-defined near age~75, where a singular increase of the nullcline was observed -- reflecting the sudden change in balance between resilience and robustness.

%why is there a tipping point?
%damage > repair
%damage - repair very sensitive to f
Within our model the tipping point occurs because robustness and resilience are degraded with increasing FI. Once damage rate exceeds repair rate the FI growth can accelerate -- which further increases damage rate and decreases repair rate. This runaway process is inhibited when the number of undamaged attributes becomes small enough to balance a higher probability of damage with a lower probability of repair. The age dependence of damage and repair determines when the transition occurs, while the FI dependence determines how sharp the transition is. We observe that the nullcline increases very suddenly and only saturates just before $f=1$. The effect is captured by the model by the critical value $\Delta_f \equiv d_f-r_f = \Delta_f^\ast \equiv 4$. 

Our dynamical analysis indicates that the tipping point emerges from gradual age- dependent changes to robustness and resilience, and that frailty increases after the tipping point result from frailty-dependent changes to robustness and resilience. This does not imply a single causal mechanism, or the absence of delayed effects of aging biology; rather, it shows that the aggregate consequences of age dependent robustness and resilience changes lead to a dynamical transition. Consistent with our observations, age-related changes are believed to underpin frailty onset \cite{Kim2024-wx,Fried2021-vh}. While a tipping point near age~75 appears reasonable, the exact biological mechanisms remain murky \cite{Kim2024-wx}. Cardiac decline is an important candidate since it is known to accelerate after age~70, \cite{Ribeiro2023-yf} and plays a major role in frailty, e.g.\ blood pressure medications reduce the FI in mice (angiotensin-converting enzyme inhibitors) \cite{Howlett2021-my}. In fact, the FI tends to be a better predictor of heart disease than traditional measures such as the Framingham risk score \cite{Farooqi2020-rk,Wallace2014-fa}. There are, however, a number of other significant biological changes near age~75. The proteome changes dramatically near ages~70-80.\cite{Lehallier2019-hd} In retrospect, our dynamical network analysis of multiple biological ages also showed a marked change in dynamical behaviour around age 80, and indicated that the central drivers were epigenetic changes and cardiometabolic decline,\cite{Pridham2024-in} that ultimately drove changes in the FI (see supplemental of \cite{Pridham2024-in}). Intriguingly, research on tissue integrity in flies and other lab animals has also identified two distinct phases to aging, with a late phase characterized by loss of tissue integrity and greatly increased risk of death \cite{Zane2024-tj, Raz2026-oh}. A widespread exhaustion of tissue robustness and resilience, including cardiomyocytes, would unify these previous observations with our own -- at least at the tissue and functional levels. If we are to understand molecular drivers, however, we need to fully unpack the biological mechanisms underlying loss of robustness and resilience, and subsequent gain of frailty. One path forward would be to combine our approach with deficit clustering and deep phenotyping \cite{Foote2025-ws}. Recent results from twin studies suggest that dynamical parameters of aging are  heritable\cite{Shenhar2026-ne}.

Previous works are conflicting on the existence of such a tipping point in age-related health. Our contribution is in providing direct dynamical evidence for this tipping point by modelling damage and repair rates as functions of both age and FI, and to show how this naturally reconciles exponential-like FI growth with a late-life tipping point. 
%with continuously declining robustness and resilience, and validating in a distinct population. We have also demonstrated how the tipping point emerges naturally from basic . We provide a model for forecasting in light of this tipping point, and a framework that could be used to probe the effects of interventions on both the tipping point and the underlying robustness and resilience. 
%This tipping point should be accounted for when forecasting health trajectories and risk categories. 

The tipping point paradigm differs from the prevailing idea that the FI is an exponential process\cite{Mitnitski2015-ia}. Exponential processes, including the FI and Gompertz' law, have a single regime in which rapidly worsening health follows directly from compounding over the entire life-course, i.e.\ damage-promotes-damage without age-promotes-damage. Our analysis of the tipping point instead indicates that something fundamentally changes near age~75, and for this to happen we must have age-promotes-damage.  Nevertheless, the damage-promotes-damage parameter ($\Delta_f$) in our model was exactly strong enough to permit a tipping point, indicating that damage-promotes-damage contributes. Previous theories from geriatric medicine tend to focus solely on damage-promotes-damage, including the frailty index\cite{Mitnitski2015-ia, Kim2024-wx} and cascade theories\cite{Nagi1976-hz, Manini2011-bw}. The frailty phenotype paradigm \cite{Fried2021-vh} proposed instead age-related critical failure, but failed to achieve a consensus about the presence or nature of the criticality. Our work clarifies that there are two key stages of aging trajectories, and that the tipping point is driven ultimately by age-related changes other than clinical health indicators. This is consistent with the geroscience hypothesis that aging arises from underlying cellular changes across health systems\cite{Seals2016-vn, Kennedy2004-tn}. In short, we observed that health deficits alone do not describe health trajectories of damage and repair, we must also know the age of the individual.
% and phenotype\cite{Fried2021-vh} paradigms

%frailty
Fried \textit{et al} proposed that frailty reflects a loss of homeostatic resilience leading to a critical transition \cite{Fried2021-vh}. Our dynamical analysis provides direct quantitative evidence that frailty onset occurs at a ``critical'' (sudden) transition in dynamical health states. Prior research has inferred that frailty is characterized by a state of enhanced vulnerability due to insufficient robustness and resilience to mitigate environmental demands, \cite{Fried2021-vh, Kim2024-wx} here it emerges naturally from modelling health trajectories. Individuals are classified as frail if their FI exceeds $0.2$ (i.e. $f>0.2$), with higher values grading more severe cases\cite{Kim2024-wx}. We directly modelled loss of robustness and resilience and can confirm this leads to a sudden increase in vulnerability, starting at approximately $f\approx 0.2$ (Figures~\ref{fig:dfdt} and S2). We observed two dynamical states at the population-level: young and old, consistent with the critical model of frailty emergence \cite{Fried2021-vh}. Indeed, FI becomes the dominant driver of health at older ages,\cite{Pridham2023, Pridham2024-in, Farrell2016-xp} which is consistent with a tipping point. %What's more, we observe the changes are driven by increasing age, consistent with changes to underlying biological mechanisms \cite{Fried2021-vh, Kim2024-wx}. 
Since not all individuals age at the same rate \cite{Jylhava2017-wc}, we speculate that the critical transition in the nullcline could also vary between individuals. 

Our results suggest that the damage-promotes-damage effect \cite{Mitnitski2015-ia, Taneja2016-bj} is driven by a changing \emph{stable} nullcline rather an instability \cite{Mitnitski2015-ia} (Figure~\ref{fig:dfdt}). Nevertheless, our parameter values indicate that robustness and resilience are both lost with increasing FI, supporting a significant degree of damage promoting further damage. In Supplemental Section~S9 we linearize our model and show that for small $f$ our results can be approximated by an age-dependent instability that emerges near age $100$. Work in mice has shown both a saturating feedback in senescent cell count \cite{Karin2019-gf} and an instability emerging at advanced ages \cite{Avchaciov2022}, consistent with this simplified picture. The approximation also helps to explain why the average FI appears exponential at the population-level and suggests the instability picture may be a reasonable approximation \cite{Mitnitski2015-ia}, particularly if it includes variables that could capture underlying biological changes\cite{Taneja2016-bj}. Note, however, that the FI increases several years \textit{after} the nullcline transitions (Figure~\ref{fig:summary}c), and therefore the FI may be a lagging indicator of underlying health changes.

The ``rectangular''\cite{Seals2016-vn} structure of the nullcline with a long healthy period, followed by rapid decline, delineates an optimal population health trajectory \cite{Seals2016-vn, Kennedy2004-tn} that has maximally compressed morbidity. While decreasing $\Delta_f$ will improve robustness and resilience, it will also extend the decline period. 
%An example of this would be treating a single chronic disease by glucose monitoring of diabetes, which does not prevent onset of other age-related diseases. Since there are hundreds of age-related diseases\cite{Katzir2021-nt}, the impact of treating any one of these conditions will be limited without a means of treating the shared, underlying age-related mechanisms. Indeed, most Medicare recipients have 3 or more chronic conditions\cite{Fabbri2015-vd}. 
Interventions that target $\Delta_f$ are sub-optimal since they act to extend the decline period. Better would be interventions that target baseline damage and repair, $\Delta_0$, or aging rate, $\Delta_t$, since either would delay the tipping point to older ages. Indeed, contemporary frailty management seeks to directly build robustness and resilience as early as possible \cite{Kim2024-wx}.
Nevertheless, healthy older individuals above the tipping point but with low FI may only drift slowly towards the stable (unhealthy) nullcline. This means that individuals past their tipping point age could still maintain good health if they can avoid or reverse initial deficit accumulation. ``Healthy aging'' \cite{Michel2017-eo} in these populations past age~75 is therefore predicted to require mitigation of environmental stressors, such as vaccination or removal of environmental fall hazards. 
%Crossing the tipping point dramatically increases risk for and accumulation of health deficits if such stressors are not reduced. 
Passing the tipping point dramatically increases the risk and accumulation of health deficits if such stressors are not reduced.

%sex effects
Stratifying by sex permitted us to explore the effects of different parameterizations since males are known to live shorter but healthier (lower FI) lives than females  \cite{Hubbard2015-tq}. Males showed better initial robustness and resilience (smaller $\Delta_0$), higher health sensitivity, $\Delta_f$, and faster aging rate, $\Delta_t$. The smaller $\Delta_0$ delays decline whereas the higher $\Delta_f$ and $\Delta_t$ steepens and hastens decline, respectively. The net effect was that young males had a lower nullcline than young females, keeping their FI low. At older ages the nullclines crossed and males had a higher nullcline causing the FI to grow faster. Importantly, males also had higher mortality rates: both at baseline and with increasing FI, explaining why they died younger. These observations are consistent with known sex differences \cite{Hubbard2015-tq}. While the sharper decline experienced by males is desirous in terms of health-span,\cite{Seals2016-vn, Kennedy2004-tn} it comes at the cost of a shorter life.

% ADDED A SUBGROUP PARAGRAPH (probably needs a rewrite)
Stratifying by sub-populations with specific chronic conditions or care settings revealed that the fitted parameters and corresponding nullclines vary systematically with baseline health and age. Groups with arthritis, diabetes, heart disease, lung disease, and those living in nursing homes showed progressively worse damage–repair differences (larger $\Delta_0$, $\Delta_t$, and $\Delta_f$), consistent with higher FI and older age in these sub-populations (Figure 6). Despite these shifts, the tipping point in age remained broadly similar, with the nullcline still rising steeply in the 70–80 year range for most groups. What changed was primarily the level and shape of the nullcline: as health status worsened, the nullcline elevated and flattened, so that higher FI values became stable over the entire age range and the tipping-point–like sharp rise became less pronounced. In the most extreme case, individuals living in nursing homes appeared effectively “beyond” the tipping point, with a nearly linear, high-lying nullcline indicating that poor health is dynamically stable even at relatively younger chronological ages. Together, these findings suggest that heterogeneity in underlying robustness and resilience is captured by systematic shifts in the damage–repair parameters and nullcline geometry, rather than by large shifts in the age at which the tipping point occurs.

%future directions
Differences between model parameters are highly interpretable since they directly affect damage, repair and hazard rates. $\Delta_0$ delays decline, $\Delta_t$ scales age, and $\Delta_f$ controls the sharpness of decline. This makes either the parameters or their differences a tool for analyzing population health, heterogeneity, and interpreting the effects of interventions or associations with desirable aging trajectories, including identifying drivers of frailty. 
%The discontinuity in the age range of 70-80 suggests this is the key discriminating age range for assessing the effectiveness of interventions that mitigate or delay decline. 
The discontinuity in the age range of 70-80 suggests this is a key range where health trajectory models are likely to fail, producing unrealistic predictions. Interventions that might mitigate or delay decline can also be probed using our model, future research should consider: (i) fitting the model to sub-populations of interest (e.g.\ treatment vs control), (ii) modifying the model to permit covariates to identify associations (such as in $\Delta_0$ and $\Delta_t$), and (iii) completely individualizing the model fits. It is interesting to consider how these parameters could help to parse known associations with the FI, such as social vulnerability\cite{Ayeni2022-ps} and physical exercise\cite{Howlett2021-my}. %\cite{Negm2019-rt}.

We note a few limitations to our study. First, we used an FI built entirely out of questionnaire data, which may not capture every dimension of health \cite{Widagdo2016-gv, Pridham2023}. We are agnostic to what constitutes repair, which could represent resolution of a deficit like a fall injury, or adapted behaviour by an individual to negate the effects of a deficit, like adding a grab bar to make a bathtub more accessible. The stochasticity of damage and repair may have incorporated heterogeneity into the dynamical rates that could blur their values, although our population and subpopulation analyses showed similar qualitative behaviour indicating that our results are robust to differences due to heterogeneity. Survival and censorship could affect our results although this is unlikely since the majority of data are in the lower mortality rate regimes and our results are consistent across populations and sub-populations with different rates of attrition. Finally, the data are sampled every 2~years, and we therefore cannot comment on robustness and resilience on significantly shorter timescales (e.g.\ with respect to acute diseases such as the flu).

%summary
Age-related health includes complex trajectories with individuals experiencing both gradual decline and major transitions in health attributes. These transitions appear to capture the effects of stochastic stressor events such as illnesses or falls. Trajectories can be quantitatively modelled using damage and repair of health attributes to jointly capture changes in both transition risk and gradual decline in robustness and resilience. Such quantitative models can be analyzed to infer general behaviour of the population of interest, enhancing our knowledge of what aging trajectories look like, and offering a framework %for better identifying underlying drivers. 
for better forecasting health needs.
Here we analyze aging Western human populations and find that age-related health has two distinct dynamical phases.
%Here we provide an analysis of natural, \textit{in situ}, aging of human populations. 
A tipping point near age~75 separates these dynamical phases -- where robustness and resilience become insufficient for stable good health and after which individuals are expected to tend towards worse health over time. 

%%%%%%%%
\section*{Acknowledgments}
ADR thanks the Natural Sciences and Engineering Research Council of Canada (NSERC) for operating Grant RGPIN-2025-04649. This work was supported in part by the Zuckerman STEM Leadership Program and the NSERC PGS-D program (GP). The HRS (Health and Retirement Study) is sponsored by the National Institute on Aging (grant number NIA U01AG009740) and is conducted by the University of Michigan. ELSA is funded by the National Institute on Aging (R01AG017644), and by UK Government Departments coordinated by the National Institute for Health and Care Research (NIHR).

\section*{Author contributions statement}
ADR and KR supervised. ADR conceived the project and contributed to analysis. GP did the primary analysis and initial draft. All authors reviewed the manuscript.

\section*{Data availability}
%\noindent All data used in the present study are publicly available. ELSA \cite{elsa} is available from the UK Data Service \url{https://ukdataservice.ac.uk/}. HRS \cite{hrs} is available from \url{https://hrs.isr.umich.edu/data-products}. 
\noindent All data used in the present study are publicly available. ELSA \cite{elsa} (study number 5050) were obtained with registration and approval from the UK Data Service at \url{https://doi.org/10.5255/UKDA-SN-5050-20}. Health and Retirement Study (HRS) \cite{hrs} were accessed via the University of Michigan, the data are available with registration and approval from \url{https://hrs.isr.umich.edu/data-products}.

\section*{Code availability}
Software for fitting the model is publicly available on GitHub at \\ \url{https://github.com/GlenPr/Damage-repair-survival-rate-model}.

\section*{Disclosures}
None directly related to this work. In the past three years KR has received honoraria for invited lectures, rounds and academic symposia on frailty from: Burnaby Family Practice, Chinese Medical Association, University of Nebraska-Omaha, the Australia New Zealand Society of Geriatric Medicine, the Atria Institute, University of British Columbia, McMaster University, and the Fraser Health Authority. KR is co-founder of Ardea Outcomes (DGI Clinical until 2021), which in the past 3 years has had contracts with pharma and device manufacturers (Danone, Hollister, INmune, Novartis, Takeda) on individualized outcome measurement. 

%None.
%\section*{Additional information}

%\clearpage

\clearpage
\bibliography{ref}

%supllmenetal

%add S in front of equation number
\renewcommand{\theequation}{S\arabic{equation}}
\renewcommand{\thefigure}{S\arabic{figure}}
\renewcommand{\thetable}{S\arabic{table}}
\renewcommand{\thesection}{S\arabic{section}}
\setcounter{equation}{0}  
\setcounter{figure}{0}  
\setcounter{table}{0}  
\setcounter{section}{0}

\newpage

%%%%%%%%%%%%%%%%%%%%%%%%%%%%%%%%%%%%%%%%

\noindent{\large\textbf{Supplemental information for: Aging health dynamics cross a tipping point near age 75}}

%%%%%%%%%%%%%%%%%%%%%%%%%%%%%%%%%%
This supplemental includes additional results and supporting evidence for the main text. We begin with additional results that were not included in the main text for want of space in Section~\ref{sec:fi:results}. We provide additional details on the specific FI variables used in Section~\ref{sec:si:fi}. Model selection is performed in Section~\ref{sec:modelselection}, where we show that the model linear in both FI and age is the best choice. We then compare population-level statistics for both ELSA and HRS using additional models in Section~\ref{sec:si:fit} (compared to Figure~1). This includes the survival curves using a proportional hazard assumption (i.e.\ $\ln{(h)}\propto f$).

In Section~\ref{sec:si:ptuning} we tune the number of health attributes, which shows that reducing the effective number of health attributes improves the fit to higher-order population-level statistics (variance, coefficient of variation and auto-correlation).

We then consider sex effects by fitting separately to males and females in Section~\ref{sec:si:sex}. While the main results are the same, the specific parameterizations of males and females are different, capturing known differences between males and females \cite{Hubbard2015-tq} (males tend to live shorter but healthier lives: the sex-frailty paradox).

Finally we disclose the mathematical derivations underlying our results. This begins with an analysis of the model nullcline in Section~\ref{sec:si:nullcline}. Next we derive the objective function, the survival-modified log-likelihood, in Section~\ref{sec:si:obj}. We include the gradient for efficient optimization, and a short proof of self-consistency between the log-likelihood formalism and Eq.~1. We derive an approximation of our model in Section~\ref{sec:si:stab} which makes it easier to compare our results to other models. Lastly, we provide the math needed to simulate our results in Section~\ref{sec:si:sim}.

\section{Additional results} \label{sec:fi:results}
\begin{figure*}[!ht]
    \centering 
        \begin{subfigure}[t]{\textwidth}
        \centering
        \includegraphics[width=\textwidth]{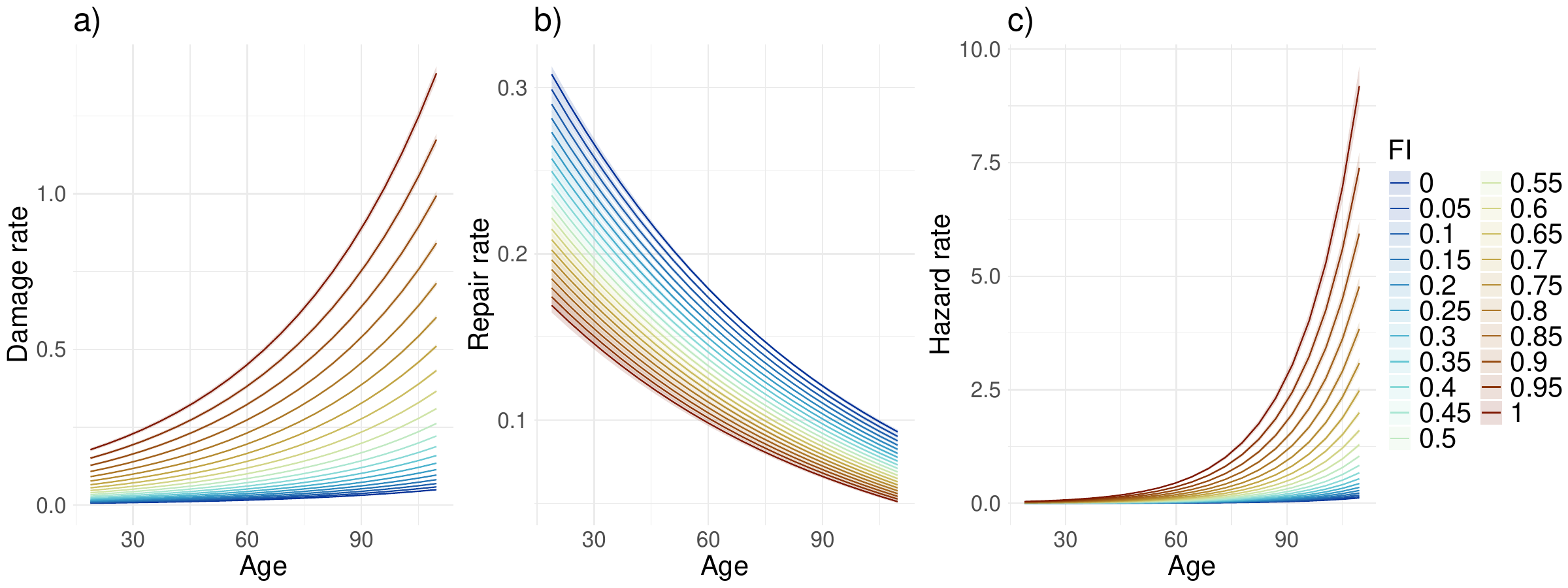} 
        \caption{HRS.}
    \end{subfigure}
    ~
    \begin{subfigure}[t]{\textwidth}
        \centering
        \includegraphics[width=\textwidth]{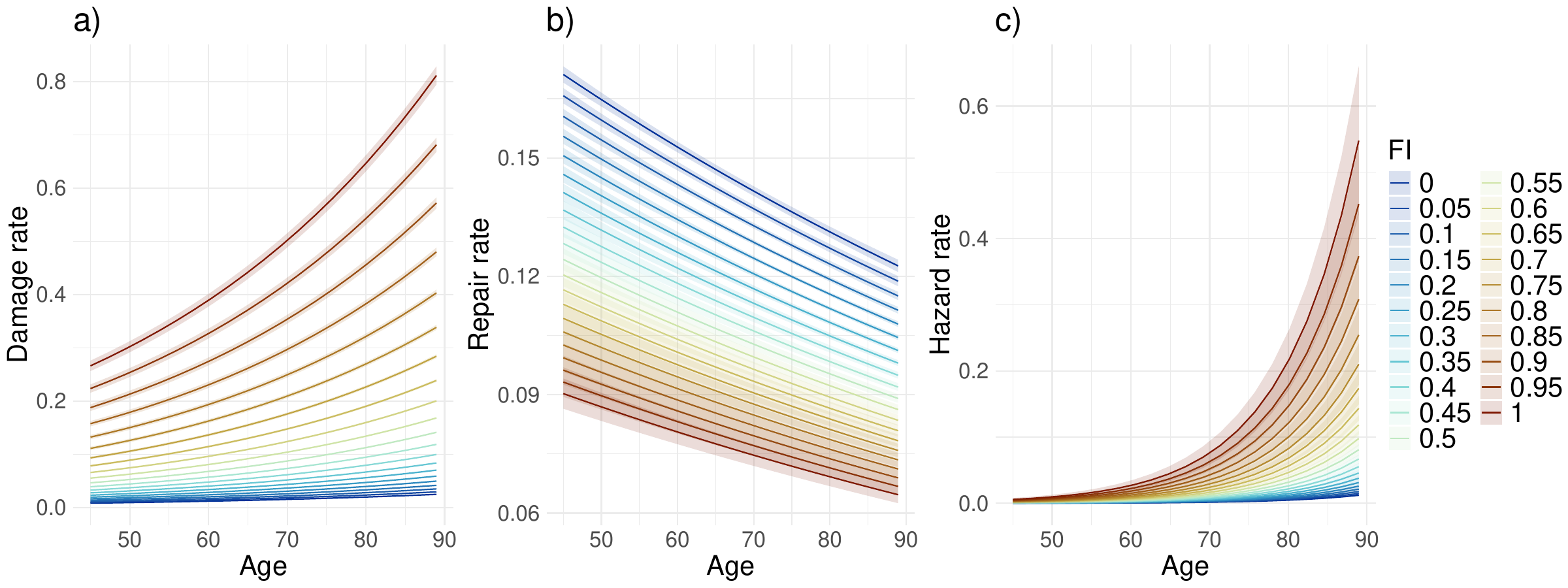}  
        \caption{ELSA.}
    \end{subfigure}
    \caption{\textbf{Robustness and resilience decrease continuously with age and FI}, and death hazard increases. ELSA survival hazard is low compared to HRS because only a fraction of deaths are recorded in the end-of-life files \cite{NatCen_Social_Research2015-nw}. Nevertheless, the age and FI dependence for survival is similar for both studies. Parameter estimates with standard errors (100 repeat bootstrap).} \label{fig:rates}
\end{figure*}

The model estimates for damage, repair and death hazard rates are plotted in Figure~\ref{fig:rates}. Observe that damage and death rates increased continuously with age and increasing FI, whereas repair rates decreased. Increasing damage rate indicates a loss of robustness whereas decreasing repair rate indicates a loss of resilience.

\begin{figure*}[!ht]
\includegraphics[width=.49\textwidth]{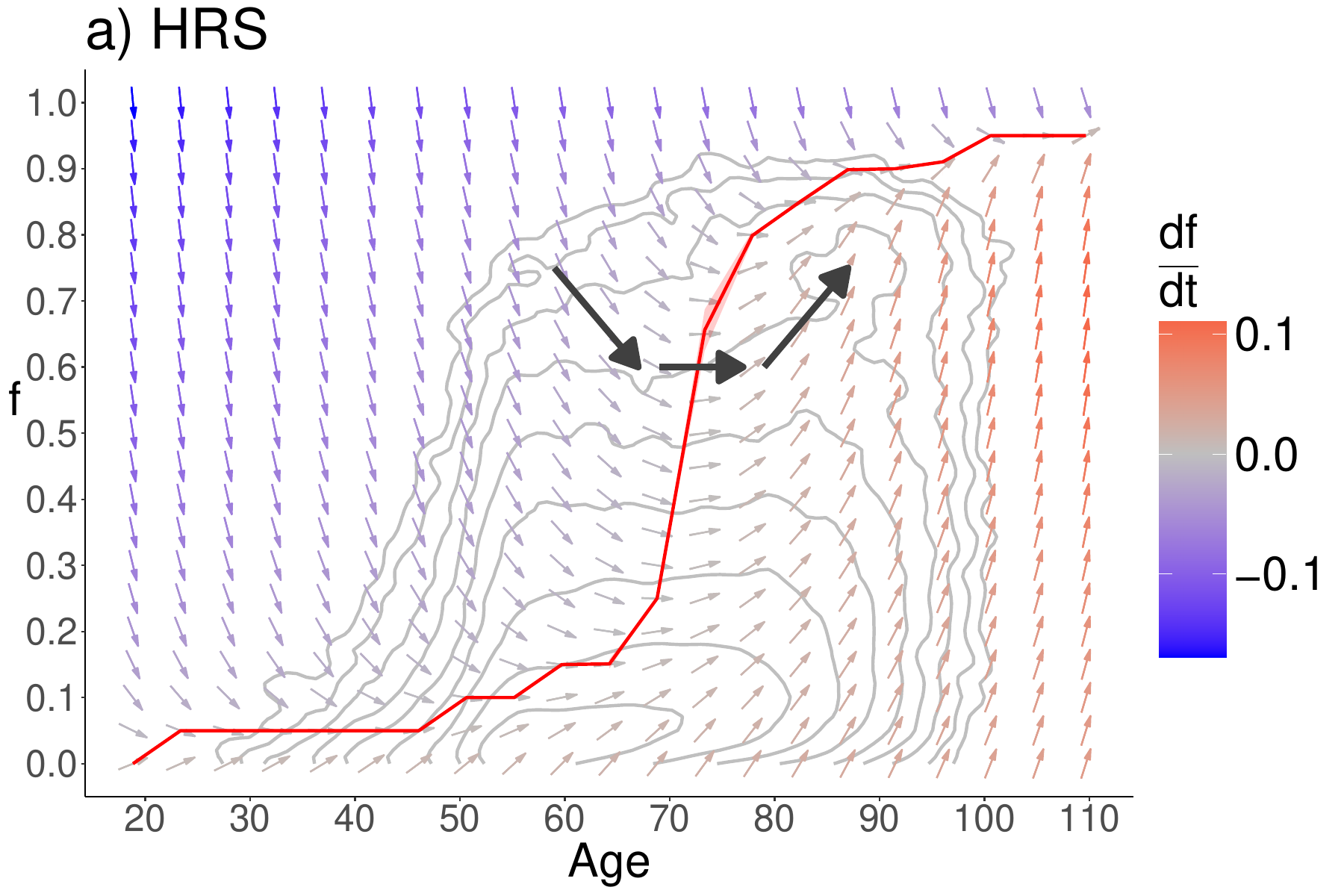} 
\includegraphics[width=.49\textwidth]{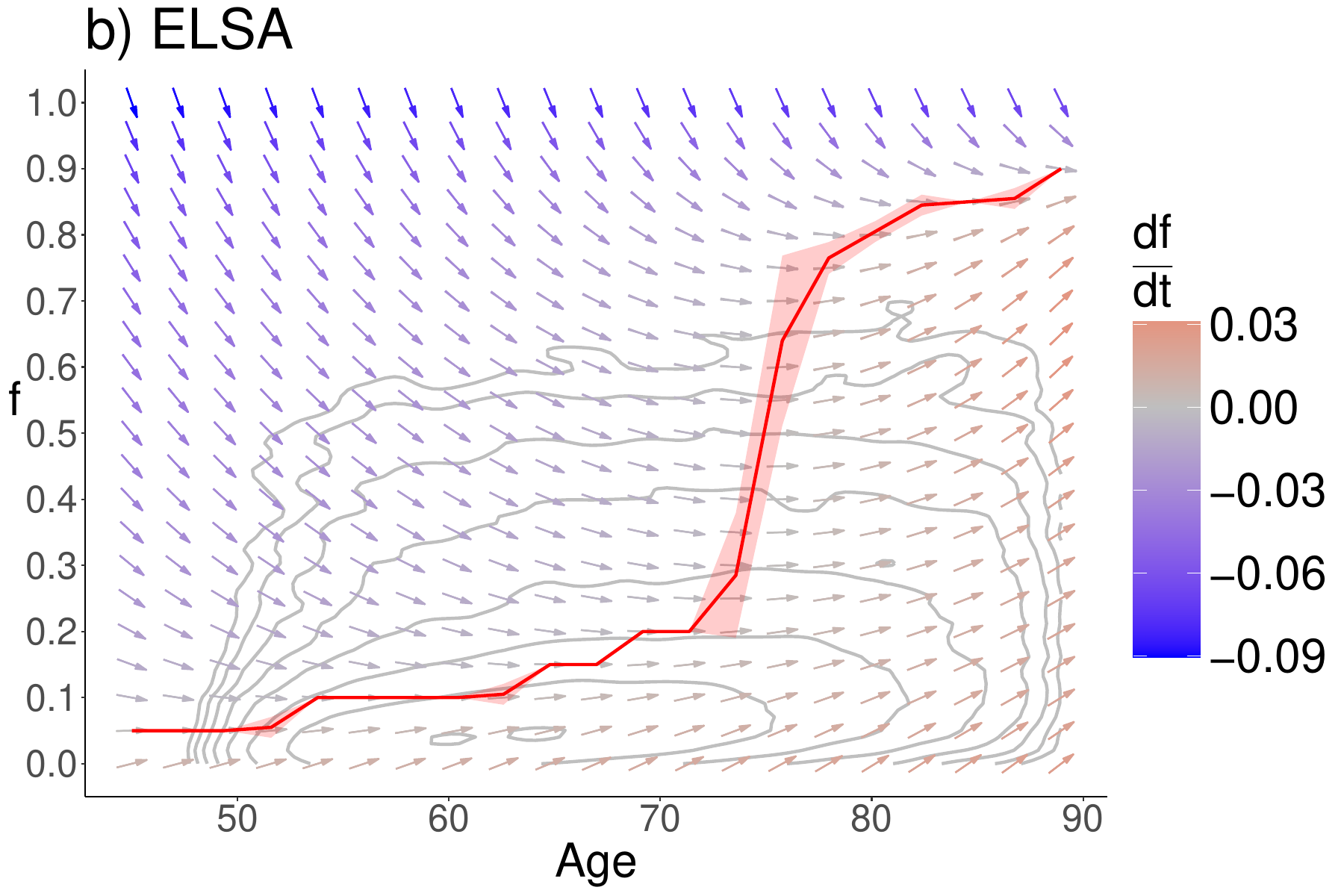} 
%    \centering 
%        \begin{subfigure}[t]{0.49\textwidth}
%        \centering
%        \includegraphics[width=\textwidth]{figures/hrs_dfdt.pdf} 
%        \caption{HRS.}
%    \end{subfigure}
%    ~
%    \begin{subfigure}[t]{0.49\textwidth}
%        \centering
%        \includegraphics[width=\textwidth]{figures/elsa_dfdt.pdf} 
%        \caption{ELSA.}
%    \end{subfigure}%
    \caption{\textbf{FI velocity field} in terms of current health and age. Higher FI, $f$, corresponds to worse health. Small arrows represent the expected flow of individuals at each point. Blue arrows will see a decrease in $f$ over time, red arrows will see an increase. The grey lines are population density contours (binned on a $\log_2$-scale). The population appears to follow the field lines in (a) as indicated by the big arrows. The nullcline (red line) is where the velocity is $0$ and hence the expected flow is no change to $f$. Uncertainties in the arrows are too small to see.} \label{fig:dfdtpop}
\end{figure*}

The velocity field, $df/dt$, is based on model predictions, and allows us to inquire what parts of the velocity field are actually occupied by observed individuals. The population densities are added as contours to the velocity fields in Figure~\ref{fig:dfdtpop}. As we can see, the population is heavily concentrated at low FIs and intermediate ages. Despite this, it is visually plausible that the population is being pushed by the velocity field in analogy to a cloud in the wind (big arrows). At the same time, stochastic effects serve to scatter the population across the velocity field –– while mortality effects prune the population at larger FIs.

\begin{figure*}[h]
    \centering 
    \includegraphics[width=.95 \textwidth]{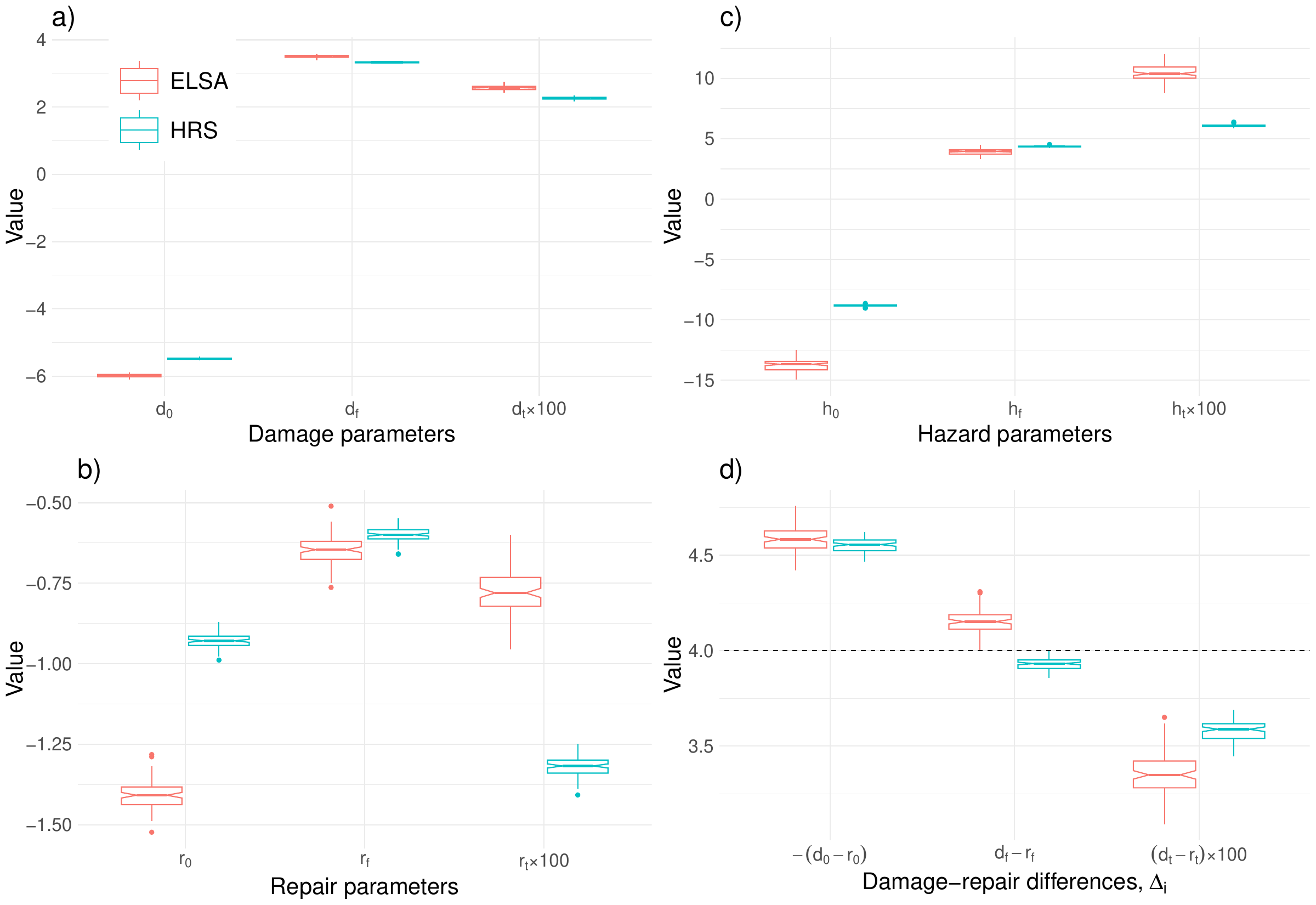} 
    %\begin{subfigure}[t]{\textwidth}
    %    \centering
    %    \includegraphics[width=\textwidth]{figures/hrs_par.pdf} 
    %    \caption{HRS.}
    %\end{subfigure}
    %~
    %\begin{subfigure}[t]{\textwidth}
    %    \centering
    %    \includegraphics[width=\textwidth]{figures/elsa_par.pdf}  
    %    \caption{ELSA.}
    %\end{subfigure}
    \caption{\textbf{Parameter estimates}. Parameter estimates from ELSA (left, red) and HRS (blue, right) are mostly comparable (a-c) -- particularly the damage-repair differences in (d) that control the nullcline position, scale, and shape.  The shape parameter $\Delta_f \equiv d_f-r_f$ is very close to $4$, where a discontinuity in the nullcline appears (see Section~\ref{sec:si:nullcline} details). Changes to robustness appear to dominate, with $|d_f| >|r_f|$ and $|d_t| >|r_t|$ (a and b).  Bar is median, notch is approximate 95\% CI for median, box is interquartile range (IQR), whiskers are 1.5$\times$IQR or furthest datum \cite{Wickham2016-kw} from 100 bootstrap replicates.} \label{fig:par}
\end{figure*}

The estimated fit parameters for 100 bootstrap replicates are presented in Figure~\ref{fig:par}. Observe that in both studies $\Delta_f \approx 4$, which is precisely the value for which there is a vertical tipping point in the nullcline.

%%%%%%%%%%%%%%%%%%%%
\FloatBarrier 

\section{FI variables} \label{sec:si:fi}
We seek to model a realistic FI and hence followed standard rules. For the FI we included 30+ variables from multiple domains \cite{Searle2008-xi}. We used the HRS rand FI developed elsewhere by Theou \textit{et al.} \cite{Theou2023-aw}. We excluded one variable: number of doctor visits in previous 2 years, since it was not obvious how to binarize in a sample-independent manner. The specific variables used are reported in Table~\ref{tab:hrsfi}. 

For ELSA, our list of FI variables is based on previous work by Rogers \textit{et al.} \cite{Rogers2017-ke}. We modified their list by excluding 8 depressive symptoms for sake of convenience. In an effort to compensate for losing these cognitive variables, we extended their list to include Parkinson's, Alzheimer's and dementia diagnoses. The specific list is given in Table~\ref{tab:elsafi}.

\begin{table}
    \centering
        \caption{HRS FI variables used} \label{tab:hrsfi}
    \begin{tabular}{llll}
        ~ & Code & Description & Encoding \\ \hline
        1 & armsa & Difficulty reaching/extending arms up &0: no, 1: yes/any \\ 
        2 & arthre & Ever diagnosed with arthritis &0: no, 1: yes \\ 
        3 & batha & Difficulty bathing or showerng &0: no, 1: yes/any \\ 
        4 & bathh & Gets help bathing, showerng &0: no, 1: yes \\ 
        5 & beda & Difficulty getting in/out of bed &0: no, 1: yes/any \\ 
        6 & bede & Uses equipment to get in/out of bed &0: no, 1: yes \\ 
        7 & bedh & Gets help getting in/out of bed &0: no, 1: yes \\ 
        8 & cancre & Ever diagnosed with cancer &0: no, 1: yes \\ 
        9 & chaira & Difficulty getting up from chair &0: no, 1: yes/any \\ 
        10 & clim1a & Difficulty climbing one stair flight &0: no, 1: yes/any \\ 
        11 & climsa & Difficulty climbing several stair flight &0: no, 1: yes/any \\ 
        12 & diabe & Ever diagnosed with diabetes &0: no, 1: yes \\ 
        13 & dimea & Difficulty picking up a dime &0: no, 1: yes/any \\ 
        14 & dressa & Difficulty dressing &0: no, 1: yes/any \\ 
        15 & dressh & Gets help dressing &0: no, 1: yes \\ 
        16 & eata & Difficulty eating &0: no, 1: yes/any \\ 
        17 & eath & Gets help eating &0: no, 1: yes \\ 
        18 & hearte & Ever diagnosed with heart problems &0: no, 1: yes \\ 
        19 & hibpe & Ever diagnosed with high blood pressure &0: no, 1: yes \\ 
        20 & homcar & Received home health care within previous 2 years &0: no, 1: yes \\ 
        21 & hosp & Had a hospital stay within previous 2 years &0: no, 1: yes \\ 
        22 & lifta & Difficulty lifting/carrying 10lbs &0: no, 1: yes/any \\ 
        23 & lunge & Ever diagnosed with lung disease &0: no, 1: yes \\ 
        24 & moneya & Difficulty managing money &0: no, 1: yes/any \\ 
        25 & nhmliv & Living in nursing home at time of interview &0: no, 1: yes \\ 
        26 & nrshom & Had a nursing home stay within previous 2 years &0: no, 1: yes \\ 
        27 & outpt & Had outpatient surgery within previous 2 years &0: no, 1: yes \\ 
        28 & phonea & Difficulty using the telephone &0: no, 1: yes/any \\ 
        29 & pusha & Difficulty pushing/pulling a large object &0: no, 1: yes/any \\ 
        30 & shlt & Self-reported health & 0: excellent-good, 1: fair-poor \\ 
        31 & shopa & Difficulty shoping for groceries &0: no, 1: yes/any \\ 
        32 & spcfac & Visited a specialized health facility within previous 2 years &0: no, 1: yes \\ 
        33 & stoopa & Difficulty stooping/kneeling/crouching &0: no, 1: yes/any \\ 
        34 & stroke & Ever diagnosed with a stroke &0: no, 1: yes \\ 
        35 & toilta & Difficulty using the toilet &0: no, 1: yes/any \\ 
        36 & toilth & Gets help using the toilet &0: no, 1: yes \\ 
        37 & walk1a & Difficulty walking one block &0: no, 1: yes/any \\ 
        38 & walkra & Difficulty  walking across rooms &0: no, 1: yes/any \\ 
        39 & walkre & Needs equipment to  walk across rooms &0: no, 1: yes \\ 
        40 & walkrh & Gets help walking across rooms &0: no, 1: yes \\ 
        41 & walksa & Difficulty walking several blocks &0: no, 1: yes/any \\ \hline
    \end{tabular}
\end{table}

\begin{table}
    \caption{ELSA FI variables used} \label{tab:elsafi}
    \centering
    \begin{tabular}{lll}
        ~ & Description & Encoding \\ \hline
        1 & Difficulty walking 100 yards & 0: no, 1: yes \\ 
        2 & Difficulty sitting for about two hours & 0: no, 1: yes \\ 
        3 & Difficulty getting up from a chair after sitting for long periods & 0: no, 1: yes \\ 
        4 & Difficulty climbing several flights of stairs without resting & 0: no, 1: yes \\ 
        5 & Difficulty climbing one flight of stairs without resting & 0: no, 1: yes \\ 
        6 & Difficulty stooping kneeling or crouching & 0: no, 1: yes \\ 
        7 & Difficulty reaching or extending arms above shoulder level & 0: no, 1: yes \\ 
        8 & Difficulty pulling pushing large objects like a living room chair & 0: no, 1: yes \\ 
        9 & Difficulty lifting carrying over 10 lbs like a heavy bag of groceries & 0: no, 1: yes \\ 
        10 & Difficulty picking up a 5p coin from a table & 0: no, 1: yes \\ 
        11 & Difficulty dressing including putting on shoes and socks & 0: no, 1: yes \\ 
        12 & Difficulty walking across a room & 0: no, 1: yes \\ 
        13 & Difficulty bathing or showering & 0: no, 1: yes \\ 
        14 & Difficulty eating such as cutting up your food & 0: no, 1: yes \\ 
        15 & Difficulty getting in or out of bed & 0: no, 1: yes \\ 
        16 & Difficulty using the toilet including getting up or down & 0: no, 1: yes \\ 
        17 & Difficulty using a map to get around in a strange place & 0: no, 1: yes \\ 
        18 & Difficulty preparing a hot meal & 0: no, 1: yes \\ 
        19 & Difficulty shopping for groceries & 0: no, 1: yes \\ 
        20 & Difficulty making telephone calls & 0: no, 1: yes \\ 
        21 & Difficulty taking medications & 0: no, 1: yes \\ 
        22 & Difficulty doing work around the house or garden & 0: no, 1: yes \\ 
        23 & Difficulty managing money eg paying bills   keeping track of expenses & 0: no, 1: yes \\ 
        24 & Self-reported general health & 0: very good--good, 1: fair--very bad \\ 
        25 & Self-reported eyesight (corrected) & 0: excellent-good, 1: fair-poor \\ 
        26 & Self-reported hearing (corrected) & 0: excellent-good, 1: fair-poor \\ 
        27 & Chronic: lung disease diagnosis & 0: no, 1: yes \\ 
        28 & Chronic: asthma diagnosis & 0: no, 1: yes \\ 
        29 & Chronic: arthritis diagnosis & 0: no, 1: yes \\ 
        30 & Chronic: osteoporosis diagnosis & 0: no, 1: yes \\ 
        31 & Chronic: cancer diagnosis & 0: no, 1: yes \\ 
        32 & Chronic: Parkinson's diagnosis & 0: no, 1: yes \\ 
        33 & Chronic: psychiatric condition diagnosis & 0: no, 1: yes \\ 
        34 & Chronic: Alzheimer's diagnosis & 0: no, 1: yes \\ 
        35 & Chronic: dementia diagnosis & 0: no, 1: yes \\ 
        36 & CVD: high blood pressure diagnosis & 0: no, 1: yes \\ 
        37 & CVD: angina diagnosis & 0: no, 1: yes \\ 
        38 & CVD: heart attack & 0: no, 1: yes \\ 
        39 & CVD: congestive heart failure diagnosis & 0: no, 1: yes \\ 
        40 & CVD: heart murmur diagnosis & 0: no, 1: yes \\ 
        41 & CVD: abnormal heart rhythm & 0: no, 1: yes \\ 
        42 & CVD: diabetes or high blood sugar diagnosis & 0: no, 1: yes \\ 
        43 & CVD: stroke diagnosis & 0: no, 1: yes \\ \hline 
    \end{tabular}
\end{table}

\FloatBarrier

\section{Model selection} \label{sec:modelselection}
We computed the Bayesian Information Criteria (BIC) and out-of-sample (test) log-likelihood across 100~bootstrap replicates. A lower BIC and higher log-likelihood indicate better performing models. In particular, the log-likelihood captures how well the model fits (train) or predicts (test) the specific health trajectories in the data. For each bootstrap iteration, the out-of-sample individuals are those who were not randomly selected, representing approximately $e^{-1}=37\%$ of the population. The in-sample are the approximately 63\% of individuals who were randomly included. The BIC is computed from the in-sample (train) log-likelihood as
\begin{align}
\text{BIC} &= 6(\text{parameters})\ln(\text{number of data points})-2\text{(in-sample log-likelihood)}.
\end{align}
For the survival component, only entries with recorded deaths are included in the number of data points. The log-likelihood is the value of the objective function, Eq.~2.

\begin{figure*}[!ht]
    \centering 
        \includegraphics[width=\textwidth]{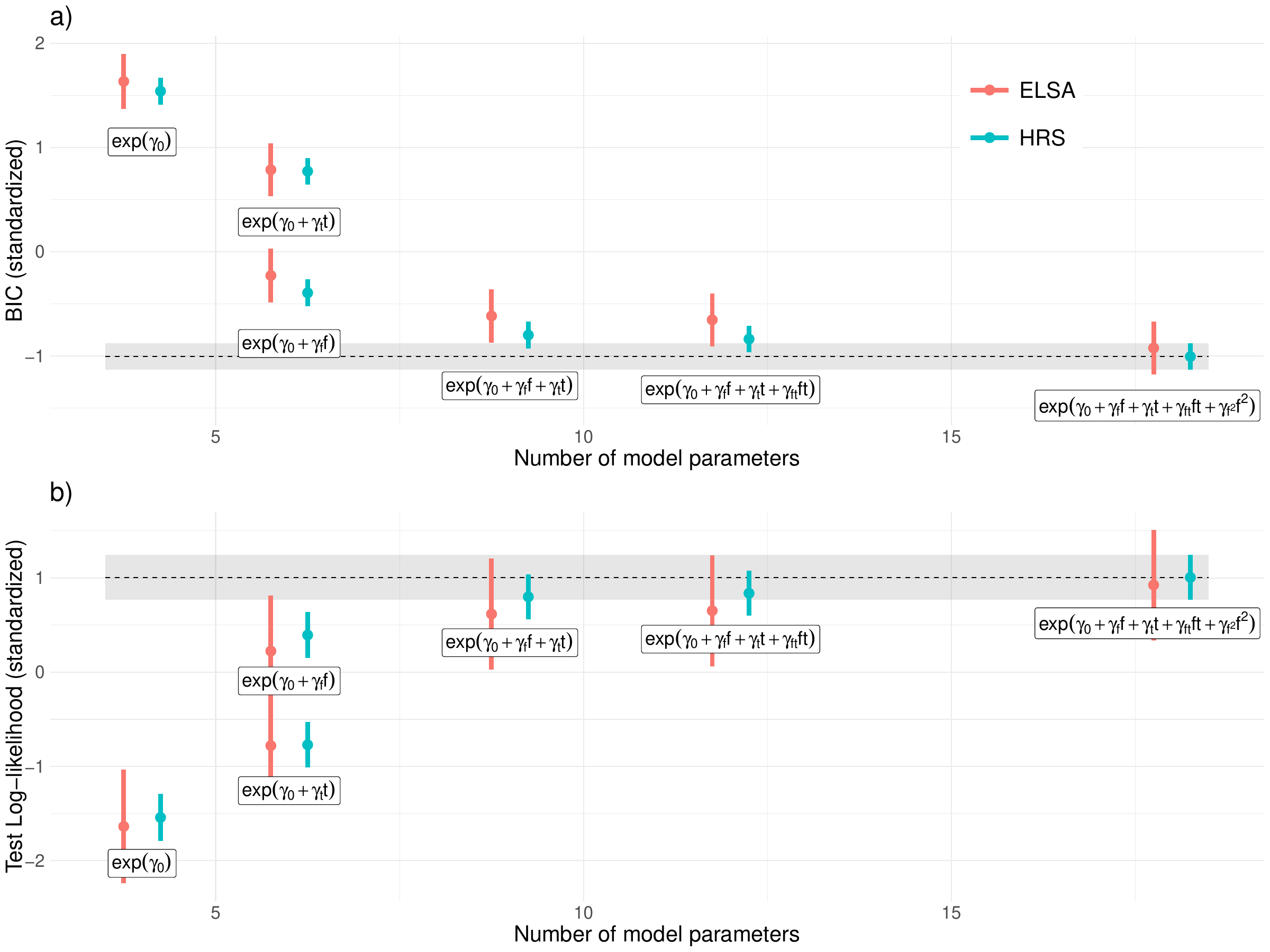} 

    \caption{\textbf{Conservative model selection} shows that linear terms for both $f$ and $t$ are present. More complex models do not appear to improve the fit. Band is best-fitting HRS model. Error bars are standard errors (bootstrap, 100~repeats). Scores have been centered to zero mean and scaled to unit variance for visualization.} \label{fig:modelselection}
\end{figure*}

In Figure~\ref{fig:modelselection} we present the BIC and test log-likelihood with bootstrap errors. We find diminishing returns for models more complex than linear in both $f$ and $t$, suggesting it is an efficient model. Note that for the constant model, $\exp{(\gamma_0)}$ we included a Gompertz term for the survival hazard ($\ln{(D)} = \exp{(d_0)}$, $\ln{(R)} = \exp{(r_0)}$, and $\ln{(h)} = \exp{(h_0+h_t t)}$), since it represents our simplest model and we know human survival is Gompertzian. All other models pick the same parametric form for all three: damage, repair and mortality rates.

%%%%%%%%%%%%%%%%%%%%%%%

\begin{figure*}[!ht]
    \centering 
        \includegraphics[width=\textwidth]{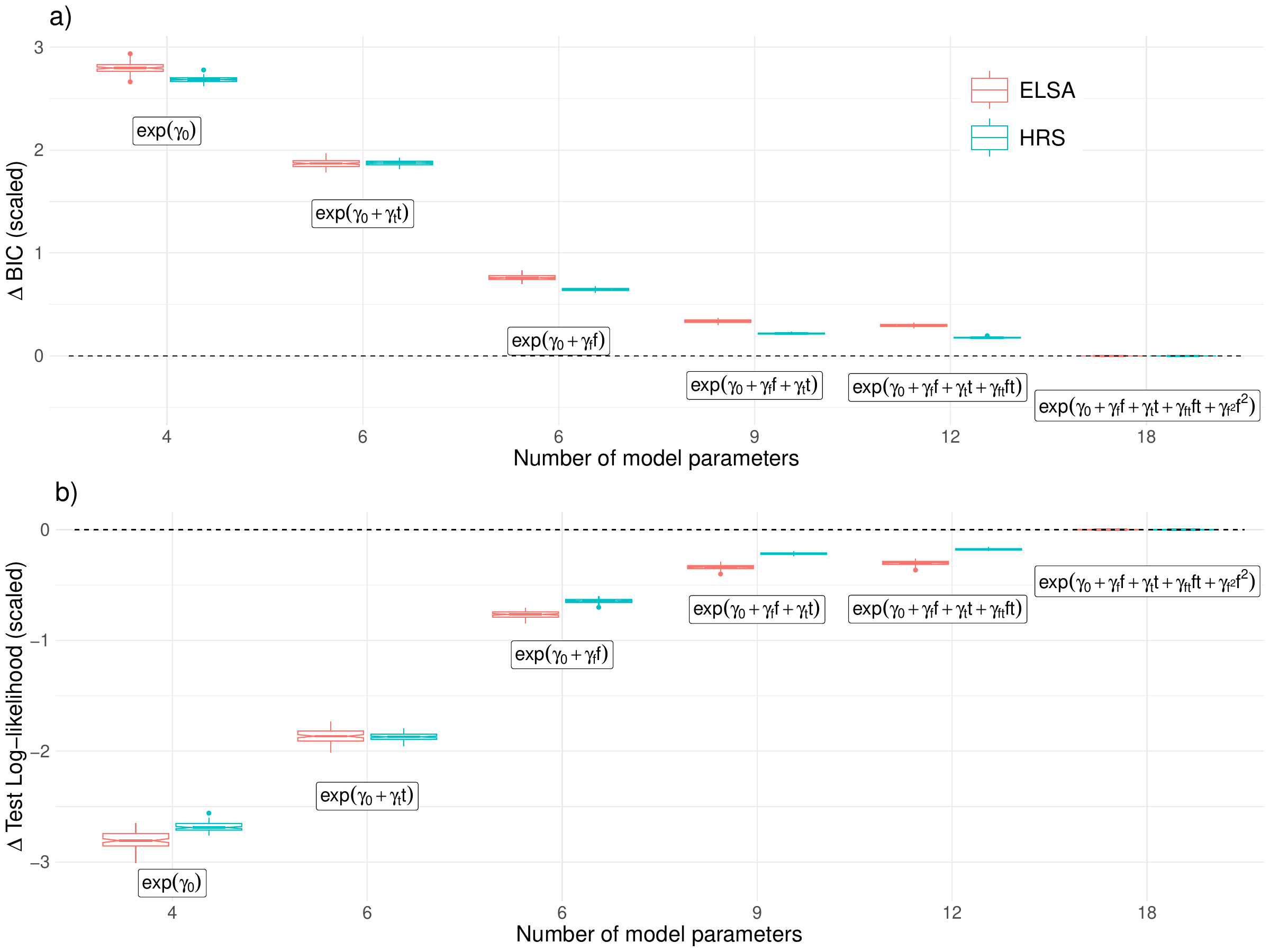} 

    \caption{\textbf{Cavalier model selection} suggests that additional terms may be present, but there is diminishing returns past the linear, $\exp{(\gamma_0+\gamma_ff+\gamma_tt)}$ model. Boxes are median and interquartile range from scores from bootstrap replicates (100~repeats). Scores have been scaled to unit variance for visualization. Bar is median, notch is approximate 95\% CI for median, box is interquartile range (IQR), whiskers are 1.5$\times$IQR or furthest datum \cite{Wickham2016-kw}.} \label{fig:modelselection2}
\end{figure*}

Figure~\ref{fig:modelselection} may be too conservative since it includes variation in the overall predictability of the population. In Figure~\ref{fig:modelselection2} we consider the \textit{differences} in BIC and test log-likelihood, where we have bootstrapped comparing the same populations between the different models. That is, we are comparing the models for 100 replicates of the population, always comparing the models after fitting to the same individuals. Again we find diminishing returns for models more complex than linear in both $f$ and $t$, although here we see a small but significant difference between the linear model and more complex models (particularly after including quadratic $f^2$).

Both the conservative and cavalier model selections indicated diminishing returns past the linear model including $f$ and $t$. The conservative model selection shows non-significant improvement for more complex models whereas the cavalier model selection appears to show significant improvements for more complex models (although they are small relative to the inter-study variability). We infer that the linear model is efficient but it may be possible to improve upon it by including additional terms. We selected the linear model since it is simple and performs as well or almost as well as the more complex models.

\FloatBarrier

\section{Fit diagnostics} \label{sec:si:fit}
\begin{figure*}[!ht]
\centering
\includegraphics[width=\textwidth]{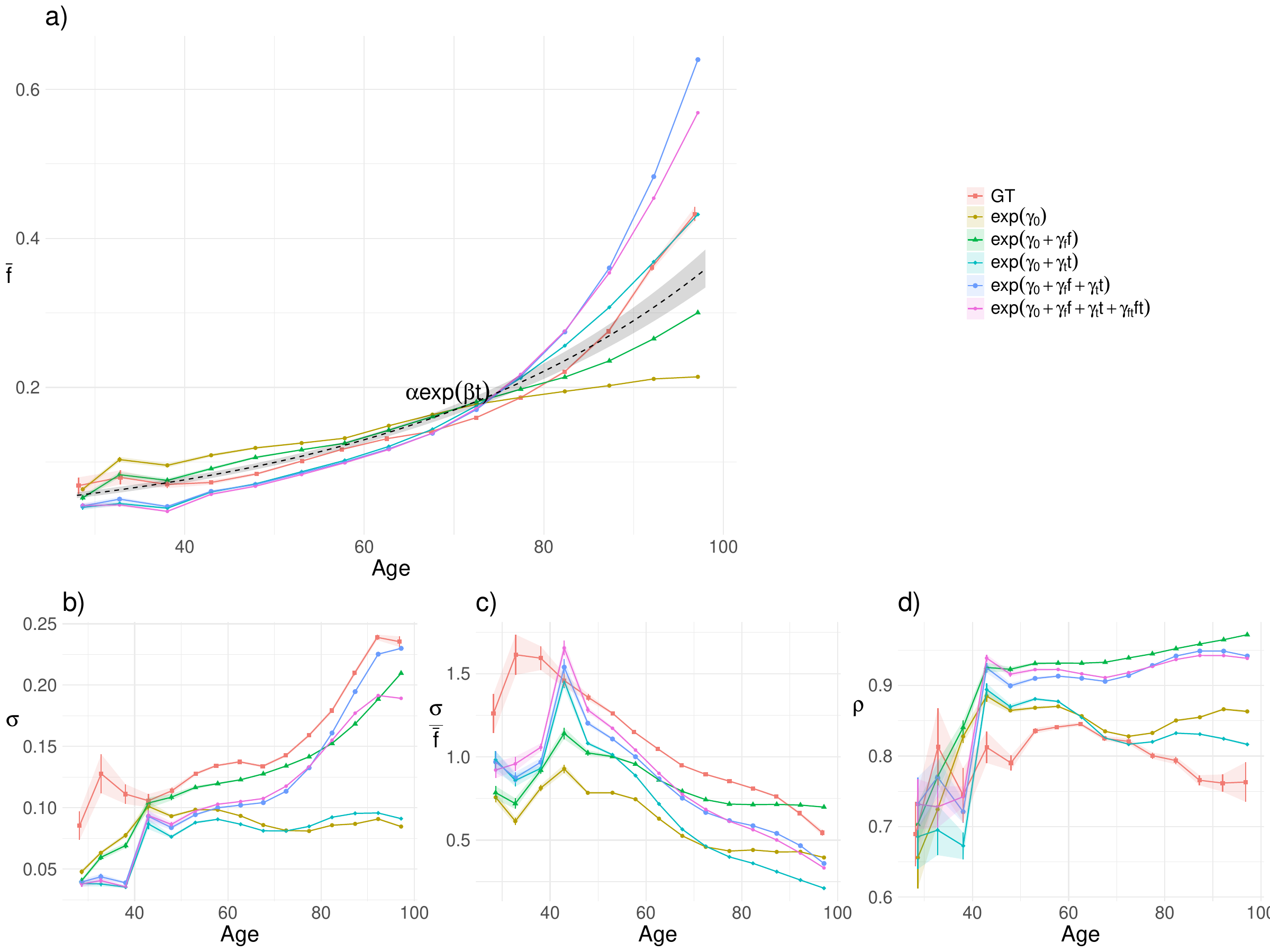} 
    \caption{\textbf{FI population-level statistics for more models} (HRS). Ground truth (GT) versus simulations using model fits. (a) mean FI, $\bar{f}$, (b) FI standard deviation, $\sigma$, (c), coefficient of variation $\sigma/\bar{f}$, and (d) FI auto-correlation, $\rho$ (lag-1). Error bars are standard errors (bootstrap, 100~repeats).} \label{fig:modelphenom2}
\end{figure*}

\begin{figure*}[!ht]
\includegraphics[width=\textwidth]{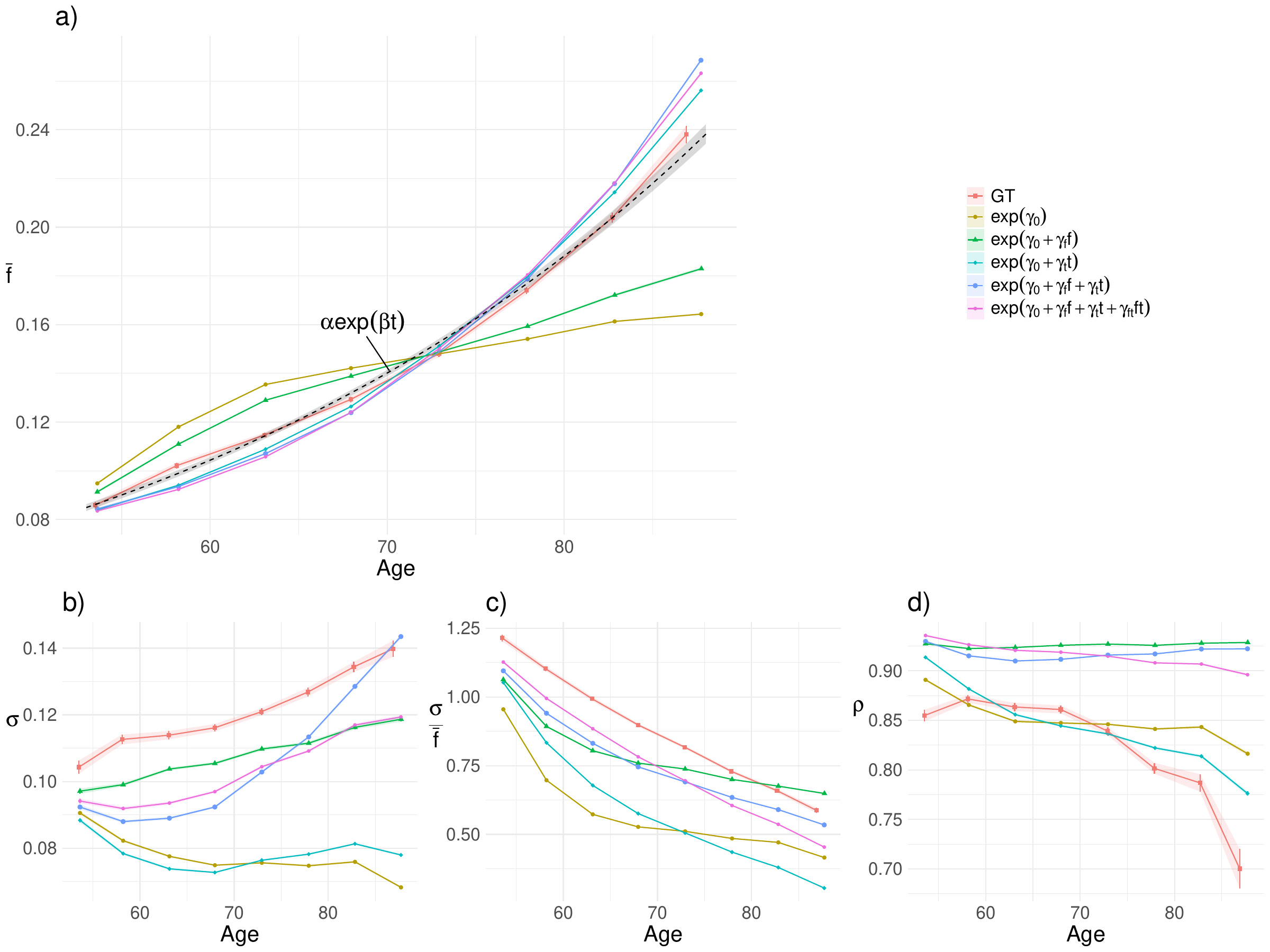} 
    \caption{\textbf{FI population-level statistics for ELSA}. Ground truth (GT) versus simulations using model fits. (a) mean FI, $\bar{f}$, (b) FI standard deviation, $\sigma$, (c), coefficient of variation $\sigma/\bar{f}$, and (d) FI auto-correlation, $\rho$ (lag-1). Error bars are standard errors (bootstrap, 100~repeats).} \label{fig:modelphenomelsa}
\end{figure*}

In the main text we demonstrate our model recapitulates population-level statistics for a representative sample of models. In Figures~\ref{fig:modelphenom2} (HRS) and \ref{fig:modelphenomelsa} (ELSA) we extend the set of models compared. It is interesting that the full Gompertz model, $\exp{(\gamma_0 +\gamma_t t)}$, recapitulates the phenomena despite having a much lower log-likelihood (Section~\ref{sec:modelselection}). This indicates that the time dependence correctly captures the population behaviour but not individual trajectories, which clearly depend on individual health via $f$. We note that consistent with Section~\ref{sec:modelselection}, we see that inclusion of an $f\cdot t$ interaction term does little to modify the model behaviour.
%consistent with the minor change observed in fit metrics (Section~\ref{sec:modelselection}). 

%Notably, the mean fits well quantitatively whereas the higher-order statistics (b-d) tend to be the right shape but wrong scale. 

%    \caption{\textbf{FI population-level statistics} for ELSA. The linear model including both the FI, $f$, and age, $t$, correctly captures the superlinear convexity of the mean FI, the complex shape of the standard deviation, the linearly decreasing CV past age 40, and the roughly constant auto-correlation past age 40. Notably, the mean fits well quantitatively whereas the higher-order statistics (b-d) tend to be the right shape but wrong scale. Note that the exponential model does not correctly capture the curvature of the mean FI: the curvature is too large at young ages and too small at old ages (black dashed line; a). The mean FI increases slowly until around age 70-80 where it starts to rapidly increase. Simulated models versus ground truth (GT). Error bars are standard errors (bootstrap, 100~repeats).} 

\begin{figure*}[!ht]
\includegraphics[width=\textwidth]{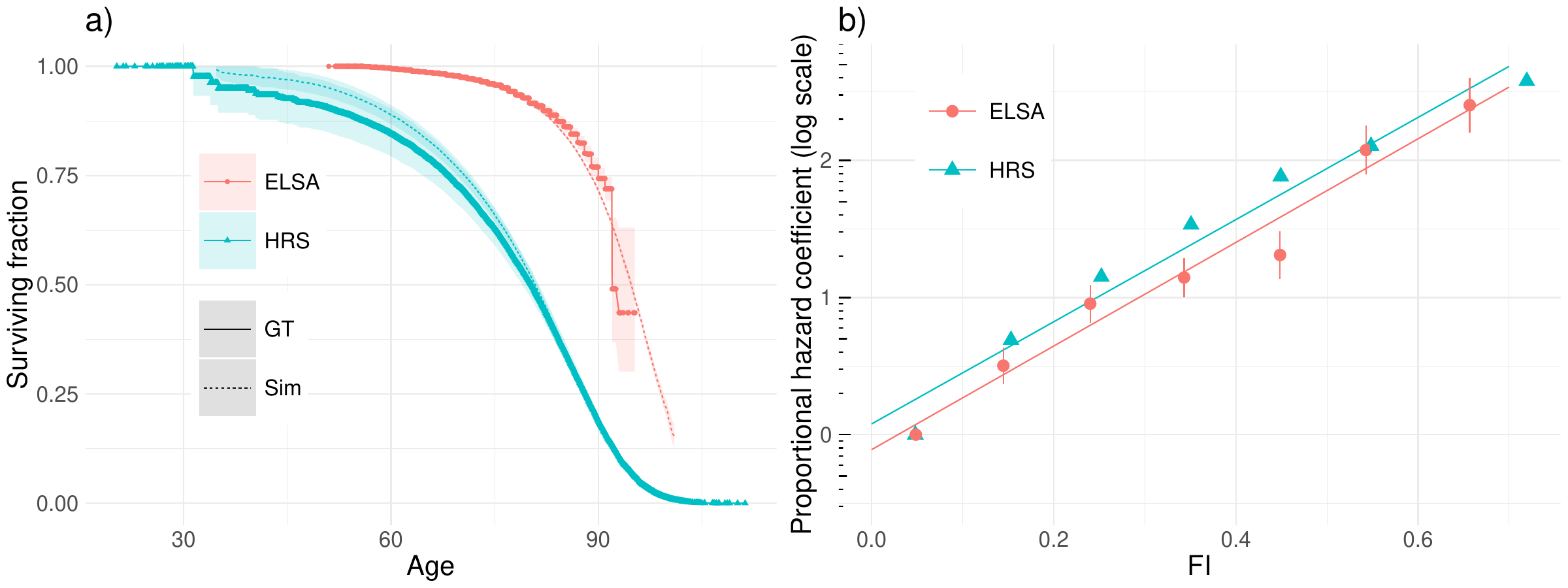}
    \caption{\textbf{Survival fit visually well}. (a) Survival simulation visually reproduced the ground truth (GT). (b) linear proportional hazard (lines) approximates the non-parametric hazard (points) well (time-dependent Cox modelling). The model (a) and assumption (b) both fit visually well. HRS appears to be sublinear in $f$, however (b). Error bars are standard errors. ELSA has lower mortality due to limited data as discussed in the text. } \label{fig:s}
\end{figure*}

Our survival model fit the data reasonably well as visualized in Figure~\ref{fig:s}. The overall population-level survival was correctly reproduced by the simulation (a), and the (linear) proportional hazard assumption for the FI visually fits reasonably well (b). Note that ELSA survival used end-of-life files which are available for only a subset of individuals, resulting in a much lower mortality rate \cite{NatCen_Social_Research2015-nw}. As demonstrated by Figures~\ref{fig:rates} and \ref{fig:par}, the age and FI dependence of mortality in both HRS and ELSA were nevertheless similar, with the primary difference being the baseline hazard (much lower for ELSA) and age-dependence (stronger for ELSA). This suggests most of the exclusions in ELSA survival were completely at random, and that our primary results remain consistent between the two studies.
%This suggests that most of the exclusions in ELSA survival were completely at random, though possibly with a weak age-dependence. 

\section{Tuning the effective number of health attributes} \label{sec:si:ptuning}
\begin{figure*}[!ht]
    \centering 
        \includegraphics[width=\textwidth]{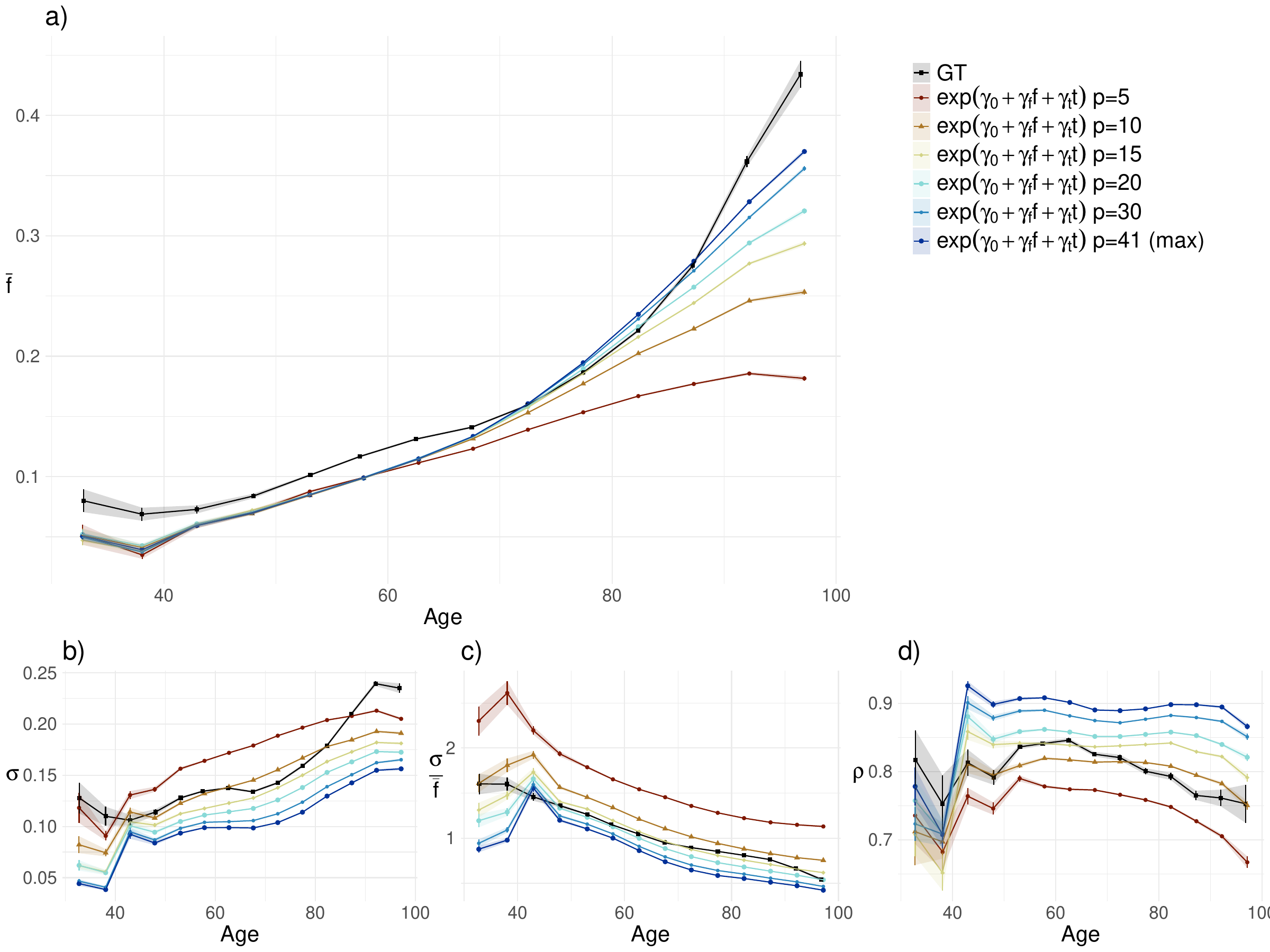} 

    \caption{\textbf{Number of health attributes strongly affects higher-order statistics}, including standard deviation (b), CV (c) and auto-correlation (d). HRS. We see good visual agreement for these statistics for between 10 (orange triangles) and 20 (teal dots) health attributes rather than the true number, 41. This is likely due to the underlying correlations between health attributes that reduces the degrees of freedom and thus effective number of independent health attributes. The mean appears to fit worse with decreasing number of attributes, however (a). Error bars are standard errors (bootstrap, 100~repeats).} \label{fig:tuning}
\end{figure*}

The effective number of health attributes depends on the underlying correlation structure, which prior research indicates contains nested domains \cite{Pridham2023}. This reduces the effective number of deficits compared to our model, which assumes conditional independence given $f$ and age (which should be sensitive to overall health but not domains). When we simulate fewer deficits we see better quantitative agreement for the higher order statistics, Figure~\ref{fig:tuning}. This comes at the cost of the mean, however, which underestimates the curvature at old ages. A plausible reason for this is that the increased variance also increase the hazard\cite{Pridham2024-kidney}, magnifying the survival misfit (Figure~\ref{fig:s}b). Alternatively, the missing nested correlation structure could be excluding mutual events at older ages. Regardless, our model with fewer effective parameters nevertheless approximates the data reasonably well.
% We infer that the true model includes additional, nested domains of correlated deficits. 
%If deficits are highly correlated then they will go deficit together, leading to a sharp increase at old ages -- as desired. 

\FloatBarrier

%%%%%%%
\section{Sex effects} \label{sec:si:sex}
\begin{figure*}[!ht]
\includegraphics[width=\textwidth]{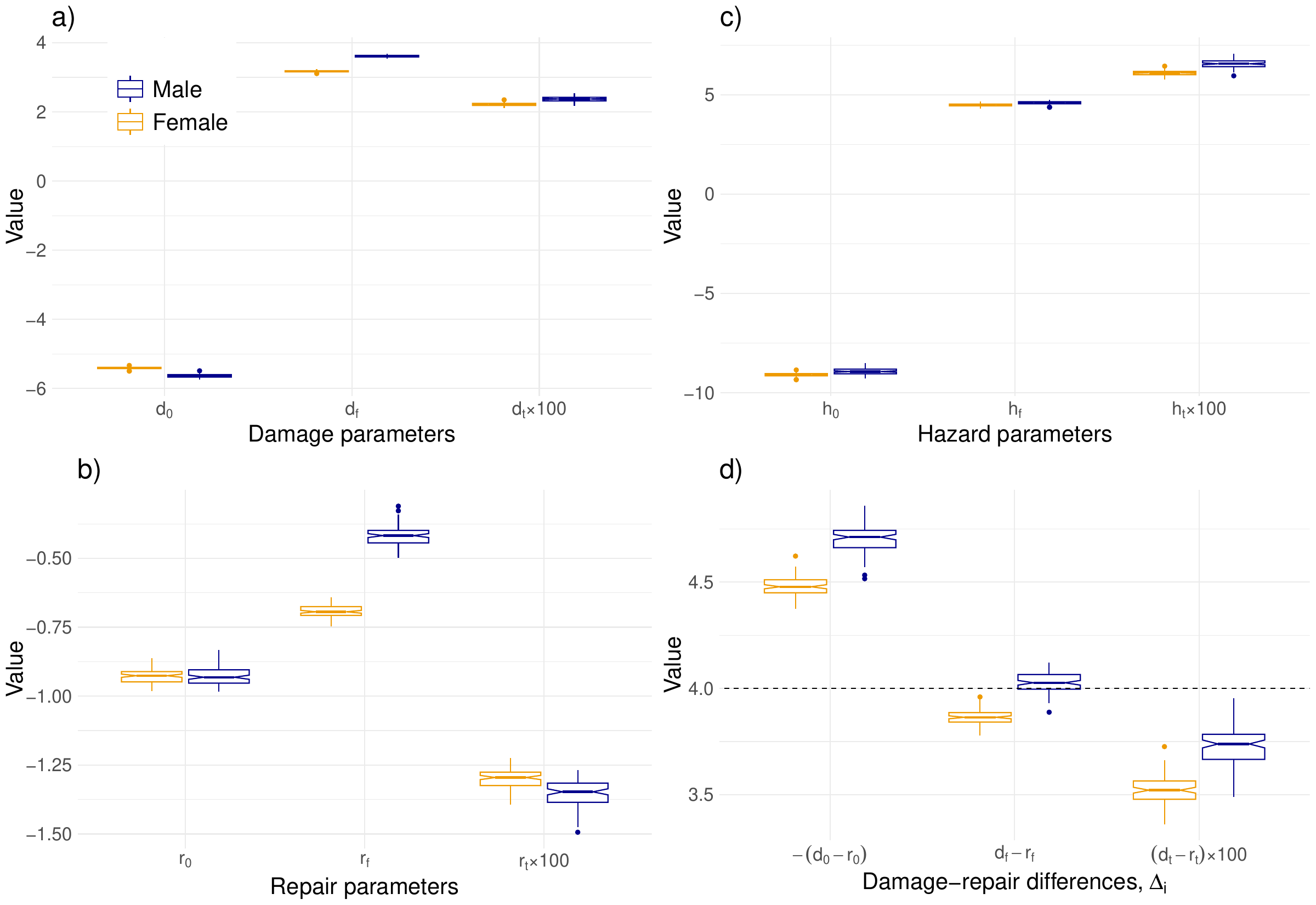} 
    \caption{\textbf{Parameter estimates by sex -- HRS}. Each point is an estimate from 100~bootstrap replicates. Males have better baseline health $\Delta_0$, but are more sensitive to poor health $\Delta_f$ and old ages $\Delta_t$ (note the signs in d). Bar is median, notch is approximate 95\% CI for median, box is interquartile range (IQR), whiskers are 1.5$\times$IQR or furthest datum \cite{Wickham2016-kw}.} \label{fig:sex_par_hrs}
\end{figure*}

\begin{figure*}[!ht]
\includegraphics[width=\textwidth]{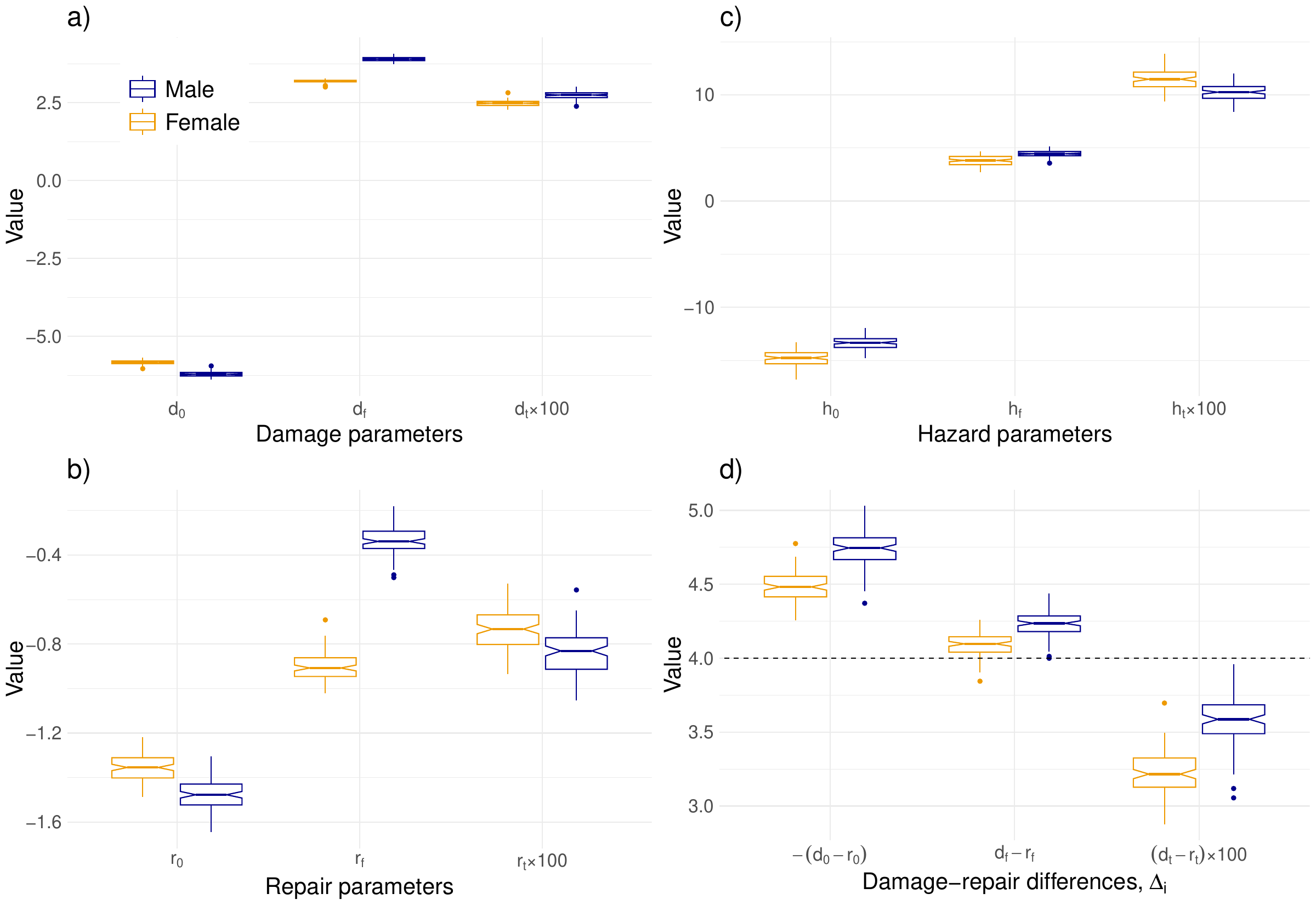} 
    \caption{\textbf{Parameter estimates by sex -- ELSA}. Each point is an estimate from 100~bootstrap replicates. As with HRS, males have better baseline health $\Delta_0$, but are more sensitive to poor health $\Delta_f$ and old ages $\Delta_t$ (note the signs in d). Bar is median, notch is approximate 95\% CI for median, box is interquartile range (IQR), whiskers are 1.5$\times$IQR or furthest datum \cite{Wickham2016-kw}.} \label{fig:sex_par_elsa}
\end{figure*}

We observed sex differences in the parameter estimates across HRS and ELSA (Figures~\ref{fig:sex_par_hrs} and \ref{fig:sex_par_elsa}, respectively). Each point in the boxplot is a bootstrap replicate, meaning that non-overlapping box notches have significantly different median parameter estimates. Males showed lower baseline damage rate $d_0$ but faster loss of robustness with worsening health $d_f$ and increasing age $d_t$. They also showed a greater loss of resilience with worsening health $r_f$, a higher baseline hazard $h_0$ and a higher hazard due to worse health $h_f$. The net cumulative effects of these parameterizations for males is: lower $\Delta_0$, higher $\Delta_f$, and higher $\Delta_t$.

\begin{figure*}[!ht]
\includegraphics[width=.49\textwidth]{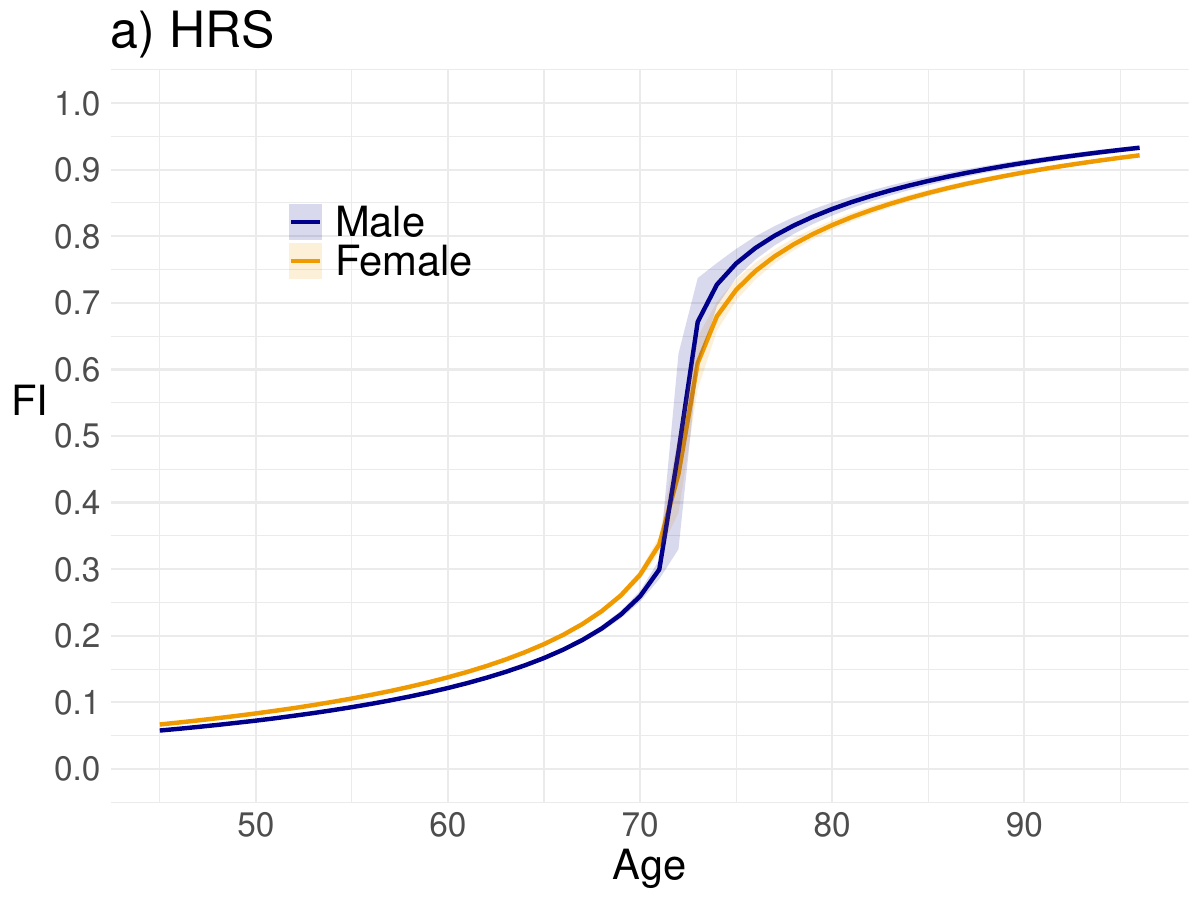}
\includegraphics[width=.49\textwidth]{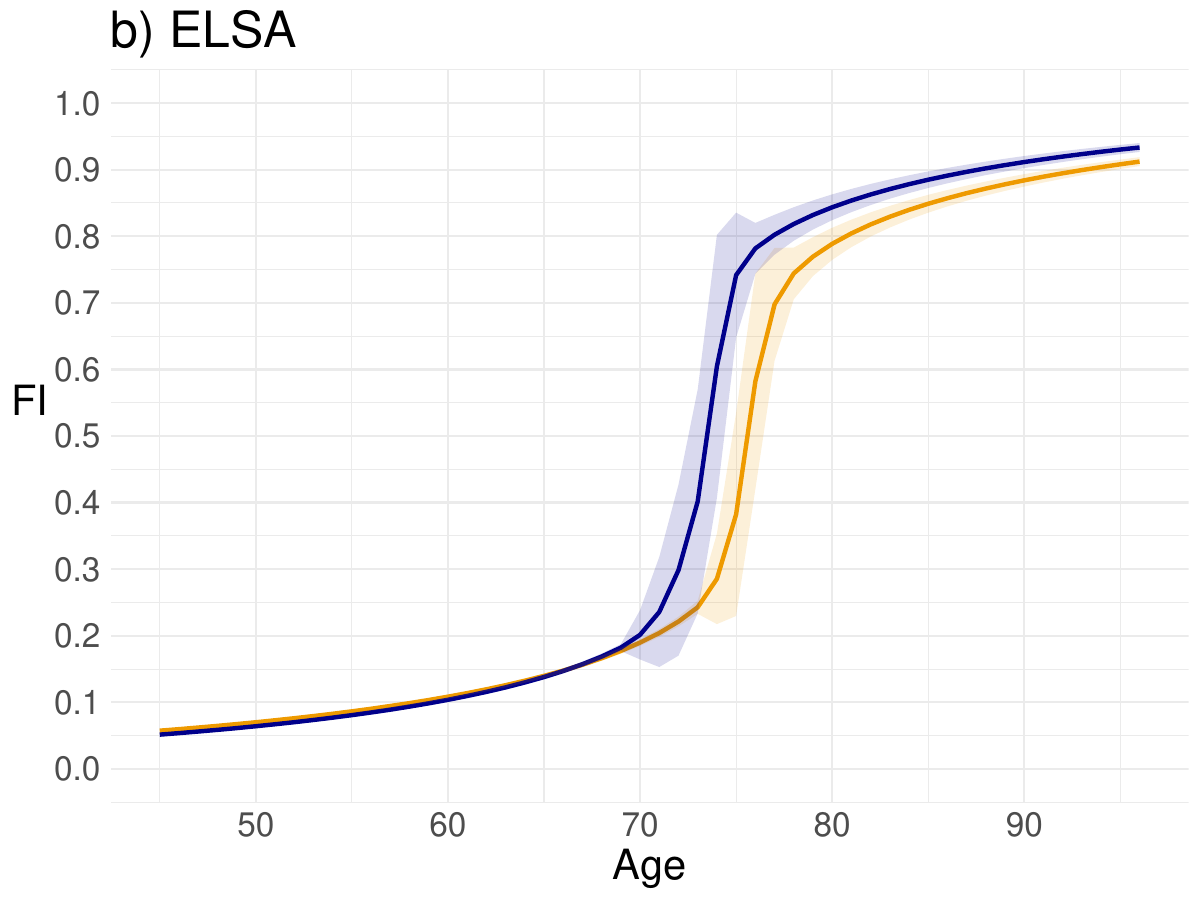} 
    \caption{\textbf{Nullcline by sex} based on parameter values. Males consistently start out lower at young ages but cross and end up higher at older ages. Error bars are standard errors (bootstrap, 100~repeats).} \label{fig:sex_nullcline}
\end{figure*}

The effect of the parameterization on the nullcline is shown in Figure~\ref{fig:sex_nullcline}. Males are stable at a lower FI at young ages but a higher FI at older ages. Males are known to have a lower FI than females but do not live as long \cite{Hubbard2015-tq}. According to our parameter estimates this is because males have lower initial damage rate, $d_0$, but are more sensitive to loss of robustness and resilience with both age and worsening health, $\Delta_f$ and $\Delta_t$. Death appears to be via higher baseline hazard $h_0$ and greater sensitivity to poor health $h_f$. These results are consistent with males reaching adulthood in a more robust state but being less tolerant to age-related decline. Males appear to age faster but from a better starting point. 
%Biological aging rate is ostensibly proportional to both $\Delta_f$ and $\Delta_t$, both of which are notably higher in men.

\FloatBarrier

\section{Nullcline analysis} \label{sec:si:nullcline}
%\textbf{There must be a saddlepoint for $\Delta_f > 4$, where is it? 0.5? prove}
Figure~4 demonstrates that for our parameter values there is a tipping point near age~75. Here we analytically show that this occurs for $\Delta_f=4$, consistent with model estimates from the data. 

Take the FI dynamics of our best model are (Section~\ref{sec:modelselection}),
\begin{align}
    \frac{d}{dt}f &= (1-f)\exp{\bigg( d_0+d_ff+d_tt \bigg)} - f\exp{\bigg( r_0+r_ff+r_tt \bigg)}
\end{align}
where $f$ is the FI.

From here there are five key steps: (i) find the nullcline where $df/dt=0$, (ii) solve for $t^*(f^*)$, (iii) find the critical points of $t^*(f^*)$, (iv) solve for $\Delta_f(f)\equiv d_f-r_f$, and finally (v) find the critical points $\Delta_f^*(f)$. The nullcline occurs when the derivative is $0$ i.e.\
\begin{align}
    \ln{(1-f^*)} + d_0+d_ff^*+d_tt^*&= \ln{(f^*)} + r_0+r_ff^*+r_tt^*.
\end{align}
It is useful to define
%\begin{subequations}
    \begin{align}
        \Delta_i \equiv d_i - r_i,
    \end{align}
%\end{subequations}
then we rearrange and solve for $t$,
\begin{align}
     t^*\Delta_t &= -\Delta_0-\Delta_ff^* + \ln{\bigg(\frac{f^*}{1-f^*}\bigg)} \nonumber \\
\implies t^* &= -\frac{\Delta_0}{\Delta_t}-\frac{\Delta_f}{\Delta_t}f^* + \frac{1}{\Delta_t}\ln{\bigg(\frac{f^*}{1-f^*}\bigg)}.
\end{align}

This defines a curve, though not necessarily a function (since more than one value of $f^*$ could have the same $t^*$). This curve has a discontinuity when $dt^*/df^* = 0$,
\begin{align}
    \frac{d t^*}{df^*}  &= \frac{-\Delta_f + \frac{1-f^*}{f^*}\bigg(\frac{1}{1-f^*}+\frac{f^*}{(1-f^*)^2}\bigg)}{\Delta_t} \nonumber \\
    &= -\frac{\Delta_f}{\Delta_t} + \frac{1}{\Delta_tf^*(1-f^*)}.
\end{align}

We can have $dt^*/df^* = 0$ only if
\begin{align}
    \Delta_f &= \frac{1}{f^*(1-f^*)}, \nonumber \\
\implies f^*(1-f^*) &= \frac{1}{\Delta_f}.
\end{align}
When can this happen? The critical points of $x(1-x)$ are
\begin{align}
    1-x^* &= x^*, \nonumber \\
\implies x^* &= 1/2
\end{align}
with $.5^2=1/4$ being the value taken at this point.  $1/4$ is the maximum of $x(1-x)$, which is symmetrical about $x=1/2$ and bounded by 0.

This means that the smallest possible value of $\Delta_f$ for which there exists a critical point is $\Delta_f=4$. At the critical point $dt^*/df^*=0$ and hence $df^*/d t^*\to\infty$ diverges leading to a discontinuity. For larger $\Delta_f$ the solutions bifurcate; for smaller there is no solution.

%The critical point occurs at time
%\begin{align}
%     t^* &= -\frac{\Delta_0}{\Delta_t}-\frac{\Delta_f}{\Delta_t}\bar{f}^* + \frac{1}{\Delta_t}\ln{\bigg(\frac{\bar{f}^*}{1-\bar{f}^*}\bigg)}
%\end{align}
%     t^* &= -\frac{\Delta_0}{\Delta_t}-\frac{1}{\Delta_t(1-\bar{f}^*)} + \frac{1}{\Delta_t}\ln{\bigg(\frac{\bar{f}^*}{1-\bar{f}^*}\bigg)} 
%&=  -\frac{\Delta_0}{\Delta_t}-\frac{2}{\Delta_t},
%for $\Delta_f=4$.
%     &=   -\frac{\Delta_0}{\Delta_t}-\frac{\Delta_f}{2\Delta_t},
We have $\Delta_f\approx 4$ and hence we know $\bar{f}^* \approx 1/2$. For $\bar{f}^* = 1/2$ the critical time, $t^*$, has a discontinuity at $72.3\pm0.3$~years for HRS and $75.0\pm0.5$~years for ELSA, consistent with what was observed in Figures~2 and 3. Note that for $\Delta_f=4$ the expression is simply $t^* = -\Delta_0/\Delta_t - 2/\Delta_t$.

%%%%%%%%%%%%%%%%%%%%%%%%%%%%%%%%%
\section{Objective function} \label{sec:si:obj}
We fit to longitudinal health deficit data and survival using the modified log-likelihood including survival and its gradient, then optimizing using the BFGS quasi-Newton's method \cite{R_Core_Team2021-uq}. The fit is repeated 100 times via bootstrapping i.e.\ random sampling of individuals from the population (with replacement). The log-likelihood is based on transition rates between states, and is shown to be consistent with the stochastic model of the FI, Eq.~1, in Section~\ref{sec:consistent}.

We consider the survival-modified likelihood using the \textit{a posteriori}
\begin{align}
    p(\text{data} | \text{parameters}) &= p(\text{parameters} | \text{data}) p(\text{data}) \nonumber \\
    &= \text{likelihood}\cdot\text{prior}.
\end{align}
We know that $p(\text{data})$ is $0$ if an individual is dead and should gradually decrease from $1$ to $0$ as an individual's health declines. We can measure this decline by the FI and hence
\begin{align}
    p(\text{data} | f, \Delta t, \delta) &= S(f,\Delta t)h(f,\Delta t)^\delta
\end{align}
where $S$ is the probability of surviving from the previous measurement to the current time point, $\delta = 1$ is the individual is dead and $0$ otherwise. The observation of each datum will have this term.

We can then write the survival-modified likelihood as
\begin{align}
    p &= \prod_{i=1}^N \prod_{j=1}^p\prod_{k=1}^{T-1}\bigg[ Pr(b_{ijk}=1 | b_{ik-1}, f_{ik-1})^{b_{ijk}}(1-Pr(b_{ijk}=1 | b_{ik-1}, f_{ik-1}))^{1-{b_{ijk}}}\bigg]^{1-\delta_{ik}}S(f_{ik}, f_{ik-1}, \Delta t_{ik})h(f_{ik},f_{ik-1})^{\delta_{ik}}.
\end{align}
On the log scale,
\begin{align}
    l &= \sum_{i=1}^N \sum_{j=1}^p\sum_{k=1}^{T-1} (1-\delta_{ik}) \bigg[b_{ijk}\ln{(Pr(b_{ijk}=1 | b_{ik-1}, f_{ik-1}))} + (1-{b_{ijk}})\ln{(1-Pr(b_{ijk}=1 | b_{ik-1}, f_{ik-1}))}\bigg] \nonumber \\
    &+\sum_{i=1}^N \sum_{k=1}^{T-1} \bigg[\ln{(S(f_{ik}, f_{ik-1}, \Delta t_{ik}))}+\delta_{ik}\ln{(h(f_{ik},f_{ik-1}))} \bigg].
\end{align}
where we have $N$ individuals, $p$ variables and $T$ time points. The first term captures health transitions and the second captures survival.
%The first term is a \textit{likelihood} whereas the last term is a survival (likelihood) which we treat as a \textit{prior}.
%\textbf{Note}: scaling up hazard/survival by $p$ doesn't work.

The likelihood has four cases for health transitions:
\begin{subequations}
    \begin{align}
    1-S_d &\approx Pr(b_{ik-1} = 0~\text{and}~b_{ik}=1)~\text{(damage),}
    \end{align}
    \begin{align}
    1-S_r &\approx Pr(b_{ik-1} = 1~\text{and}~b_{ik}=0)~\text{(repair),}
    \end{align}
    \begin{align}
    S_r &\approx Pr(b_{ik-1} = 1~\text{and}~ b_{ik}=1)~\text{(`survived' repair),}
    \end{align}
    \begin{align}
    S_d &\approx Pr(b_{ik-1} = 0 ~\text{and}~ b_{ik}=0)~\text{(`survived' damage),}
    \end{align}
\end{subequations}
where $S_d$ and $S_r$ are defined as the survival probability for damage and repair, respectively. Specifically,
\begin{subequations} \label{eq:sdsr} 
\begin{align}
    S_d &\equiv \exp{\bigg( -\int_{t_{k-1}}^{t_{k}} D(f,t) dt \bigg)}
\end{align}
\begin{align}
    S_r &\equiv \exp{\bigg( -\int_{t_{k-1}}^{t_{k}} R(f,t) dt \bigg)}  
\end{align}
\end{subequations}
where $D$ and $R$ are the damage and repair hazards, respectively.

There are an infinite number of paths connecting each time point pair, and in principle variables may have an infinite number of damage/repair cycles between measurements. These cycles are regulated by $\Delta t$ e.g.
\begin{align}
    Pr(b_{ik-1} = 0~\text{and}~b_{ik}=0) &= S_d + (1-S_d)(1-S_r) + ... \nonumber \\
    &\approx 1 - D \Delta t + D R\Delta t^2 + O(\Delta t^3) \nonumber\\
    &\approx S_d. \label{eq:cycles}
\end{align}
%\begin{align}
%    Pr(b_{ik-1} = 0~\text{and}~b_{ik}=1) &= 1-S_d + (1-S_d)(1-S_r)(1-S_d) + ... \nonumber \\
%    &\approx D \Delta t + D^2R\Delta t^3 + O(\Delta t^5) \nonumber\\
%    &\approx 1-S_d
%\end{align}
In general, the lowest order ($\Delta t$) for each of the four cases is given by $S_r$, $S_d$, $1-S_r$ or $1-S_d$. Thus for sufficiently small $\Delta t$ we can concern ourselves entirely with the survival probabilities. 
%Furthermore, for small $\Delta t$ we have at most one event per time interval meaning that we can assume piece-wise constant $f$ between time points. 
%The true underlying model is much more complex and requires a path integral, which we will treat later using both simulation and the differential equation formalism.

The likelihood in terms of $S_r$ and $S_d$ is thus
\begin{align}
    l &\equiv \sum_{i=1}^N \sum_{j=1}^p\sum_{k=1}^{T-1} (1-\delta_{ik}) \bigg[ \nonumber \\
    &\phantom{+}b_{ijk}(1-b_{ijk-1})\ln{(1-S_d)}  \nonumber \\
    &+(1-b_{ijk})b_{ijk-1}\ln{(1-S_r)} \nonumber \\
    &+b_{ijk}b_{ijk-1}\ln{(S_r)} \nonumber \\
    &+ (1-{b_{ijk}})(1-{b_{ijk-1}})\ln{(S_d))}\bigg] \nonumber \\
    &+\sum_{i=1}^N \sum_{k=1}^{T-1} \bigg[\ln{(S(f_{ik}, f_{ik-1}, \Delta t_{ik}))}+\delta_{ik}\ln{(h(f_{ik},f_{ik-1}))} \bigg],
\end{align}
which becomes exact in the small $\Delta t$ limit (small enough that multiple events between time points becomes negligible).

\section{Gradient}
The gradient is needed for efficient optimization. Assume that the model parameters for $D$ are $\vec{\theta}_d$, $R$ are $\vec{\theta}_r$, and $h$ are $\vec{\theta}_h$ with no overlap between the three. The gradient in general form is then
\begin{subequations}
    \begin{align}
    \frac{\partial}{\partial \theta_{dn}}l &= \sum_{i=1}^N \sum_{j=1}^p\sum_{k=1}^{T-1} (1-\delta_{ik}) \bigg[ \nonumber \\
    &\phantom{+}b_{ijk}(1-b_{ijk-1})\frac{-S_d}{(1-S_d)}\bigg( -\int_{t_{k-1}}^{t_{k}} \frac{\partial}{\partial\theta_{dn}} D(f,t) dt \bigg)  \nonumber \\
    &+ (1-{b_{ijk}})(1-{b_{ijk-1}})\bigg( -\int_{t_{k-1}}^{t_{k}} \frac{\partial}{\partial\theta_{dn}} D(f,t) dt \bigg)\bigg],
    \end{align}
    \begin{align}
    \frac{\partial}{\partial \theta_{rn}}l &= \sum_{i=1}^N \sum_{j=1}^p\sum_{k=1}^{T-1}  (1-\delta_{ik}) \bigg[ \nonumber \\
    &+(1-b_{ijk})b_{ijk-1}\frac{-S_r}{(1-S_r)}\bigg( -\int_{t_{k-1}}^{t_{k}} \frac{\partial}{\partial\theta_{rn}} R(f,t) dt \bigg) \nonumber \\
    &+b_{ijk}b_{ijk-1}\bigg( -\int_{t_{k-1}}^{t_{k}} \frac{\partial}{\partial\theta_{rn}} R(f,t) dt \bigg),~\text{and}
    \end{align}
    \begin{align}
    \frac{\partial}{\partial \theta_{hn}}l &= \sum_{i=1}^N \sum_{k=1}^{T-1}\bigg( -\int_{t_{k-1}}^{t_{k}} \frac{\partial}{\partial\theta_{hn}} h(f,t) dt \bigg)+\delta_{ik}\frac{1}{(h(f_{ik},f_{ik-1}))}\bigg( \frac{\partial}{\partial\theta_{hn}} h(f,t)  \bigg) \bigg]. \label{eq:dl}
    \end{align}
\end{subequations}

\textbf{Rate Models}\\
An important consideration for the hazards is non-negativity, which is preferably hard-coded by choice of model. The most general but solvable exponential model is a generalized additive model with Gompertz term,
\begin{subequations}
    \begin{align}
        D(f,t) &\equiv \exp{\bigg(\sum_j d_j \phi_j(f) + t \sum_j d_{tj} \phi_j(f) \bigg)},
    \end{align}
    \begin{align}
        R(f,t) &\equiv \exp{\bigg(\sum_j r_j \phi_j(f) + t \sum_j r_{tj} \phi_j(f) \bigg)},~\text{and}
    \end{align}
    \begin{align}
        h(f,t) &\equiv \exp{\bigg(\sum_j h_j \phi_j(f) + t \sum_j h_{tj} \phi_j(f) \bigg)},
    \end{align}
\end{subequations}
where $\phi_j(f)$ is any desired function of $f$, for our purposes we use a constant ($1$) or polynomial. The derivatives are
\begin{subequations} \label{eq:grad_exp}
    \begin{align}
        \frac{\partial}{\partial d_n}  D(f,t) &= D(f,t) \phi_n(f),
    \end{align}
    \begin{align}
        \frac{\partial}{\partial d_{tn}}  D(f,t) &= t D(f,t) \phi_n(f),
    \end{align}
    \begin{align}
        \frac{\partial}{\partial r_n}  R(f,t) &= R(f,t) \phi_n(f),
    \end{align}
    \begin{align}
        \frac{\partial}{\partial r_{tn}}  R(f,t) &= t R(f,t) \phi_n(f),
    \end{align}
    \begin{align}
        \frac{\partial}{\partial h_n}  h(f,t) &= h(f,t) \phi_n(f).
    \end{align} 
    \begin{align}
        \frac{\partial}{\partial h_{tn}}  h(f,t) &= th(f,t) \phi_n(f).
    \end{align} 
\end{subequations}
The gradient (Eq.~\ref{eq:dl}) depends on the integral of these over a time interval, which are
\begin{subequations} \label{eq:grad_int_exp}
    \begin{align}
       -\int_{t_{k-1}}^{t_{k}} \frac{\partial}{\partial \alpha_d}  D(f,t) dt &= \frac{1}{\alpha_d^2} D(f,t_{k-1})\big(\alpha_d t_{k-1} - 1 + (1-\alpha_d t_{k}) \exp{(\alpha_d \Delta t)} \big),
    \end{align}
    \begin{align}
        -\int_{t_{k-1}}^{t_{k}} \frac{\partial}{\partial d_n}  D(f,t) dt &= \frac{1}{\alpha_d} D(f,t_{k-1})\big(1 - \exp{(\alpha_d \Delta t)} \big)\phi_n(f),
    \end{align}
    \begin{align}
        -\int_{t_{k-1}}^{t_{k}} \frac{\partial}{\partial \alpha_r}  R(f,t) dt &= \frac{1}{\alpha_r^2} R(f,t_{k-1})\big(\alpha_r t_{k-1} - 1 + (1-\alpha_r t_{k}) \exp{(\alpha_r \Delta t)} \big),
    \end{align}
    \begin{align}
        -\int_{t_{k-1}}^{t_{k}} \frac{\partial}{\partial r_n}  R(f,t) dt &= \frac{1}{\alpha_r} R(f,t_{k-1})\big(1 - \exp{(\alpha_r \Delta t)} \big)\phi_n(f),
    \end{align}
    \begin{align}
        -\int_{t_{k-1}}^{t_{k}} \frac{\partial}{\partial \alpha_h}  h(f,t) dt  &= \frac{1}{\alpha_h^2} h(f,t_{k-1})\big(\alpha_h t_{k-1} - 1 + (1-\alpha_h t_{k}) \exp{(\alpha_h \Delta t)} \big),~\text{and}
    \end{align}
    \begin{align}
        -\int_{t_{k-1}}^{t_{k}} \frac{\partial}{\partial h_n}  h(f,t) dt &= \frac{1}{\alpha_h} h(f,t_{k-1})\big(1 - \exp{(\alpha_h \Delta t)} \big)\phi_n(f).
    \end{align}
\end{subequations}
Where
\begin{subequations}
    \begin{align}
        \alpha_d &\equiv \sum_j d_{tj} \phi_j(f),
    \end{align}
    \begin{align}
        \alpha_r &\equiv \sum_j r_{tj} \phi_j(f),~\text{and}
    \end{align}
    \begin{align}
        \alpha_h &\equiv \sum_j h_{tj} \phi_j(f).
    \end{align}
\end{subequations}
If we wish to take the Gompertz term $\alpha\to0$ we can simply substitute $(1-e^{\alpha\Delta t})/\alpha \to -\Delta t$.

We can then use chain rule to get the derivatives in terms of model parameters,
\begin{subequations} 
    \begin{align}
    \frac{\partial \alpha_d}{\partial d_{tn}} &= \phi_n,
    \end{align}
    \begin{align}
    \frac{\partial \alpha_r}{\partial r_{tn}} &= \phi_n,~\text{and}
    \end{align}
    \begin{align}
    \frac{\partial \alpha_h}{\partial h_{tn}} &= \phi_n,
    \end{align}
\end{subequations}
%    \begin{align}
%       \frac{\partial}{\partial d_{tn}}  D(f,t) &= \frac{\partial}{\partial \alpha_{d}}  D(f,t) \frac{\partial \alpha_d}{\partial d_{tn}}
%    \end{align}
hence we simply multiply by $\phi_n$.

\section{The objective function is well-posed} \label{sec:consistent}
Our derived model is a stochastic model of $f$, Eq.~1. To fit the model to the data we derived a likelihood function based on transition rates. Here we show that in the mean field approximation we recover Eq.~1 and hence the objective function is well-posed.

Assuming transition rate models for damage, $D(f,t)$, and repair, $R(f,t)$, the probability of observing $p_d$ deficits in $p$ attributes is a Markov model of the form
\begin{align}
    Pr(p_d(t+\Delta t) | p_d(t)) &= \sum_{r=0}^{p_d(t)} Pr (\text{r repairs}) Pr(\text{d damage}) \nonumber \\
    &= \sum_{r=0}^{p_d(t)} \binom{p_d(t)}{r}\big( 1 - S_r \big)^r\big( S_r \big)^{p_d(t)-r} \binom{p-p_d(t)}{d}\big( 1 - S_d \big)^d\big( S_d \big)^{p-p_d(t)-d}
\end{align}
where $d=p_d(t+\Delta t)-p_d(t)+r = \Delta p_d + r$ is constrained. Also note that $S_r$ is the probability of not repairing and $S_d$ is the probability of not damaging during the time step $\Delta t$. 

If we take the limit $\Delta t\to 0$ we find
\begin{align}
Pr(p_d(t+dt) | p_d(t)) = 
\begin{cases}
  1-p_d(t)R(f,t) dt - (p-p_d(t))D(f,t)dt  & \text{if $\Delta p = 0$} \\
  p_d(t)R(f,t)dt & \text{if $\Delta p = -1$} \\
  (p-p_d(t))D(f,t)dt & \text{if $\Delta p = 1$} \\
  0 & \text{if $|\Delta p| > 1$} \label{eq:PrNd}
\end{cases}
\end{align}
where all $\mathcal{O}(dt^{2})\to 0$ (and higher powers). Keep in mind that $p_d(t)=pf(t)$ (by definition of $f$) and $0\leq p_d(t+dt) \leq p$ will constrain certain values. Note that due to the general rule for marginalizing, $Pr(p_d(t+dt)) = \langle Pr(p_d(t+dt) | p_d(t)) \rangle_{p_d(t)}$, thus we can use Eq.~\ref{eq:PrNd} to find the marginal average using $\langle \langle p_d(t+dt) \rangle_{|p_d(t)} \rangle_{p_d(t)}$ (i.e. average first over $Pr(p_d(t+dt) | p_d(t))$ then over $p_d(t)$). To be clear, $\langle x \rangle_y$ denotes averaging $x$ over all possible $y$.

Since the FI is defined by $p_d/p$, for constant $p$, we can use Eq~\ref{eq:PrNd} to compute the average FI, this yields the differential equation
\begin{align}
    \frac{d}{dt}  \bar{f} &= \langle (1-f)D(f) \rangle  - \langle f R(f) \rangle 
\end{align}
which to mean-field approximation (zeroth order) is
\begin{align}
    \frac{d}{dt} \bar{f} &\approx  (1 - \bar{f} ) D(\bar{f},t) - \bar{f}R(\bar{f},t) \label{eq:fmft}
\end{align}
which is exact when $\phi_i(f)\equiv 1$ i.e. $D$ and $R$ are constant; $\bar{f}\equiv \langle f \rangle$. Observe that this is exactly Eq.~1 with substitution $f\to\bar{f}$.

The next highest order includes additional terms,
\begin{align}
    \frac{d}{dt} \bar{f} &\approx  \bigg[1 - \bar{f} - \sum_i d_i \frac{d \phi_i (\bar{f})}{df} \text{Var}(f) \bigg] D(\bar{f},t) - \bigg[ \bar{f} + \sum_i r_i \frac{d \phi_i (\bar{f})}{df} \text{Var}(f) \bigg] R(\bar{f},t).
\end{align}
For sufficiently large number of attributes, the variance term will become small relative to the mean and we can ignore the higher-order corrections. Hence our objective function is well-posed as it approximates our desired dynamical equation.

\section{Stability analysis} \label{sec:si:stab} %new
We can summarize the approximate behaviour of the model by assuming small $f$ (relative to both $d_f$ and $r_f$). This permits us to better compare our results to other researchers'. The approximation is justified by the observed small parameter values (Figure~\ref{fig:par}), the small nullcline at young ages (Figure~4), and the population-level density that shows most individuals are measured at small $f$ (Figure~\ref{fig:dfdtpop}) -- all of these are evidence that the data are predominantly of low $f$ individuals.

Eq.~1 with the selected model is
\begin{align}
    \frac{df}{dt} &= (1-f)e^{d_0+d_f f + d_t t} - f e^{r_0 + r_f f + r_t t} \nonumber \\
    &\approx e^{d_0+ d_t t} - (e^{r_0 + r_t t} + e^{d_0+ d_t t} - d_f e^{d_0+ d_t t})f
\end{align}
to linear order in $f$. Thus the general form is
\begin{align}
    \frac{df}{dt} &= \gamma(t) + \alpha(t) f
\end{align}
for small $f$. $\gamma(t)$ represents unmitigated damage, which increases with age, and $\alpha(t)$ captures the stability of the feedback and also increases with age. For $\alpha(t) < 0$ the system is stable and $f$ will tend to recover from perturbations that increase it. We expect a homeostatic system to be stable and thus have $\alpha(t) < 0$, which we have previously observed consistently using a different analysis \cite{mallo}. For $\alpha(t) > 0$ the system is unstable and perturbations that increase $f$ will compound, driving the system to higher values of $f$. At $\alpha(t)=0$ the system is marginally-stable and is driven entirely by $\gamma(t)$. Importantly, we are able to estimate at which age $\alpha(t)$ will change sign and therefore stability. For HRS we estimate the age is $103.6\pm0.5$~years-old. These ages are considerably older than our estimates for the tipping point, where the small $f$ approximation begins to become unrealistic. Nevertheless the approximation still qualitatively captures the behaviour of the full model. Whereas the approximate model becomes unstable, the full model instead saturates at a value close to $f\approx 1$.

%add figures????

A loss of stability with age has been recapitulated by other researchers. Avchaciov \textit{et al.} inferred a transition from stable to unstable near the lifespan of mice \cite{Avchaciov2022}. Karin \textit{et al.}\cite{Karin2019-gf} and subsequent work by that lab \cite{Yang2023-xj} inferred a transition from stable damage regulation to saturated damage removal in mice senescent cells and E.~coli cell membranes, respectively. All of these works also performed dynamical analysis of aging data and reach a similar conclusion: that there is a stable phase of good health at young ages that ends with an unstable phase near the end of life -- consistent with our results.

\section{Simulation} \label{sec:si:sim}
We simulate using inverse-transform sampling (Section~\ref{sec:its}). We seed the simulation using the complete case data from the initial wave for each study, including each individual's starting deficits and baseline age. The simulation then generates artificial waves sampled with the same average frequency as the observed data.

\section{Inverse-transform sampling} \label{sec:its}
In inverse-transform sampling, a probability density function is exactly sampled by sampling from a uniform distribution on $(0,1)$ and then transforming using the associated inverse cumulative probability function. The sampling function is thus (e.g.\ see \cite{Bender2005-ag}),
\begin{subequations} \label{eq:expsample}
    \begin{align}
        \tau(f,t_k,t_{k-1}) &= \frac{1}{\alpha}\ln{\bigg( 1 - \frac{\alpha}{\Gamma(f,t_{k-1}) }\ln{(u)} \bigg)} + t_{k-1},~\text{where}
    \end{align}
    \begin{align}
        u \sim \text{uniform}(0,1).
    \end{align}
\end{subequations}
%    \begin{align}
        %u_{min} =
        %\begin{cases}
         %   0 ~ \text{if $\alpha > 0$, or}\\
         %   \exp{\bigg(\frac{\Gamma(f,t_{k-1})}{|\alpha|}\bigg)}~\text{if $\alpha < 0$}.
%        \end{cases}
%    \end{align}
%u_{max}$ strictly enforces that the event must have occurred after $t_{k-1}$. 
Accept-reject sampling \cite{Robert2010-ne} is then used to determine if a death event occurs within the sampling interval $t_{k}$ and $t_{k+1}$. For $\alpha <0$ all samples of $u < \exp{(\frac{\Gamma(f,t_{k-1})}{|\alpha|})}$ will give negative $\exp{(\tau)} < 0$, which are discarded assuming no event to preserve the correct distribution.

If $\alpha\equiv 0$ we have the exponential survival model \cite{Bender2005-ag} as a special case
\begin{subequations} \label{eq:expsample2}
    \begin{align}
        \tau(f,t_k,t_{k-1}) &= -\frac{\ln{(u)}}{\Gamma(f,t_{k-1})} + t_{k-1},~\text{where}
    \end{align}
    \begin{align}
        u \sim \text{uniform}(0,1). 
    \end{align}
\end{subequations}

Damage and repair events were sampled using the survival functions ($S_d$ and $S_r$, Eq.~\ref{eq:sdsr}). We sampled a random variable from a uniform distribution from $0$ to $1$, $u\in[0,1]$ and recorded an event if the random variable exceeded the respective survival function, $u_d <S_d$ or $u_r < S_r$.

The algorithm proceeds as follows:
\begin{itemize}
    \item For each individual, initialize set of binary variables each as $0$ or $1$.
    \item while $t < t_{max}$
    \begin{itemize}
        \item Increment $t_{k} = t_{k-1}+\Delta t$.
        \item Carry forward all previous values $\vec{b}(t_{k-1})$ and conditions (alive/dead).
        \item For each individual sample $\tau_h$. If $\tau_h < t_{k+1}$ kill that individual and set their time of death at $\tau_h$.
        \item For each individual and each deficit variable ($b=1$) sample $u_r\in[0,1]$ for repair. If $u_r > S_r(t_{k})$ then repair i.e. set $b=0$.
        \item For each individual and each repaired variable ($b=0$) sample $u_d\in[0,1]$ for damage. If $u_d > S_d(t_{k})$ then damage i.e. set $b=1$. 
    \end{itemize}
\end{itemize}

This algorithm is not formally exact since it doesn't permit multiple repair/damage cycles between time steps. However, as discussed in Eq.~\ref{eq:cycles}, these cycles are all order $(\Delta t)^2$ or higher, meaning that if $\Delta t$ is sufficiently small the algorithm becomes arbitrarily close to the true model. 
%The (exact) alternative is to use Eq.~\ref{eq:expsample} iteratively sample the next event, although this is much slower -- particularly for large populations which inevitably have rare individuals with many events -- and must then be retroactively converted to wave-format (i.e.\ measurements at specific sampling rates).

The lowest-order correction is proportional to $D R\Delta t^2$ compared to $1-D\Delta t$ or $1-R\Delta t$ thus the relative error (for small $\Delta t$) for these terms are
\begin{subequations}
\begin{align}
    \text{relative error} &= \frac{DR\Delta t^2}{{1-R\Delta t + DR\Delta t^2}} \approx DR\Delta t^2,~\text{and}
\end{align}
\begin{align}
    \text{relative error} &=\frac{DR\Delta t^2}{{1-D\Delta t + DR\Delta t^2}} \approx DR\Delta t^2.
\end{align}
\end{subequations}
E.g. if $DR\Delta^2 t=0.01$ then the algorithm has approximately a 1\% relative error for damage and repair effects (specifically $b(t_{k-1})=0 \to b(t_k)=0$ and $b(t_{k-1})=1 \to b(t_k)=1$ transitions). $D$ and $R$ are typically small, $\ll 1$ (Figure~\ref{fig:rates}). We found that $\Delta t = 0.2~\text{years}$ is small enough that the simulation no longer depends on the step size (not shown). We thus simulated at $\Delta t = 0.2~\text{years}$ and saved every approximately 10 iterations to produce simulated waves that emulate the observed data.

\section{The hazards for damage, repair and mortality are log-linear in age supporting our model assumptions.}

We modeled damage, repair and mortality as random event rates. Model selection (Section~\ref{sec:modelselection}) showed that the rates were fit well by log-linear relationships with age and the FI. Here we show that the a Cox proportional hazard model with flexible age and FI dependence produces log-linear dependence on age and FI for all three rates (damage, repair and mortality).

\begin{figure*}[!ht]
\centering
\includegraphics[width=0.99\textwidth]{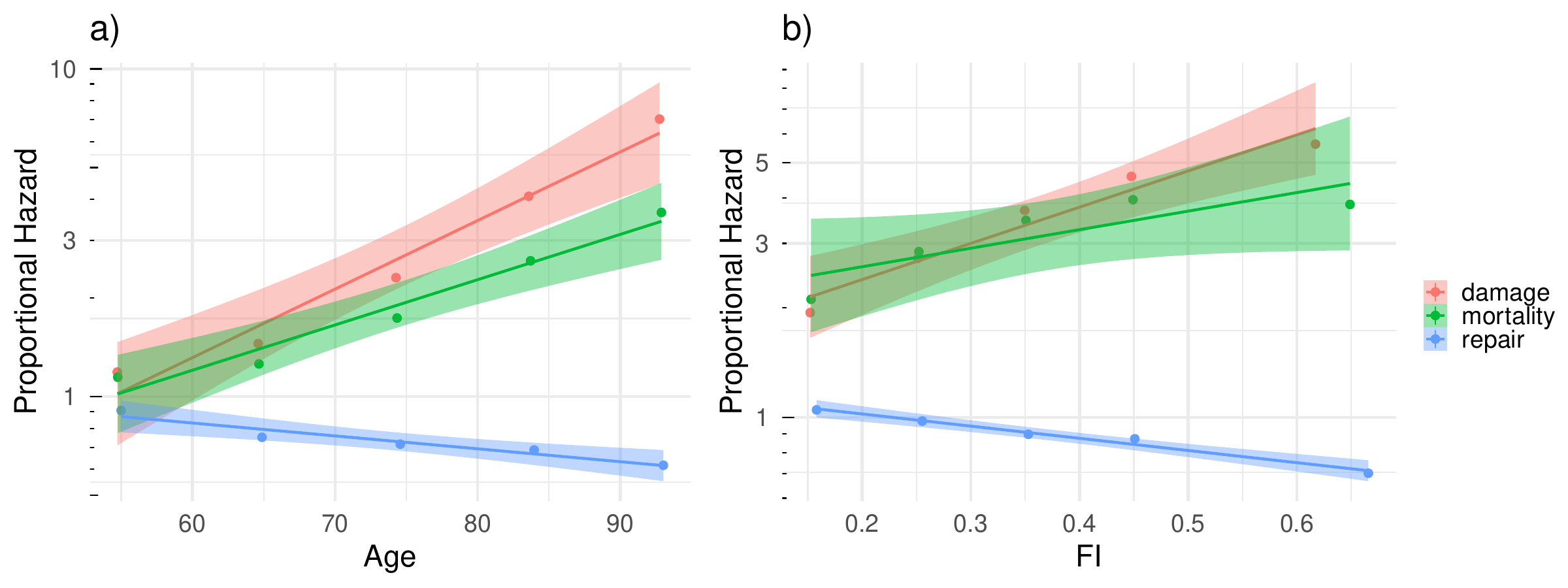} 
    \caption{\textbf{Cox proportional hazard shows rates are log-linear in age and FI -- HRS}. a) Hazards are log-linear with respect to age. b) Hazards are log-lineat with respect to FI.} \label{fig:fi_age_cox}
\end{figure*}

We fitted Cox proportional hazard models for binned age and FI then graphically investigated if the dependence was log-linear. We used age cuts  age cuts 40-50,50-60,60-70,70-80 and 90$+$. We used FI cuts <0.1,0.1-0.2,0.2-0.3,0.3-0.4,0.4-0.5 and 0.5$+$. Figure~\ref{fig:fi_age_cox} demonstrates that the dependence is close to log-linear in all cases, consistent with our model choice.

\section{Logistic regression also shows a tipping point around age 75, demonstrating the tipping point is not a model artifact} \label{sec:si:cp}
\begin{figure*}[!ht]
\centering
\includegraphics[width=0.99\textwidth]{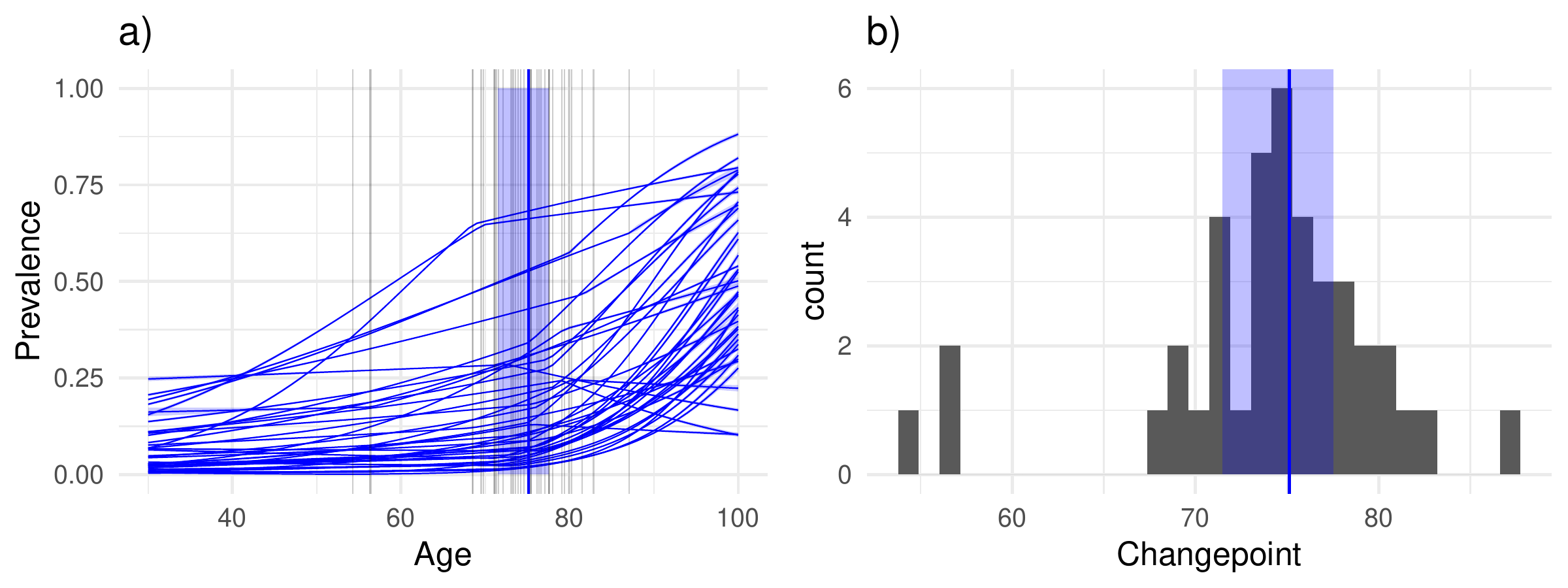} 
    \caption{\textbf{Logistic regression changepoint model for prevalence vs age shows breakpoints cluster at age 75 -- HRS}. Vertical blue line is median changepoint age, band is interquartile range. a) Prevalence curve estimates for each deficit (curves) together with changepoint estimates where the logistic-linear slope changes (vertical lines). b) Histogram of changepoint estimates show that they cluster around age 75. Changepoint models used BIC model selection (40/41 variables showed a changepoint.)} \label{fig:si:cp}
\end{figure*}

Here we show that the tipping point is not an artifact of our model by showing that an alternative, changepoint logistic regression model recapitulates the tipping point behaviour. Using the health attribute deficit prevalence curves we fitted a logistic-linear model of age while allowing a single change in slope at any age using the \texttt{segmented} package in \texttt{R}\cite{muggeo2008segmented}. We used the BIC to determine if the changepoint was justified by the data: comparing the uniform slope model to one changepoint model (40/41 variables showed a changepoint). We see that the changepoints cluster around age 70-80, Figure~\ref{fig:si:cp}. Outliers included living in a nursing home ($54\pm1$, nhmliv), had a nursing home stay within last 2 years ($56\pm1$, nrshom), had a hospital stay within last 2 years ($56\pm1$, hosp), and difficulty stooping/kneeling/crouching ($87\pm1$, stoopa). We estimate the changepoint location at $75\pm4$ years old (median$\pm$mean absolute deviation).

\section{The tipping point flattens but does not move among sub-populations of progressively worse health. Individuals living in a nursing home may already be past their tipping point.}
Here we use both our primary model (described in the main text) and the logistic regression changepoint model (described in Section~\ref{sec:si:cp}) to fit sub-populations and investigate whether and how the tipping point changes. We find that the tipping point location changes little between sub-populations, always in the range 70-80, but it does flatten with worsening health and advanced age -- as shown by the lung disease and nursing home sub-populations that are older and have higher FI.

\begin{figure*}[!ht]
\centering
\includegraphics[width=0.99\textwidth]{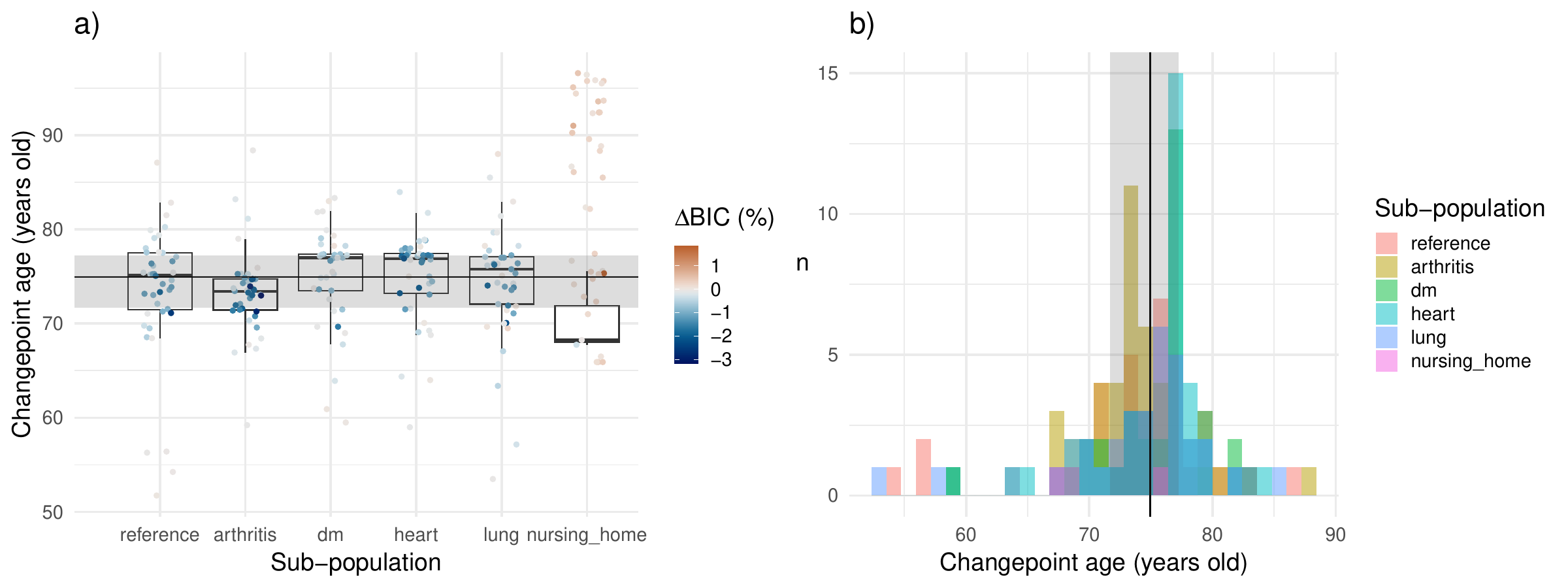} 
    \caption{\textbf{Logistic regression changepoint model shows tipping point age doesn't move with age but may be entirely gone in nursing home sub-population -- HRS}. a) All variable changepoint locations (points) with boxplot of only those that improved model performance (Delta BIC < 0). Changepoint models with worse performance have Delta BIC > 0 and are indicative of no changepoint being found. The nursing home sub-population had only 3 of the 41 variables (self-reported health, DM status and gets help toileting) that showed a changepoint, supporting the interpretation that this sub-population is already broadly past their tipping point. b) Histogram of changepoints with Delta BIC < 0. Black line and band are median and interquartile range of all changepoints with Delta BIC < 0. People living in nursing homes have notably fewer changepoints in deficit prevalences, consistent with our dynamical model.} \label{fig:si:cpsubpop}
\end{figure*}

In Figure~\ref{fig:si:cpsubpop} we use the logistic regression changepoint model and observe that the age is consistently near 75 regardless of the sub-population. We do, however, notice that the changepoint is not supported by the data in almost all of the nursing home variables, as indicated by BIC. This supports a picture of the population living in nursing homes as being an end-state and having generally already crossed this tipping point (median age: 87).

\begin{figure*}[!ht]
\centering
\includegraphics[width=0.85\textwidth]{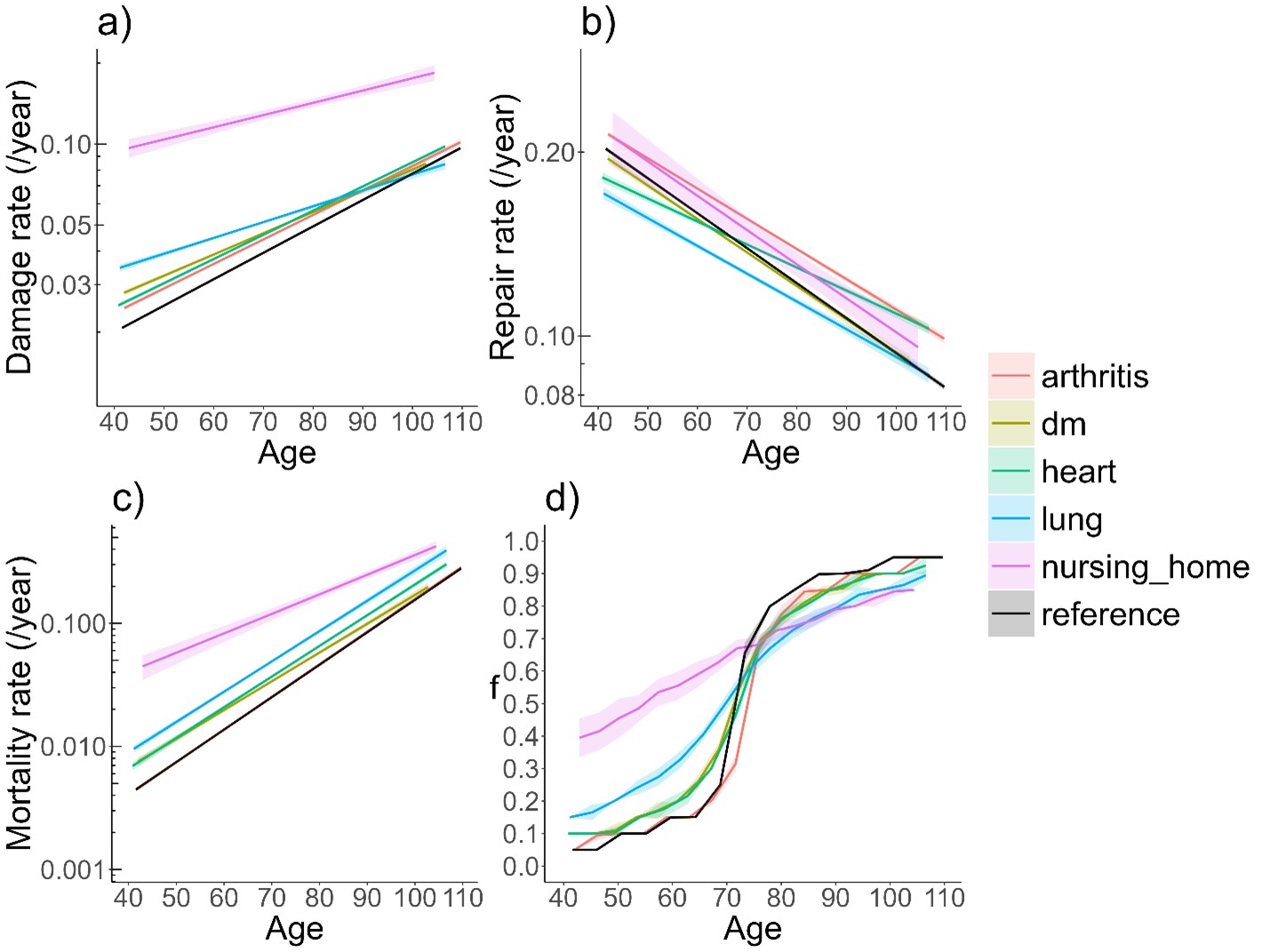} 
    \caption{\textbf{Rate estimates for sub-populations of interest -- HRS}, compared to the reference fit using all individuals. Living in a nursing home (nhmliv=1) is associated with higher damage rate (a), mortality rate (c), and a flattened nullcline that removes the tipping point shape in (d). Curves are for reference FI=0.20 a typical value seen in all sub-populations). Bands are error bars from bootstrap replicates (100 for reference population, 25 for all other populations).} \label{fig:si:subpop}
\end{figure*}

In Figure~\ref{fig:si:subpop} we use our primary model and see that the tipping point location doesn't move but does flatting, such that it is completely removed for individuals living in a nursing home -- instead the nullcline increases linearly with age. This further supports the interpretation of the nursing home being past the tipping point.

\section{Repairable variables have markedly higher repair rates but qualitatively similar behaviour to the overall variable set}
Here we consider the effect of fitting to only repairable health variables, this shows qualitatively similar behaviour but different parameter estimates particularly for the repair rate. We included both repairable and unrepairable health variables in our model to reflect how the FI is typically created in practice\cite{Theou2023-aw}. For example, several variables are ``ever diagnosed with...'' that cannot repair by definition, it can only reverse due to a noisy response. In this section we report fit estimates after excluding all ``ever...'' diagnoses in addition to all ``...within previous 2 years'' variables to exclusively look at variables that can repair within any time frame, using HRS.

\begin{figure*}[!ht]
\centering
\includegraphics[width=0.85\textwidth]{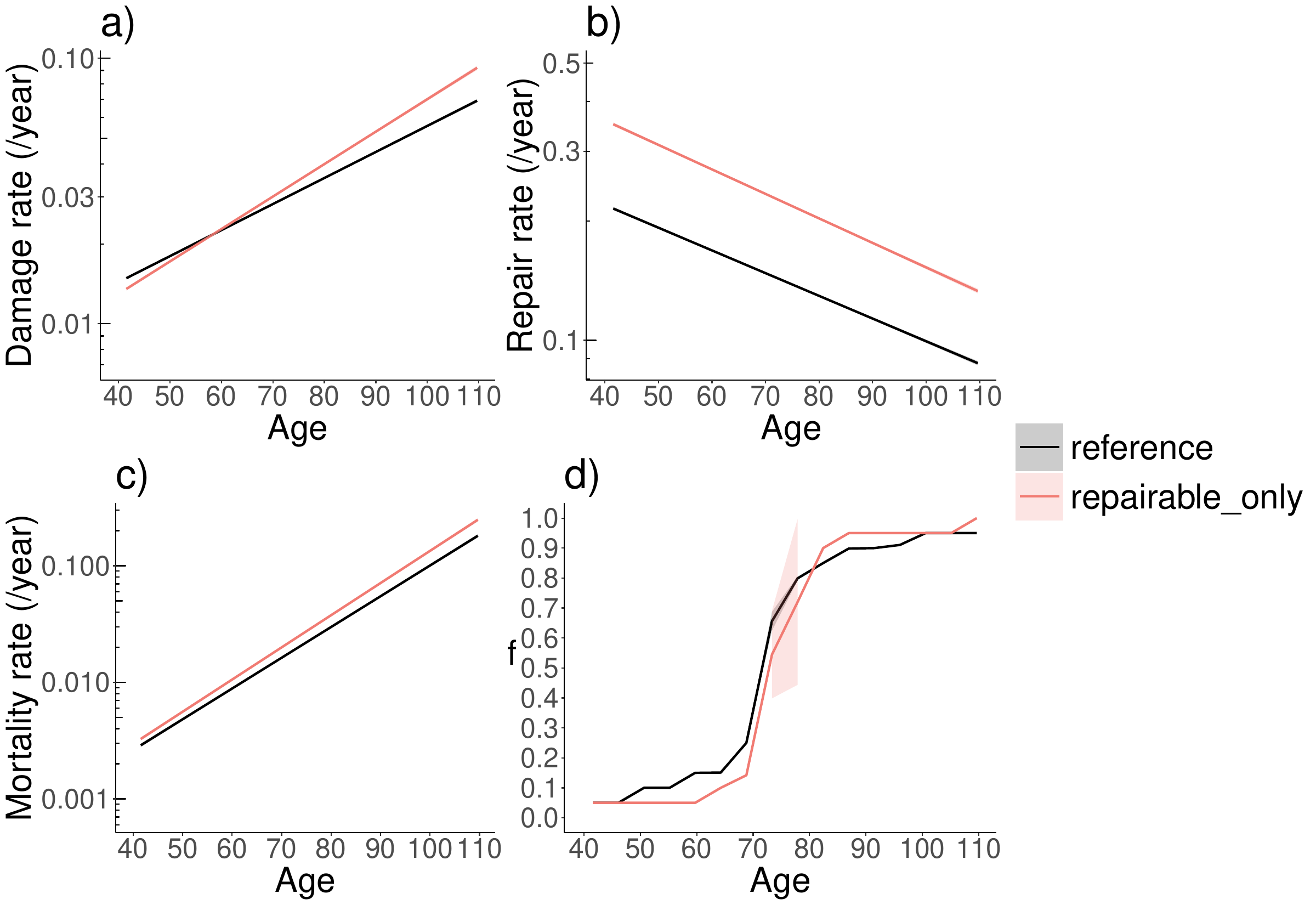} 
    \caption{\textbf{Parameter estimates for exclusively repairable health attributes -- HRS}, compared to the reference fit using all variables in Table~\ref{tab:hrsfi}. We see strong agreement in a) damage rate, c) mortality rate, and d) tipping point behaviour between the two variable sets. Notably, the repair rate is much higher in the reparaible-only set (red) compared to the complete set (black). Curves are for reference FI=0.10 (typical value; median=). Bands are error bars from 100~bootstrap replicates (too small to see in a-c).} \label{fig:si:repariable}
\end{figure*}

In Figure~\ref{fig:si:repariable} we compare the rate curves for the repairable-only variable set versus the default set used in the main text. We see that the damage and mortality rates, and the nullcline, are very similar between the two variable sets. The repair rate, however, is much larger in the repairable-only variable set, by a roughly constant amount. This is consistent with repairs occurring uniformly more often with age, that makes sense given there are relatively more deficits that can repair in this variable set.

\end{document}